\documentclass{aa}  
\usepackage{graphicx}
\usepackage{txfonts}
\usepackage{verbatim} 
\usepackage[normalem]{ulem}
\usepackage{hyperref}

\usepackage[]{xcolor}

\newcommand{\mlpd}[2]{\bgroup\markoverwith{\textcolor{red}{\rule[0.5ex]{2pt}{1pt}}}\ULon{#1} \textcolor{blue} {#2}}

\def\halpha{H$\alpha$}

\def\hdu18{HD\,189733\,b}
\def\hd20{HD\,209458\,b}
\def\gj34{GJ\,3470\,b}
\def\gju12{GJ\,1214\,b}
\def\w69{WASP-69\,b}
\def\wa76{WASP-76\,b}
\def\hatp32{HAT-P-32\,b}

\def\co2{CO$_2$}
\def\ch4{CH$_4$}
\def\h2{H$_{2}$}
\def\h2o{H$_2$O}
\def\hep{He$^{+}$}
\def\hes{He(1$^{1}$S)}
\def\het{He(2$^{3}$S)}
\def\he_rat{He\,{\sc i}$_{\lambda 10833}$/He\,{\sc i}$_{\lambda 10832}$}
\def\mlr{$\dot M$}
\def\lya{Ly$\alpha$}
\def\ha{H$\alpha$}

\def\kms{km\,s$^{-1}$}

\def\rp{$R_{\rm P}$}
\def\rj{$R_{\rm Jup}$\ }
\def\mj{$M_{\rm Jup}$\ }
\def\rxuv{$R_{\rm {XUV}}$}

\def\gs{g\,s$^{-1}$}

\begin{document} 

\title{
Characterisation of the upper atmospheres of HAT-P-32\,b, WASP-69\,b, GJ\,1214\,b, and WASP-76\,b
through their He {\sc i} triplet absorption}
\titlerunning{Upper atmospheres of HAT-P-32\,b, WASP-69\,b, GJ\,1214\,b, and WASP-76\,b.}
\author{
M.~Lamp{\'o}n\inst{1}, 
M.~L\'opez-Puertas\inst{1}, 
J.~Sanz-Forcada\inst{2}, 
S.~Czesla\inst{3},  
L.~Nortmann\inst{4},
N. Casasayas-Barris\inst{5},
J. Orell-Miquel\inst{6,7},
A.~S\'anchez-L\'opez\inst{5}, 
C.~Danielski\inst{1}, 
E.~Pall{\'e}\inst{6,7}, 
K.~Molaverdikhani\inst{8,9,10}, 
Th.~Henning\inst{9},
J.\,A.~Caballero\inst{2},
P.\,J.~Amado\inst{1},
A.~Quirrenbach\inst{10}, 
A.~Reiners\inst{5}, and 
I.~Ribas\inst{11,12} 
}
\institute{Instituto de Astrof{\'i}sica de Andaluc{\'i}a (IAA-CSIC), Glorieta de la Astronom{\'i}a s/n, 18008 Granada, Spain
\and
Centro de Astrobiolog{\'i}a (CSIC-INTA), ESAC, Camino bajo del castillo s/n, 28692 Villanueva de la Ca{\~n}ada, Madrid, Spain
\and
Thüringer Landessternwarte Tautenburg, Sternwarte 5, D-07778 Tautenburg, Germany
\and
Institut f{\"u}r Astrophysik, Georg-August-Universit{\"a}t, Friedrich-Hund-Platz 1, 37077 G{\"o}ttingen, Germany
\and
Leiden Observatory, Leiden University, Postbus 9513, 2300 RA, Leiden, The Netherlands
\and
Instituto de Astrof{\'i}sica de Canarias (IAC), Calle V{\'i}a L{\'a}ctea s/n, 38200 La Laguna, Tenerife, Spain
\and
Departamento de Astrof{\'i}sica, Universidad de La Laguna, 38026  La Laguna, Tenerife, Spain
\and
Universit{\"a}ts-Sternwarte, Ludwig-Maximilians-Universit{\"a}t München, Scheinerstrasse 1, D-81679 M{\"u}nchen, Germany
\and 
Exzellenzcluster Origins, Boltzmannstrasse 2, 85748 Garching, Germany
\and
Landessternwarte, Zentrum f{\"u}r Astronomie der Universit{\"a}t Heidelberg, K{\"o}nigstuhl 12, 69117 Heidelberg, Germany
\and
Institut de Ci\`encies de l'Espai (CSIC-IEEC), Campus UAB, c/ de Can Magrans s/n, 08193 Bellaterra, Barcelona, Spain
\and
Institut d'Estudis Espacials de Catalunya (IEEC), 08034 Barcelona, Spain
}
\authorrunning{M. Lampón et al.}
\date{Received 08 December 2022 / Accepted 17 March 2023}
 
\abstract{
 Characterisation of atmospheres undergoing photo-evaporation is key to understanding the formation, evolution, and diversity of planets.
However, only a few upper atmospheres that experience this kind of hydrodynamic escape have been characterised.  
Our  aim is to characterise the upper atmospheres of the hot Jupiters \hatp32 and \w69,
the warm sub-Neptune \gju12, and the ultra-hot Jupiter \wa76 through high-resolution observations of their He\,\textrm{I} triplet absorption.
  In addition, we also reanalyse the warm Neptune \gj34 and the hot Jupiter \hdu18.
We used a spherically symmetric 1D hydrodynamic model coupled with a non-local thermodynamic equilibrium model for calculating the He\,\textrm{I}\, triplet distribution along the escaping outflow. Comparing synthetic absorption spectra with observations, we constrained the main parameters of the upper atmosphere of these planets and classify them according to their hydrodynamic regime.
Our results show that 
\hatp32\ photo-evaporates at (130\,$\pm$\,70)$\times$10$^{11}$\,\gs\ with a  
hot (12\,400\,$\pm$\,2900\,\,K) upper atmosphere; 
\w69\   loses its atmosphere at (0.9\,$\pm$\,0.5)$\times$10$^{11}$\,\gs\ and 5250\,$\pm$\,750\,K; and \gju12, with a relatively cold outflow of  3750\,$\pm$\,750\,K,  
photo-evaporates at (1.3\,$\pm$\,1.1)$\times$10$^{11}$\,\gs. For \wa76, its weak absorption prevents us from constraining its temperature and mass-loss rate significantly; we obtained ranges of 6000-17\,000\,K and 
23.5\,$\pm$\,21.5\,$\times$10$^{11}$\,\gs.
Our reanalysis of \gj34\ yields colder temperatures, 3400\,$\pm$\,350\,K, 
but practically the same mass-loss rate as in our previous results. 
Our   reanalysis of \hdu18 yields a slightly higher mass-loss rate, (1.4\,$\pm$\,0.5)$\times$10$^{11}$\,\gs, and temperature, 12\,700\,$\pm$\,900\,K compared to previous estimates.
We also found that \hatp32, \w69, and \wa76 undergo hydrodynamic escape in the recombination-limited regime, and 
that \gju12\ is in the photon-limited regime. Our results support that photo-evaporated outflows tend to be very light, H/He\,$\gtrsim$\,98/2. 
The dependences of the mass-loss rates and temperatures of the studied planets on the respective system parameters (X-ray and ultraviolet stellar flux, gravitational potential) are well explained by the current hydrodynamic escape models.}

\maketitle
%

\section{Introduction} \label{Intro}

It seems common that planets, including Solar System terrestrials,
experience hydrodynamic atmospheric escape at some stages of their lifetime   \citep[see e.g.][]{Watson1981,Yelle_2004,Garcia_munoz_2007,Lammer_2020a,Lammer_2020b}.
This process, by which massive outflows of gas escape from the planet's upper atmosphere, is called photo-evaporation when it is triggered by stellar irradiation, and it plays a central role in planetary evolution \citep[see e.g.][]{Tian2005,jackson_2012,Owen_2017}. Photo-evaporation can be responsible for shaping features such as the radius desert and the radius valley, significantly impacting the observed planetary demography and diversity
\citep[e.g.][]{Owen_2013,Lopez_2013,Malsky_2020}.

Thus, the atmospheric characterisation of planets undergoing hydrodynamic escape
provides information about the planetary properties and the escaping mechanism \cite[e.g.][]{Lampon_2021b}, and hints about their formation history and evolutionary pathways 
\citep[see e.g.][]{Jin_2018,Owen_2020,Mordasini2020}. 
This characterisation 
requires constraining the main parameters of the planetary upper atmospheres, namely  the mass-loss rate (\mlr), the temperature of the upper atmosphere ($T$), H/He number ratio (hereafter called ratio), and ionisation profile, which are frequently largely degenerate.   
Among these parameters, it is especially important to constrain the H/He ratio. Knowledge of this
ratio (i) significantly reduces the $T$--\mlr\ degeneracy \citep[see e.g.][]{Lampon2020}; (ii) 
gives hints to the atmospheric evolution and about the composition and global processes of the middle-lower atmosphere of the planet \cite[see e.g.][]{Hu_2015,Malsky_2020,Salpeter1973,Stevenson1975,Stevenson1980,Wilson2010}; and (iii) 
contributes to understanding the detectability of the He I triplet absorption \citep{Lampon2020,Lampon_2021a,Fossati_2022}. 
In addition, constraining the H/He ratio is essential for supporting or excluding the hypothesis that the outflow of planets undergoing photo-evaporation tend to have higher H/He ratios than the usually expected solar-like $\sim$\,90/10 value \citep{Lampon_2021b}. 
To  date, the H/He ratio of the upper atmosphere has been derived for only five planets, of which four show a H/He ratio of $\gtrsim$ 97/3: \hd20\ \citep{Lampon2020,Khodachenko_2021b}, \hdu18\ \citep{Lampon_2021a,Rumenskikh_2022}, \gj34  \citep{Shaikislamov_2021,Lampon_2021a}, WASP-52\,b \citep{dongdong_2022}. Only for WASP-107\,b has a solar-like value   been derived \citep{Khodachenko_2021}.
Moreover, \cite{Dos_santos_2021} and \cite{Fossati_2022} point out that HAT-P-11\,b and WASP-80\,b, respectively, have upper atmospheres with very high H/He ratios.    

The hydrodynamic escape regime is another important aspect of atmospheres experiencing photo-evaporation
\citep{Murray_Clay_2009,Owen_2016,Lampon_2021b}. 
General properties of hydrodynamic atmospheric escape are subsumed into three different regimes: energy-limited (EL), recombination-limited (RL), and photon-limited (PL) \cite[see e.g.][]{Owen_2016,Lampon_2021b}.  
Although these regimes were theoretically predicted by \cite{Murray_Clay_2009} and \cite{Owen_2016}, evidence backed by observational results is very recent \citep{Lampon_2021b}. Consequently, so far only a few planets have been classified into these regimes: the hot Jupiters \hd20\ and \hdu18, which are in the EL and RL regimes, respectively \citep{Lampon_2021a}, and  the warm Neptune \gj34 and the sub-Neptunes HD\,63433\,b, HD\,63433\,c, and \gju12,\ which are in the PL regime \citep[][]{Lampon_2021a,Zhang_2022a,Orell-Miquel_2022}.

To date, the main problem of atmospheric characterisation is the scarcity of observations. 
Since the first time this process was observed, in \hd20\ by \cite{VidalMadjar2003}, signatures of hydrodynamic escape have been observed in only a few more exoplanets.
The observations are limited to neutral H lines (mainly \lya\ and \ha),   measured in \hd20, \hdu18, GJ\,436\,b, \gj34, KELT-20\,b, KELT-9\,b, and \hatp32\, \citep[][]{VidalMadjar2003,Lecavelier_des_Etangs_2012,Bourrier2018,Ehrenreich_2011,Kulow_2014,Casasayas_2018,Yan_2018,Wyttenbach_2020, Czesla_2022}; some ultraviolet metal lines (e.g. O I and C II)   in \hd20 and  \hdu18  \citep[][]{VidalMadjar2004,Ben_Jaffel_2013,Garcia_Munoz2021};
some lines in the near-UV (e.g. Fe II and Mg II)   in WASP-12\,b, WASP-121\,b, and \hd20\,\citep{Fossati_2010,Haswell_2012,Sing_2019,Cubillos_2020};
and   the metastable He I 2$^3$S–2$^3$P lines,\footnote{At vacuum wavelengths of 10 832.06, 10 833.22, and 10 833.31\,\AA, often referred to by their air wavelength as the He 10 830\,\AA\ triplet.} 
hereafter \het, in 
WASP-107\,b\ \citep{Spake_2018,Allart_2019,Kirk_2020,Spake_2021}, 
\w69\ \citep{Nortmann2018,Vissapragada_2020,Khalafinejad_2021}, 
HAT-P-11\,b\ \citep{Allart_2018,Mansfield_2018}, 
\hd20\ \citep{Alonso2019}, 
\hdu18\ \citep{Salz2018,guilluy_2020}, \gj34\ \citep{Palle2020,Ninan_2020}, 
WASP-52\,b\ \citep{Vissapragada_2020,Kirk_2022}, 
\wa76\ \citep{Casasayas_2021},
HAT-P-18\,b\ \citep{Paragas_2021},
\hatp32\ \citep{Czesla_2022},  
\gju12\ \citep{Orell-Miquel_2022},
TOI\,560\,b, TOI\,1430.01, TOI\,1683.01, and TOI\,2076\,b 
\citep{Zhang_2022b,Zhang_2022c}.
However, the observations in \wa76\ and \gju12 are still to be confirmed
(see \cite{Casasayas_2021} for \wa76\, and \cite{Spake_2022} for \gju12).

The ultra-hot Jupiter \wa76, the hot Jupiter \hatp32, and the warm sub-Neptune \gju12,  
are suitable for photo-evaporation studies as they show a \het\ absorption (tentative in the case of \gju12\ and \wa76) compatible with extended atmospheres.
Moreover, these planets are especially interesting due to their special features: \wa76\ is the only ultra-hot Jupiter where \het\ has been detected, \hatp32 is the planet with the largest \het\ absorption
observed  to date, and \gju12
(together with the recently measured TOI\,560\,b, TOI\,1430.01, TOI\,1683.01, and TOI\,2076\,b) 
is the smallest planet with observed \het\ absorption.
\cite{Czesla_2022} and \cite{Orell-Miquel_2022} modelled their respective \het\ absorption,  
but not in much detail. Thus, a more comprehensive study of the parameter space is required in order to infer more accurate constraints on the main parameters of their upper atmospheres.

The hot Jupiter \w69\ is another interesting exoplanet with an extended upper atmosphere, as shown by the \het\ high-resolution spectra measured by \cite{Nortmann2018}. 
Recently, \cite{Vissapragada_2020}\ also observed \het\ in this planet, but using a photometric technique. 
\cite{Vissapragada_2020} and \cite{Wang_2021} modelled the \het\ in order to constrain some of the main parameters of its upper atmosphere.
However, an extended study exploring the temperature and mass-loss rate ranges, including the H/He ratio and the derivation of its hydrodynamic escape regime, has not been performed yet.

In this work our aim is to characterise the upper atmosphere of the exoplanets \hatp32, \w69, \gju12, and \wa76  by analysing their \het\ high-resolution spectra.
We used the same methods as in \cite{Lampon2020,Lampon_2021a,Lampon_2021b}, but here we calculate synthetic spectra more accurately (as shown in Sect.\,\ref{model}). 
Further, as the X-ray and ultraviolet (XUV) irradiation of \gj34 and \hdu18\ have been recently revised, we also reanalyse these planets.  
We used a spherically symmetric 1D hydrodynamic model coupled with a non-local thermodynamic equilibrium (non-LTE) model to calculate the \het\ distribution. We should note that our model, as it is a 1D model, does not account for other parameters that might affect the strength of the \het\ absorption signal, being the stellar wind the most important \citep[see e.g.][]{Vidotto2020,Khodachenko_2021,Fossati_2022}. An estimation of this effect has, nevertheless, been performed (see Sect.\,\ref{sw}). 
Applying a high-resolution radiative transfer code that includes thermal, turbulent, and wind broadening, we compute synthetic absorption spectra, and by comparing with observations we  
constrain the main parameters of the upper atmosphere. 
In addition, by analysing the hydrogen recombination and advection processes, we determine the hydrodynamic escape regime of these exoplanets. 
Further, in order to infer more general properties, we compare the results for these planets, including the hot Jupiter \hd20, which also undergoes photo-evaporation and has been previously studied with the same method.

\begin{figure*}[htbp]
\includegraphics[angle=90, width=1.\columnwidth]{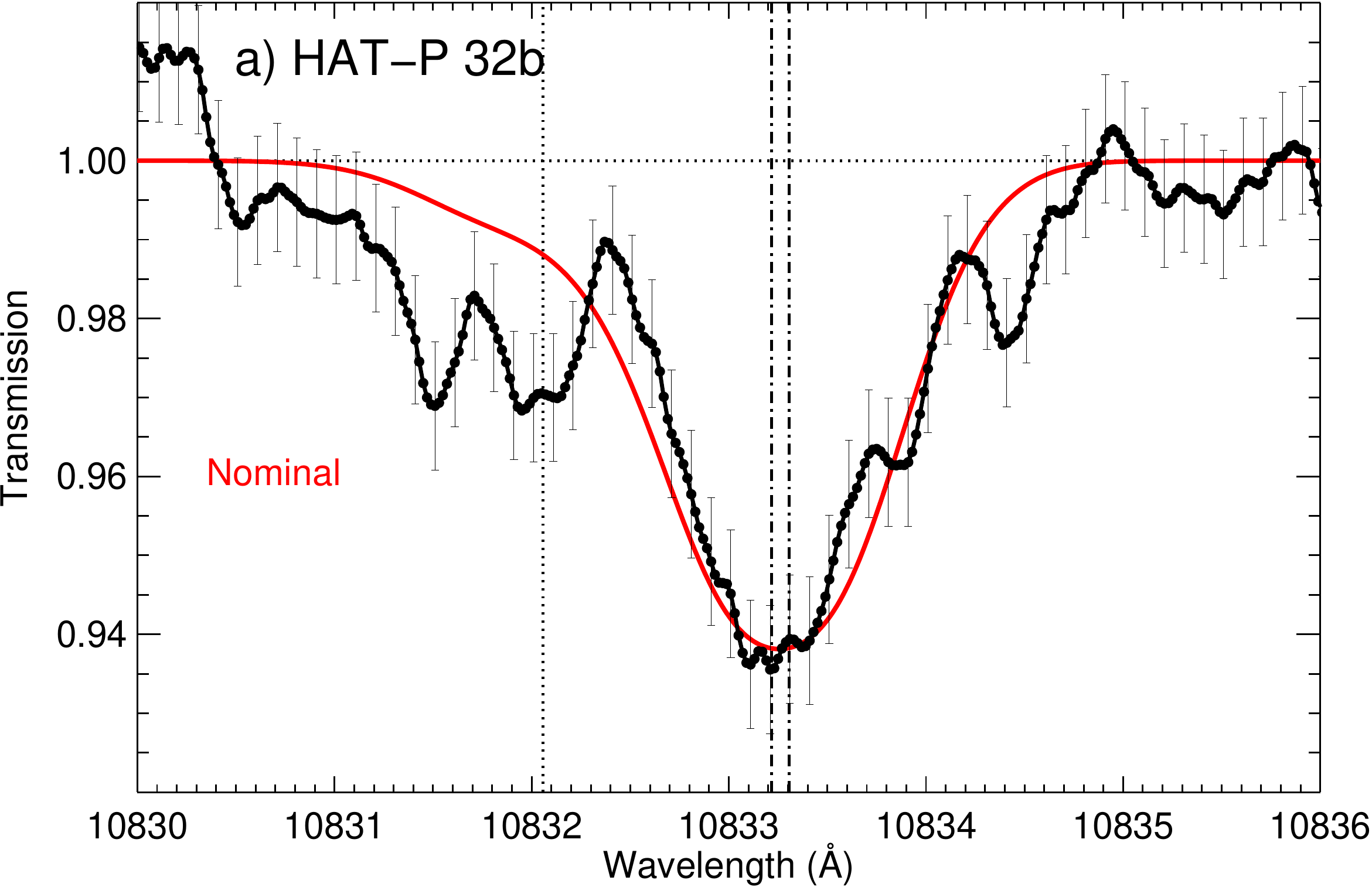}
\includegraphics[angle=90, width=1.\columnwidth]{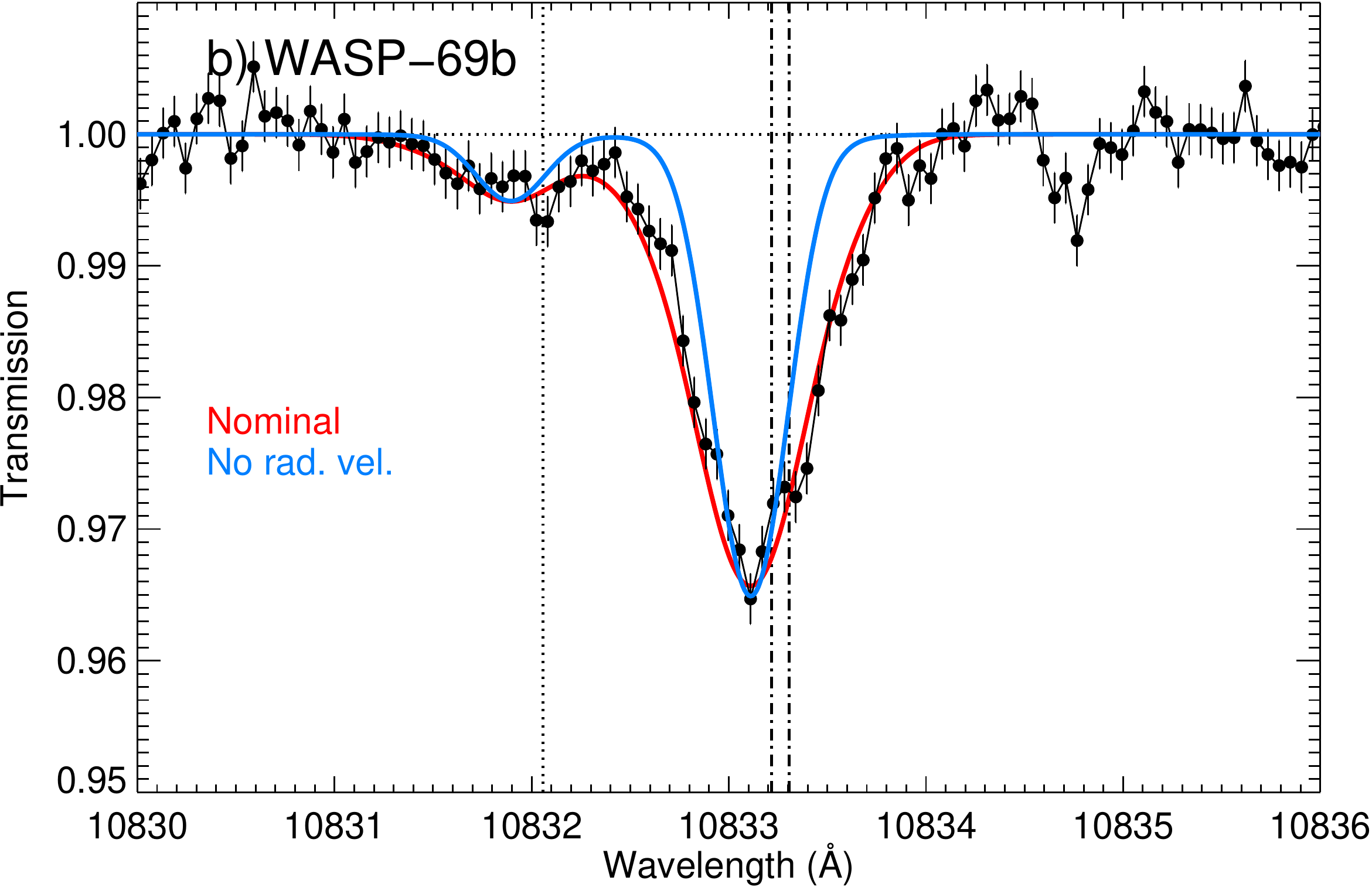}\\[0.2cm]
\includegraphics[angle=90, width=1.\columnwidth]{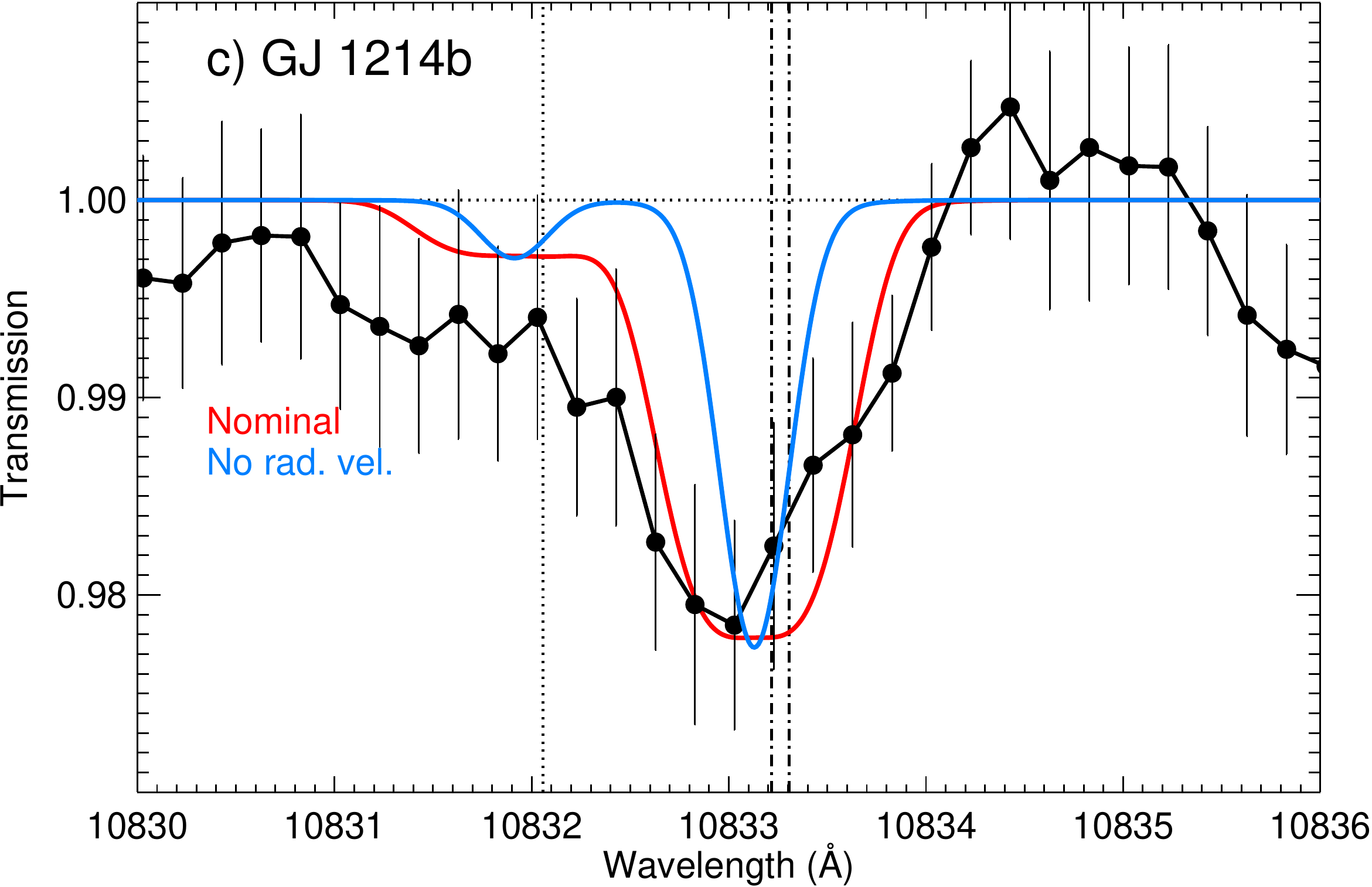} 
\includegraphics[angle=90, width=1.\columnwidth]{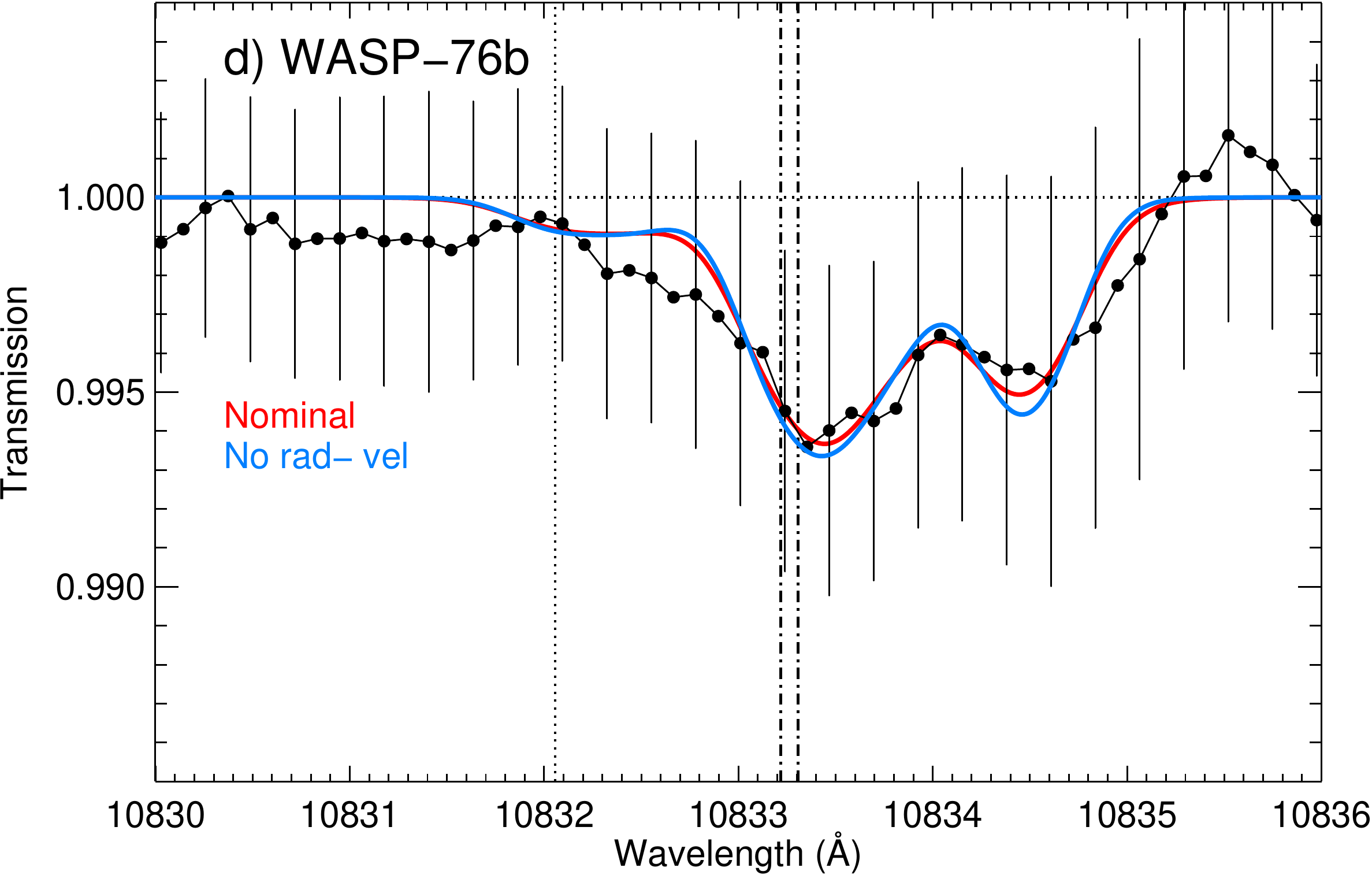} 
\caption{
Spectral transmission of the He triplet  
for several planets. For \hatp32 (a), \w69 (b), \gju12 (c), and \wa76 (d) (with different y-axis scales). Data points and their respective error bars are shown in black (adapted from \cite{Czesla_2022}, \cite{Nortmann2018}, \cite{Orell-Miquel_2022}, and \cite{Casasayas_2021}, respectively). Wavelengths are given in vacuum. The best-fit simulations are shown with red curves. For \hatp32, the best fit corresponds to a temperature of 12\,500\,K, a mass-loss rate of 1.4\,$\times$10$^{13}$\,\gs,\ and a H/He mole-fraction of 99/1; for \w69, the fit corresponds to a temperature of 5375\,K, $\dot{M}$=$10^{11}$\,\gs\ and H/He=98/2; for \gju12, a temperature of 3625\,K, $\dot{M}$=$6.3\times 10^{10}$\,\gs,\ and H/He=98/2; and for \wa76, a temperature of 9500\,K, $\dot{M}$=$1.2\times 10^{12}$\,\gs,\ and H/He=98/2. Other transmission models are shown (in blue) when the gas radial outflow velocity was not included. 
The positions of the helium lines are given by vertical dotted (weak) and dash-dotted (strong) lines.}  
\label{absorption}
\end{figure*}

The paper is organised as follows. Section \ref{Obs} summarises the \het\ observations of \hatp32, \w69, \gju12, and \wa76. Section \ref{model} briefly describes the methods we used for modelling the \het\ distribution,   computing the synthetic absorption spectra, and constraining the main parameters of the upper atmospheres. Section \ref{results} shows and discusses the results obtained. The main conclusions are summarised in Sects. \ref{sec_comparison} and \ref{summary}.

\section{Observations of the He {\sc i} triplet absorption} 
\label{Obs}

In this section we summarise the \het\ measurements that we use for our analysis of  \hatp32, \w69, \gju12, and \wa76.
The original observations are described by \cite{Czesla_2022}, \cite{Nortmann2018}, \cite{Orell-Miquel_2022}, and \cite{Casasayas_2021}, respectively. 
All these measurements were taken with the high-resolution
spectrograph CARMENES,\footnote{Calar Alto high-Resolution search for M dwarfs with Exoearths with
Near-infrared and optical Échelle Spectrographs, at the 3.5 m Calar Alto
Telescope \citep{Quirrenbach14}.}
and are shown in  Fig.\,\ref{absorption}. 

For \hatp32, the analysed \het\ transmission spectrum corresponds to
the combination of two mid-transit spectra.  
It shows a He excess absorption at the level of $\sim $6\%, which is the highest observed in a planet to date.
In contrast to most planets, 
the stronger \het\ line, formed by the two unresolved lines centred at 10\,833.22\,\AA\ and 10\,833.31\,\AA, does not show any significant shift in this planet. The weaker \het\ line, centred at 10\,832.06\,\AA, is very deep at mid-transit (see Fig.\,\ref{absorption}a), but appears substantially reduced or absent in other transit phases \citep{Czesla_2022}.
In addition, the mid-transit spectrum shows a significant excess absorption near 10831.5\,\AA, which corresponds to a blueshift of\,$\sim$\,50\,\kms.
However, this signal does not appear during the pre-transit, ingress, and start, and is faint in the egress \citep{Czesla_2022}, which then raises questions about its nature. 
As shown by \cite{Czesla_2022}, the spectra show an important redshifted absorption of $\sim$\,25\,\kms\ in the pre-transit, which increases in the ingress phase and disappears in the egress. Absorption in the egress is much weaker than during ingress, and no post-transit is observed. These features show a complex spatial distribution of the gas, compatible with a redshifted component in front of the trajectory of the planet, as suggested by \cite{Czesla_2022}.

For \w69, we study the \het\ transmission spectrum averaged over the time-resolved spectra taken between the second and third transit contacts (T2 and T3, respectively) from two different transits \citep{Nortmann2018}.
The stronger \het\ line shows a He excess absorption at the level of 3.59\,$\pm$\,0.19\% with a blueshift of $-$\,3.58\,$\pm$\,0.23\,\kms. The weaker \het\ line is shallow, at the level of $\sim$\,0.4\,\%, as shown in Fig.\,\ref{absorption}b.  
\cite{Nortmann2018} also observed a faint ingress, 
$\sim$\,1\,\%, with a redshift of 1.4\,$\pm$\,0.9\,\kms, and a stronger post-transit, 
$\sim$\,1.8\,\%, with an averaged blueshift of $-$10.7$\pm$1.0\,\kms. 
However, they did not observe pre-transit absorption. Interestingly, these features are compatible with a blueshifted tail behind the trajectory of the planet, which is just the opposite of the \hatp32\ scenario.

In the case of \gju12, we analysed the \het\ transmission spectrum averaged between the first and the fourth contacts (T1 and T4, respectively) from one transit.
The stronger \het\ line shows a He excess absorption at the level of 2.1\,$_{-0.5}^{+0.45}$\% with a blueshift of $-$\,4\,$\pm$\,4\,\kms\ (see Fig.\,\ref{absorption}c). The weaker \het\ line is at the level of $\sim$\,0.6\,\%.
The pre- and post-transits do not show any significant absorption, although the signal-to-noise ratio (S/N)
could be too low for such a detection.
This detection has to be confirmed, as only one transit was measured in good telluric conditions \citep{Orell-Miquel_2022}.

For \wa76, we analyse the \het\ transmission spectrum averaged between T1 and T4 from two different transits.
The stronger \het\ line is 0.52\,$\pm$\,0.12\% deep. Absorption is very broad and largely redshifted, 10$\pm$5\,\kms\ (see Fig.\,\ref{absorption}d). 
The low S/N prevents us from measuring pre- and post-transit absorption in this planet.
Due to the telluric contamination and the low  S/N,  these observations need confirmation \citep{Casasayas_2021}.

\section{Methods}\label{model}

\subsection{Modelling the He {\sc i} triplet density distribution}
\label{He_density}

We computed the \het\ density distribution using the model by \cite{Lampon2020}, which we briefly describe here.
We used a spherically symmetric 1D hydrodynamic model coupled with a non-LTE model for simulating the hydrodynamic atmospheric escape in the substellar direction (the one that connects the star-planet centres). 
As the solutions are degenerated with respect to the main parameters of the escape (e.g.   \mlr, $T$, and the H/He ratio), it is important to fully explore the parameter spaces. 
While 3D magneto-hydrodynamic models are necessary for reproducing the complex and detailed spatial distribution of the gas, their high computational cost usually prevents us from carrying out studies with a large number of simulations. Such studies are currently feasible using 1D spherically symmetric approaches, which allow  the main parameters of the escape to be explored reasonably well
\citep[see e.g.][]{Murray_Clay_2009,Stone_2009,Tripathi_2015,Owen_2020}.

In order to simplify the hydrodynamic calculations, we assumed that 
{the speed of sound of the outflow},
$v_s = \sqrt{k\,T(r)/\mu(r)}$, does not depend on the radius. In that equation $k$ is the Boltzmann constant, $T(r)$ is the temperature, and
$\mu$(r) is the mean molecular weight. We note that this does not necessarily imply that $T(r)$ and $\mu(r)$ are independent of $r$, but their ratio is.
The constant speed of sound is given by $v_{s,0}$\,=\,$\sqrt{k\,T_0/\bar \mu}$, where 
$T_0$ has a constant value, close to the maximum of the temperature profile obtained when solving the energy budget equation \citep[see e.g. Fig.\,3 in][]{Lampon2020}, and $\bar \mu$ is the average of $\mu$(r), which is iteratively calculated in the model \cite[see Appendix A in][]{Lampon2020}.
This assumption lets us decouple the hydrodynamic momentum equation from the energy budget equation and hence obtain an analytical solution analogous to the isothermal Parker wind approximation \citep{Lampon2020}.

Given the bulk parameters of the planetary system   (i.e. the mass, radius, and orbital separation of the planet, listed in Table\,\ref{table.parameters}, and the XUV stellar flux), \mlr\ (including all species considered in the model), $T_0$, and the H/He ratio are the input parameters of the model.
Among other outputs, the model yields the density distribution of species: neutral and ionised hydrogen, H$^0$ and H$^+$, respectively; the helium singlet, \hes; ionised helium, He$^+$; and the \het\ abundance.
We neglected the He$^{++}$ concentration in our calculations and assumed that the electron density is entirely produced by the H$^0$ ionisation. Production and loss processes are listed in Table\,2 of \cite{Lampon2020}, which are a minor extension of those considered by \cite{Oklopcic2018}.

Nominally, we set up the lower boundary at 1.02\,\rp\ with a total gas density of 10$^{14}$\,cm$^{-3}$, which is large enough to absorb all the stellar XUV and near-UV radiation.
In this work we also assumed a rather high upper boundary of the planet's atmosphere, close to the stellar diameter. This was motivated by the very extended atmospheres of some planets (e.g. \gju12 and \gj34). In this way it is guaranteed that we account for practically all the absorption along the line of sight (LOS) for all paths subtended by the stellar disc in every phase of the transit (see Sect.\,\ref{sec_absorption}).

\subsection{Stellar winds} \label{sw}

It has been shown in previous studies that there are also other parameters, among which the stellar wind (SW) is the most important, that could significantly affect the \het\ signal  \citep[see e.g.][]{Khodachenko_2021, Vidotto2020,Fossati_2022}. A detailed study of the SW and its effect on the planetary wind is beyond the scope of this paper as it requires a 3D model. Nevertheless, and despite the very large uncertainties in the SW parameters, we estimated its potential effects on our nominal (i.e. no SW considered) $T-$\mlr\ ranges. 
To evaluate this effect we repeated our simulations assuming that the atmosphere, still spherical, is shortened 
and consider that it spans only up to the ionopause (i.e. the region where the planetary and stellar winds pressures are in equilibrium).
The location of the ionopause was estimated following
\cite{Khodachenko_2019}, considering only the substellar direction.  
As the SW parameters (density profile, temperature, and velocity) are highly dependent on unknown factors (e.g. the stellar magnetic field), we chose those characterising the fast solar wind \cite[see Fig.\,3 in][]{Johnstone_2015}, which actually has a larger impact on the planetary wind than the slow solar wind. They were scaled to the distance of the corresponding planetary orbit. Furthermore, we use another approximation considering the same temperature and velocity as for the fast solar wind, but scaling the density to the stellar mass-loss rate. We used Eq.\,7 in \cite{Johnstone_2015} to estimate stellar mass-loss rates, except for GJ\,1214, for which we used the relationship $\dot M_{\star}$\,$\propto$\,$M_{\star}^{1.3}$, as this star is in the saturated regime \cite[see e.g.][]{Johnstone_2015,Johnstone_2015b}, and   HAT-P-32 and WASP-76, for which we used a value of three times the solar value as we do not have   the stellar rotational period (needed in Eq. 7  of  
\cite{Johnstone_2015}).  
See Table\,\ref{table.sw} for the stellar wind parameters used in this analysis. 

\begin{table}[htbp]
\centering
\caption{\label{table.sw}
Parameters of the stellar winds used in this work.}
\begin{tabular}{l c c c r r} 
\hline  \hline  \noalign{\smallskip}
Star & $\dot M_{\star}/ \dot M_{\sun}$& T$_{swp}$ & v$_{swp}$ & n$_{swp}$ & n$_{e,swp}$ \\
        & & (MK)   & (\kms) & (cm$^{-3}$) & (cm$^{-3}$) \\ 
\noalign{\smallskip} \hline \noalign{\smallskip}
\noalign{\smallskip}
HAT-P-32 & 3.0 & 2.4 & 470 & 5700 & 17\,100 \\
WASP-69 & 1.6 & 2.0 & 540 & 2800 & 4480\\
GJ\,1214 & 0.1 & 2.6 & 330 & 48\,000 & 4800 \\
WASP-76 & 3.0 & 2.4 & 470 & 6100 & 18\,300 \\
GJ\,3470 & 3.7 & 2.4 & 470 & 5500 & 20\,000\\
HD\,189733 & 3.4 & 2.4 & 470 & 6000 & 20\,400 \\
\noalign{\smallskip}
\hline
\noalign{\smallskip}
\end{tabular}
\tablefoot{
$\dot M_{\star}$ and $\dot M_{\sun}$ are the stellar and the solar mass-loss rate \cite[$\sim 2\times 10^{-14} M_{\sun}\,yr^{-1}$, ][]{Johnstone_2015}, respectively.
T$_{swp}$, v$_{swp}$, and n$_{swp}$ are the temperature, velocity, and density of the fast solar wind at the orbit of the planet. n$_{e,swp}$ is the density of the fast solar wind scaled to the stellar mass-loss rate, also at the planetary orbit.  
}
\end{table}

We find  that, in general, it might have a significant impact on the \het\ absorption for planets with very extended atmospheres when considering strong SW. However, it only slightly changes our nominal T-\mlr\ ranges. In particular, we found no significant effects (either using the fast or the scaled SW), for the derived \mlr--$T$ of \hatp32, \w69, and \wa76. For the other planets the results are discussed in Sects.\,\ref{sec_gj12} and \ref{revisiting_gj34}.

In those estimations we   assumed that \het\ is depleted at the ionopause altitude. However, under some scenarios, 
it is possible that the electron density supplied by the SW produces a \het\ enhancement \cite[see][]{Shaikislamov_2021}. For these cases the effects would be to lower the SW estimations.

\begin{table}[htbp]
\setstretch{0.85}
\centering
\caption{\label{table.parameters}
System parameters.}
\begin{tabular}{l c l} 
\hline  \hline  \noalign{\smallskip}
Parameter & Value & Ref  \\
\noalign{\smallskip} \hline \noalign{\smallskip}
\multicolumn{3}{l}{\em HAT-P-32} \\
\noalign{\smallskip}
$d$ & 286.221\,$^{+1.679}_{-1.679}$\,pc & GA\\
\noalign{\smallskip}
$R_{\star}$ &  1.219\,$\pm$\,0.016\,$R_\sun$ & HR\\
\noalign{\smallskip}
$M_{\star}$ & 1.160\,$\pm$\,0.041\,$M_\sun$    & HR   \\
\noalign{\smallskip}
$T_{\rm eff}$ & 6269\,$\pm$\,64\,K      & ZH \\
\noalign{\smallskip}
$[\rm Fe/H]_{\star}$ & -0.04\,$\pm$\,0.08       & HR   \\
\noalign{\smallskip}
$a$     & 0.0343\,$\pm$\,0.0004\,au     & HR \\
\noalign{\smallskip}
$i$ & 88.9\,$\pm$\,0.4\, deg & HR \\
\noalign{\smallskip}
$R_{\rm P}$ & 1.789\,$\pm$\,0.025\,\rj\     & HR  \\
\noalign{\smallskip}
$M_{\rm P}$     & 0.585\,$\pm$\,0.031\,\mj\     & CZ   \\
\noalign{\smallskip}
\hline
\noalign{\smallskip}
\multicolumn{3}{l}{\em WASP-69} \\
\noalign{\smallskip}
$d$ &   
50.287\,$^{+0.043}_{-0.043}$\,pc & GA\\
\noalign{\smallskip}   
$R_{\star}$ &   
0.813\,$\pm$\,0.028\,\,$R_\sun$  & AN    \\
\noalign{\smallskip}
$M_{\star}$ & 
0.826\,$\pm$\,0.029\,$M_\sun$     & AN     \\
\noalign{\smallskip}
$T_{\rm eff}$ & 
4715\,$\pm$\,50\,K      & AN     \\
\noalign{\smallskip}
$[\rm Fe/H]_{\star}$ & 
+0.144\,$\pm$\,0.077     & AN     \\
\noalign{\smallskip}
$a$             & 
0.04525\,$\pm$\,0.00053\,au     & AN \\
\noalign{\smallskip}
$i$ & 86.7\,$\pm$\,0.2\, deg & AN \\
\noalign{\smallskip}
$R_{\rm P}$     & 
1.057\,$\pm$\,0.047\,\rj\       & AN     \\
\noalign{\smallskip}
$M_{\rm P}$     & 
0.26\,$\pm$\,0.017\,\mj\     & AN     \\
\noalign{\smallskip}
\hline
\noalign{\smallskip}
\multicolumn{3}{l}{\em GJ\,1214} \\
\noalign{\smallskip}
$d$ &   
14.6416\,$^{+0.0139}_{-0.0139}$\,pc & GA\\
\noalign{\smallskip}   
$R_{\star}$ &   
0.216\,$\pm$\,0.012\,\,$R_\sun$  & HP    \\
\noalign{\smallskip}
$M_{\star}$ & 
0.150\,$\pm$\,0.011\,$M_\sun$     & HP     \\
\noalign{\smallskip}
$T_{\rm eff}$ & 
3026\,$\pm$\,150\,K      & HP     \\
\noalign{\smallskip}
$[\rm Fe/H]_{\star}$ & 
+0.39\,$\pm$\,0.15     & BR     \\
\noalign{\smallskip}
$a$             & 
0.01411\,$\pm$\,0.00032\,au     & HP \\
\noalign{\smallskip}
$i$ & 88.2\,$\pm$\,0.5\, deg & HP \\
\noalign{\smallskip}
$R_{\rm P}$     & 
0.24463\,$\pm$\,0.00473\,\rj\       & CL     \\
\noalign{\smallskip}
$M_{\rm P}$     & 
0.02571\,$\pm$\,0.00135\,\mj\     & CL     \\
\noalign{\smallskip}
\hline
\noalign{\smallskip}
\multicolumn{3}{l}{\em WASP-76} \\
\noalign{\smallskip}
$d$ &   189.039\,$^{+2.952}_{-2.952}$\,pc & GA\\
\noalign{\smallskip}   
$R_{\star}$ &   1.756\,$\pm$\,0.071\,\,$R_\sun$  & EH    \\
\noalign{\smallskip}
$M_{\star}$ & 1.458\,$\pm$\,0.021\,$M_\sun$     & EH     \\
\noalign{\smallskip}
$T_{\rm eff}$ & 6329\,$\pm$\,65\,K      & EH    \\
\noalign{\smallskip}
$[\rm Fe/H]_{\star}$ & +0.366\,$\pm$\,0.053     & EH     \\
\noalign{\smallskip}
$a$             & 0.0330\,$\pm$\,0.0002\,au     & EH \\
\noalign{\smallskip}
$i$ & 88.0\,$^{+1.3}_{-1.6}$\,deg & WE \\
\noalign{\smallskip}
$R_{\rm P}$     & 1.854\,$^{+0.077}_{-0.076}$\,\rj\       & EH     \\
\noalign{\smallskip}
$M_{\rm P}$     & 0.894\,$^{+0.014}_{-0.013}$\,\mj\     & EH     \\
\noalign{\smallskip}
\hline
\end{tabular}
\tablefoot{
GA: \cite{Gaia_EDR3} (Gaia Early Data Release 3);   
HR: \citet{Hartman_2011};
ZH: \citet{Zhao_2014};
CZ: \citet{Czesla_2022};
AN: \citet{Anderson_2014};
HP: \citet{Harpsoe_2013};
BR: \citet{Berta_2011};
CL: \citet{Cloutier_2021};
EH: \citet{Ehrenreich_2020};
WE: \citet{West_2016}.
}
\end{table}

\subsection{Stellar fluxes} \label{fluxes}

The XUV stellar flux, $F_{\rm XUV}$, can ionize neutral H (below 912~\AA) and He (below 504~\AA)  atoms. Good knowledge of the spectral stellar energy distribution (SED) in this range is essential for interpreting the evaporation effects on the planet's atmosphere. However, the H in the interstellar medium absorbs much of the extreme ultraviolet (EUV, $\sim$100--920 \AA) stellar flux, hampering a correct evaluation of the stellar flux in most cases. Stellar XUV lines are formed in the transition region and corona, at temperatures of $\log\,T$(K)\,$\sim$\,4--7.5. Thus, a correct coronal model should be able to predict the XUV emission coming from   spectral lines and continuum. For this work, following \citet{san11}, we used this approach to generate a synthetic SED in the XUV spectral range, extended up to 1600\,\AA, 2100\,\AA, 1200\,\AA,\ and 1500\,\AA\ for HAT-P-32, WASP-69,  GJ\,1214, and WASP-76, respectively.
Further, we use the atomic database APEC \citep{smith}, which has some limitations beyond 1200\,\AA,\ but it is quite accurate in the XUV \citep{chad15}.  
The coronal models used in this paper are detailed in Sanz-Forcada et al. (2023, in prep.). 
In the case of WASP-76 only an upper limit could be used for the stellar XUV fluxes, since its general flux level is based on an {\it XMM-Newton} detection with a significance lower than 3$\sigma$.

\begin{figure}[htbp]
\includegraphics[angle=90.0,width=1\columnwidth]{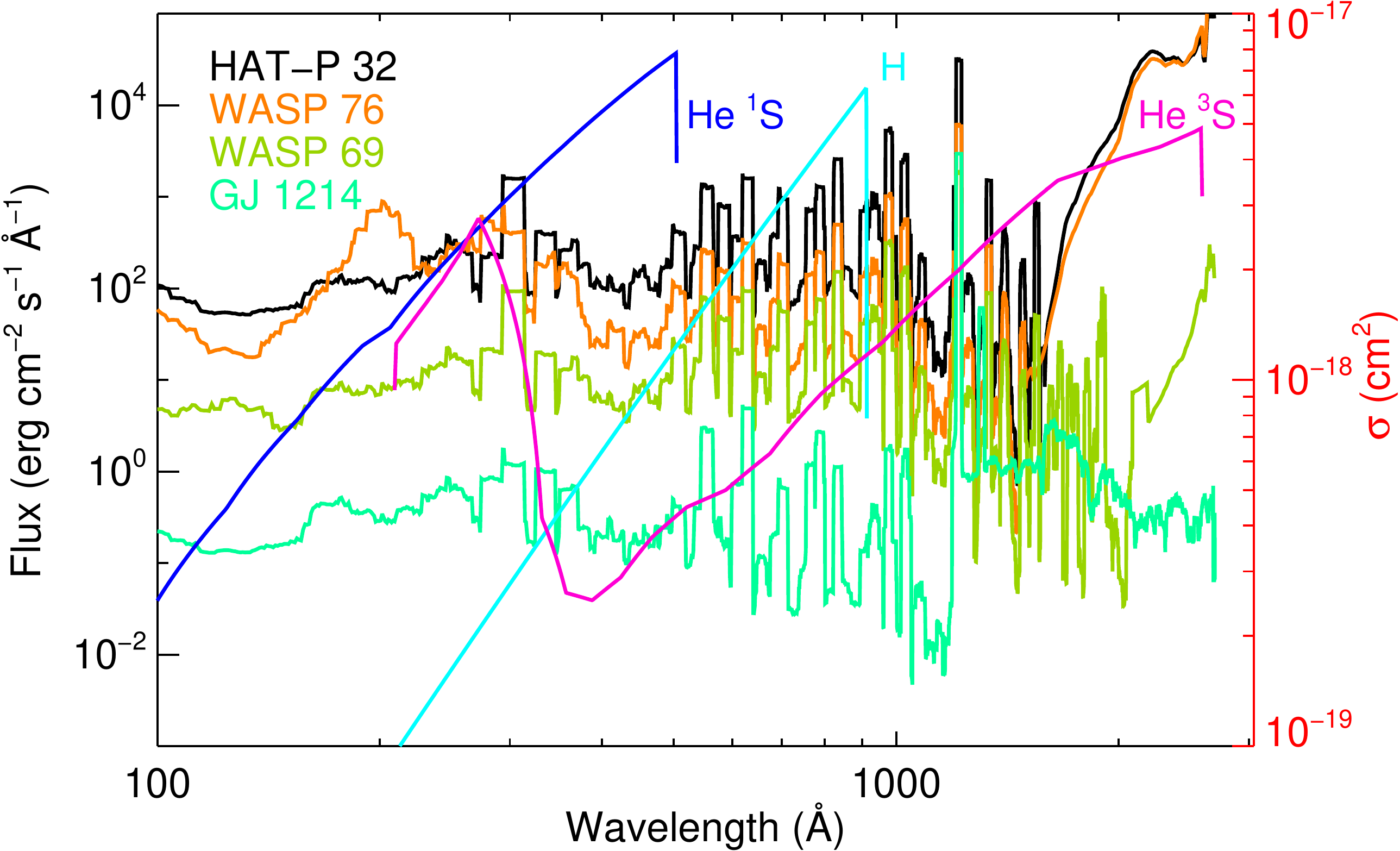}\hspace*{0.35cm}
\caption{
Flux density (left y-axis) for HAT-P-32 (black), WASP-76 (orange), WASP-69 (green), and GJ\,1214 (sea green) at the respective planet's distance plotted at a resolution of 10\,\AA. In the case of WASP-76, the plotted values are an upper limit. The H, He singlet, and He triplet ionisation cross-sections ($\sigma$, right y-axis) are also shown.} 
\label{flux_xsec} 
\end{figure}

In order to extend the SED of the stars to 2600\,\AA\ to cover the \het\ absorption,   
for HAT-P-32, WASP-69, and WASP-76 we used
the stellar atmospheric model of \cite{Castelli-Kurucz_2003} scaled
to their temperature, surface gravity, and metallicity 
(see Table\,\ref{table.parameters}).
For GJ\,1214 we used the Hubble/COS spectra for the 1200--1650\,\AA\ range, and the Hubble/STIS spectra for 1650--2600\,\AA.
The SEDs for the extended spectral range of
5--2600\,\AA\ for the planets at their respective orbital separations
are shown in Fig.\,\ref{flux_xsec}.

\begin{figure}[htbp]
\includegraphics[angle=90.0,width=1\columnwidth]{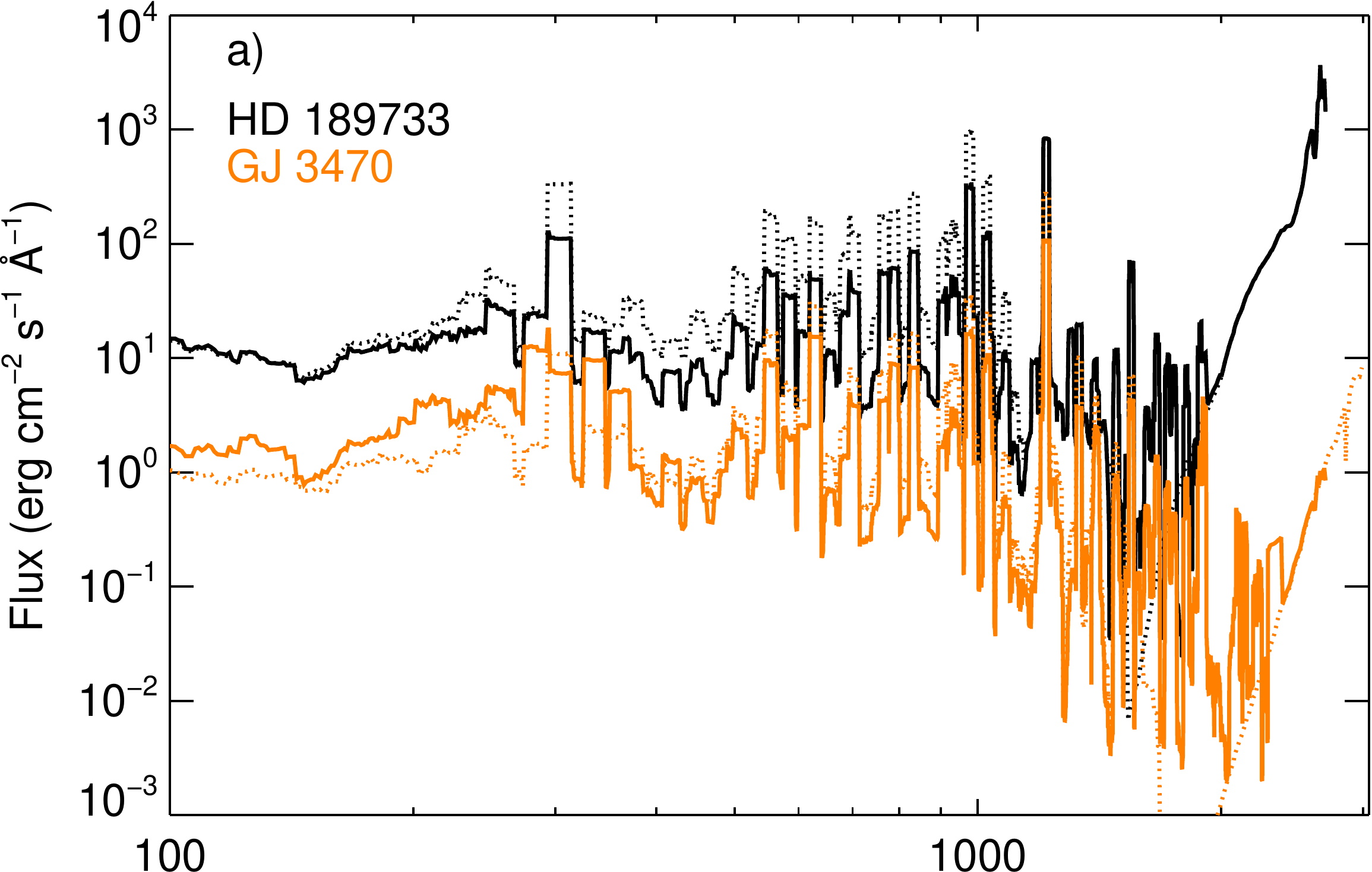}\vspace*{0.2cm}
\includegraphics[angle=90.0,width=1\columnwidth]{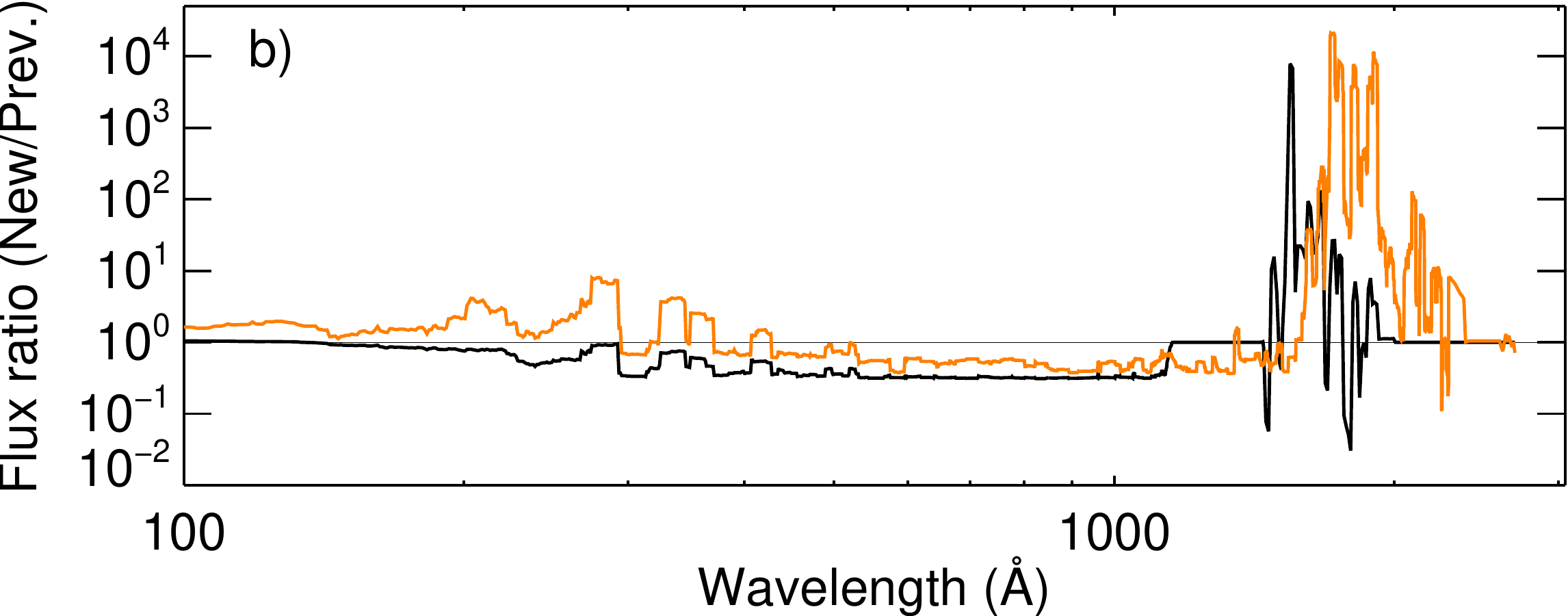}\hspace*{0.35cm}
\caption{Flux densities for HD\,189733 and GJ\,3470. a)  Previous (dotted) and new (solid) flux densities at the respective planet's distance, plotted at a resolution of 10\,\AA. b) Ratios of the new to the previous fluxes.
} 
\label{flux_new} 
\end{figure}

The XUV data of HD\,189733 and GJ\,3470 were revised after an improved treatment of the coronal abundances of the stars in Sanz-Forcada et al. (2023, in preparation). Their SEDs at the planetary distance are shown in Fig.\,\ref{flux_new}. As we see in the figure, the flux density for HD\,189733 is significantly reduced, by a factor from two to three at wavelengths between $\sim$250\,\AA\ and 1200\,\AA. In the case of GJ\,3470 the flux density is increased by an average factor of close to three at wavelengths below $\sim$400\,\AA, and it is reduced by a factor of close to two between 400\,\AA\ and 1200\,\AA.
We also extended our coronal model up to 2300\,\AA\ for GJ\,3470 and from 1450\,\AA\ to 1900\,\AA\ for HD\,189733. For longer wavelengths, up to 2600\,\AA,  
we plugged in the model of \cite{Castelli-Kurucz_2003} \cite[see][for details about the stellar fluxes of both planets]{Lampon_2021a}.

\subsection{Spectral absorption} \label{sec_absorption}

The \het\ absorptions of the four planets were computed using a radiative transfer code for the transit geometry \citep{Lampon2020}. The code inputs are the bulk parameters of the 
planetary systems, the \het\ abundances and the gas radial outflow velocities computed with the model described above.
Similarly to our previous works, the spectroscopic data for the three metastable helium lines are  from the NIST Atomic Spectra Database.\footnote{\tt https://www.nist.gov/pml/atomic-spectra-database.} In the radiative transfer code, broadenings and shifts of the lines can be included from three major sources. First, we account for the Doppler broadening caused by the microscopic thermal velocities dictated by the local kinetic temperature. Second, by assuming turbulent velocities a line widening can be incorporated \citep{Salz2018}, which also depends on the kinetic temperature and is given by $v_{\rm turb}$\,=\,$\sqrt{5kT/3m}$, where $m$ is the mass of a He atom \cite[see Eq.\,16 in][]{Lampon2020}.
Third, we also incorporated the broadening produced by the component of the radial outflow velocity of the gas along the line of sight. This effect can be notable in planets with extended atmospheres, where the absorption of the uppermost layers, usually moving out at the larger velocities, is significant. The gas radial outflow velocities can also produce a weaker absorption at the core of 
the stronger lines (compared to that in their wings)
in atmospheres that are very compressed, that move at high velocities, and with a \het\ concentration profile that peaks just above the lower boundary. Moreover, where necessary, we can also include in the absorption averaged wind calculations (e.g. day-to-night, super-rotation winds) and planetary rotation  \cite[see e.g.][and Eq.\,15 in \citeauthor{Lampon2020}, \citeyear{Lampon2020}]{Salz2018,Seidel2020},      here referred to as `non-radial' winds, complementary to the hydrodynamic radial winds.

We computed the mid-transit synthetic spectrum for \hatp32 and the phase-averaged synthetic spectrum for \w69, \gju12, and \wa76,
mimicking the \het\ observations (see Sect.\,\ref{Obs}), as 
suggested by \cite{Dos_santos_2021}.  
The inputs related to the transit geometry, the planetary and stellar radii, and the transit impact parameter are listed in Table\,\ref{table.parameters}.

\subsection{Constraining the main parameters of the upper atmosphere}
\label{grid}

Following the methods of \cite{Lampon2020,Lampon_2021a} we carried out a parameter study to constrain the main parameters of the upper atmosphere of \hatp32, \w69, \gju12, and \wa76. 
For a given H/He ratio, we performed a grid of simulations with a wide range of T and \mlr\ (hereafter  $T$\,-\mlr\ grid). We calculated the reduced $\chi^2$ contour map of the $T$\,-\mlr\ grid by comparing the synthetic spectrum of every simulation with the measured \het\ absorption profile \cite[see Sect.\,3.4 in][]{Lampon_2021a}. 
The selected spectral ranges for the fitting of spectra are 10830.5--10835\,\AA, 10831--10834.5\,\AA, 10831--10834.5\,\AA, and 10831--10836\,\AA, for \hatp32, \w69, \gju12, and \wa76, respectively. The number of fitted data points is taken as the number of independent measurements taken by CARMENES in the selected spectral range, that is, the spectral range divided by the spectral resolution of CARMENES in this channel, $\mathcal{R}$\,=\,80~400. This yields 36, 29, 31, and 44 spectral points for the respective planets. The number of degrees of freedom equals the number of spectral points minus two, the number of fitted variables \cite[temperature and  mass-loss rate; see Sect.\,3.4 in][]{Lampon_2021a}.

 Figure\,\ref{chi2} shows  the simulations for the different $T$\,-\mlr\ grids (black dots),  the best fits within the 95\% confidence of the $\chi^2$ (large symbols, hereafter the constrained $T$\,-\mlr\ range), and  the best fits for the rest of the temperatures and mass-loss rates (dotted lines, hereafter the extended $T$\,-\mlr\ range). 
In addition to $T$, \mlr, and the  H/He ratio, turbulence and non-radial winds
could significantly affect the width of the \het\ absorption line  \citep{Lampon_2021a}, impacting the  constrained $T$\,-\mlr\ range.
In the case of very extended atmospheres, the upper boundary can also influence the results. The reason is that 
the uppermost layers, which are rather dense (Sect.\,\ref{He_density}) and out-flowing at large velocities (Fig.\,\ref{vel}), could produce additional absorption and broadening of the line. 
In Sect.\,\ref{results} we discuss how such parameters affect the constrained $T$\,-\mlr\ range of the studied planets.

The H/He ratio can be constrained by comparing the H$^0$ density profiles determined from the \het\ observations to the H$^0$ abundances derived from \lya\ measurements \citep{Lampon2020, Lampon_2021b}. 
For the planets studied here, unfortunately, there are no available \lya\ absorption measurements.
Nevertheless, we   tried to constrain the H/He ratio (see Sect.\,\ref{results}), for the case of \hatp32, by using 
the simultaneous \ha\ measurements \citep{Czesla_2022}, and for the case of \gju12  
by analysing the relationship between the heating efficiency and the H/He ratio.

To constrain the mean heating efficiency of the 
upper atmosphere, $\eta$ (hereafter heating efficiency), we followed \cite{Lampon_2021a}. We used the energy-limited approximation, $\dot M_{\rm EL}$, \citep{Watson_1981,Erkaev_2007}, together with the relationship \mlr/$\dot M_{\rm EL}$\,=\,4/5 derived by \cite{Salz_2015}, to obtain
\begin{align} 
\dot M = \frac{4}{5}\, \frac {4 \pi\, R_{\rm XUV}^{2}\,F_{\rm XUV}}             {K(\xi)\, \Phi}\, \eta~, 
\label{eq:energy_lim}
\end{align}
where 
$\Phi = G\,M_{\rm p}/R_{\rm p} $ is the gravitational potential,
$M_{\rm p}$ and \rp\, are respectively the planetary mass and radius, and $G$ is the gravitational constant; $R_{\rm XUV}$ is the effective absorption radius, i.e. the altitude where the XUV optical depth is unity; $K(\xi)$\,=\,1--1.5\,$\xi+0.5\,\xi^{3}$ is the potential energy reduction factor, with $\xi$\,=\,$\left(M_{\rm P}/M_{\star} \right)^{1/3}\left(a/R_{\rm P}\right)$, where $a$ is the planetary orbital separation and $M_{\star}$ the stellar mass.
In this way, we calculated $\eta$ with \mlr\ and $R_{\rm XUV}$ obtained from our model, the system's parameters from Table\,\ref{table.parameters}, and the $F_{\rm XUV}$ from Sect.\,\ref{fluxes} listed in Table\,\ref{eqw}.

\subsection{Classification of planets by their hydrodynamic escape regimes}
\label{sec_RH}

We classified the studied planets according to their corresponding hydrodynamic escape regime following the method described by \cite{Lampon_2021b}. Briefly, regimes can be distinguished by the production and losses of H$^0$, and additionally by the heating efficiency of the outflow. 
In the recombination-limited regime, production of H$^0$ by recombination dominates over advection in practically the entire upper atmosphere (i.e. when $P_{rec}/P_{adv}$\,$\gg$\,1, where $P_{rec}$ and $P_{adv}$ are the recombination rate and the advection  rate, respectively) \cite[see Eq.\,1 and definitions in][]{Lampon_2021b}. 
In this regime, the ionisation front (IF), the region where the atmosphere transitions from essentially neutral to mostly ionised, is confined to a narrow region. 
Additionally, the heating efficiency is very low, with values considerably lower than 0.1, as radiative cooling is an important fraction of the absorbed stellar energy.

In the photon-limited regime, advection dominates the production of H$^0$, and then $P_{rec}/P_{adv} \ll 1$. In this case the IF extends to practically the entire upper atmosphere, and the heating efficiency is relatively high, with values close to or higher than 0.1, as the radiative cooling is moderate or negligible. In addition, the heating efficiency is nearly constant with respect to \mlr, as \mlr\,$\propto\,R^{2}_{XUV}$ in this regime (see Eq.\,\ref{eq:energy_lim}).

In the energy-limited regime recombination and advection are non-negligible in most of the upper atmosphere. The IF is wide, although it does not occupy the whole upper atmosphere. Moreover, the heating efficiency is close to or higher than 0.1, as in the photon-limited regime. However, the heating efficiency is not constant with respect to \mlr.

\section{Results} \label{results}

\begin{figure*}
\begin{tabular}{c c}
\includegraphics[angle=90, width=1.\columnwidth]{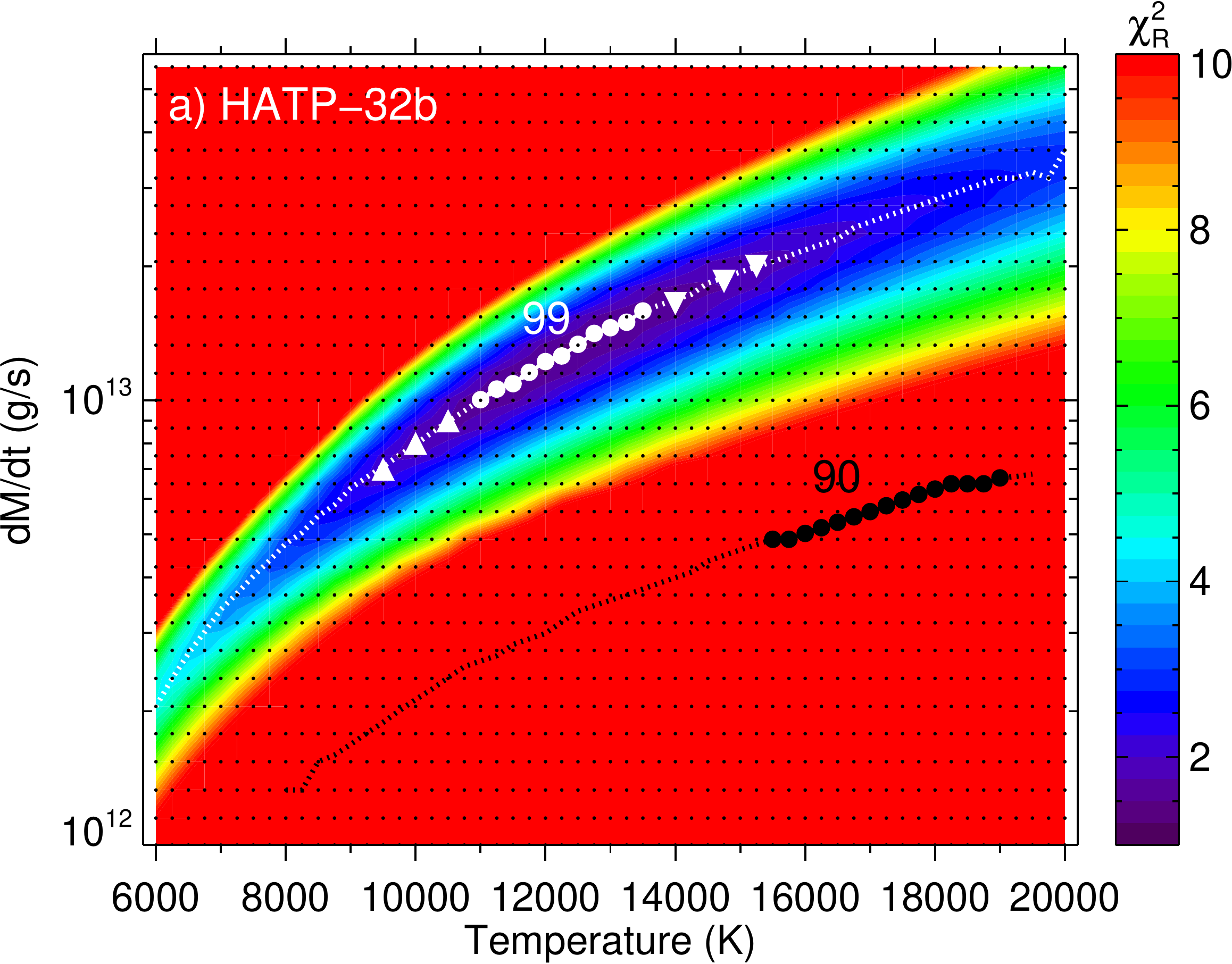} & 
\includegraphics[angle=90, width=1.\columnwidth]{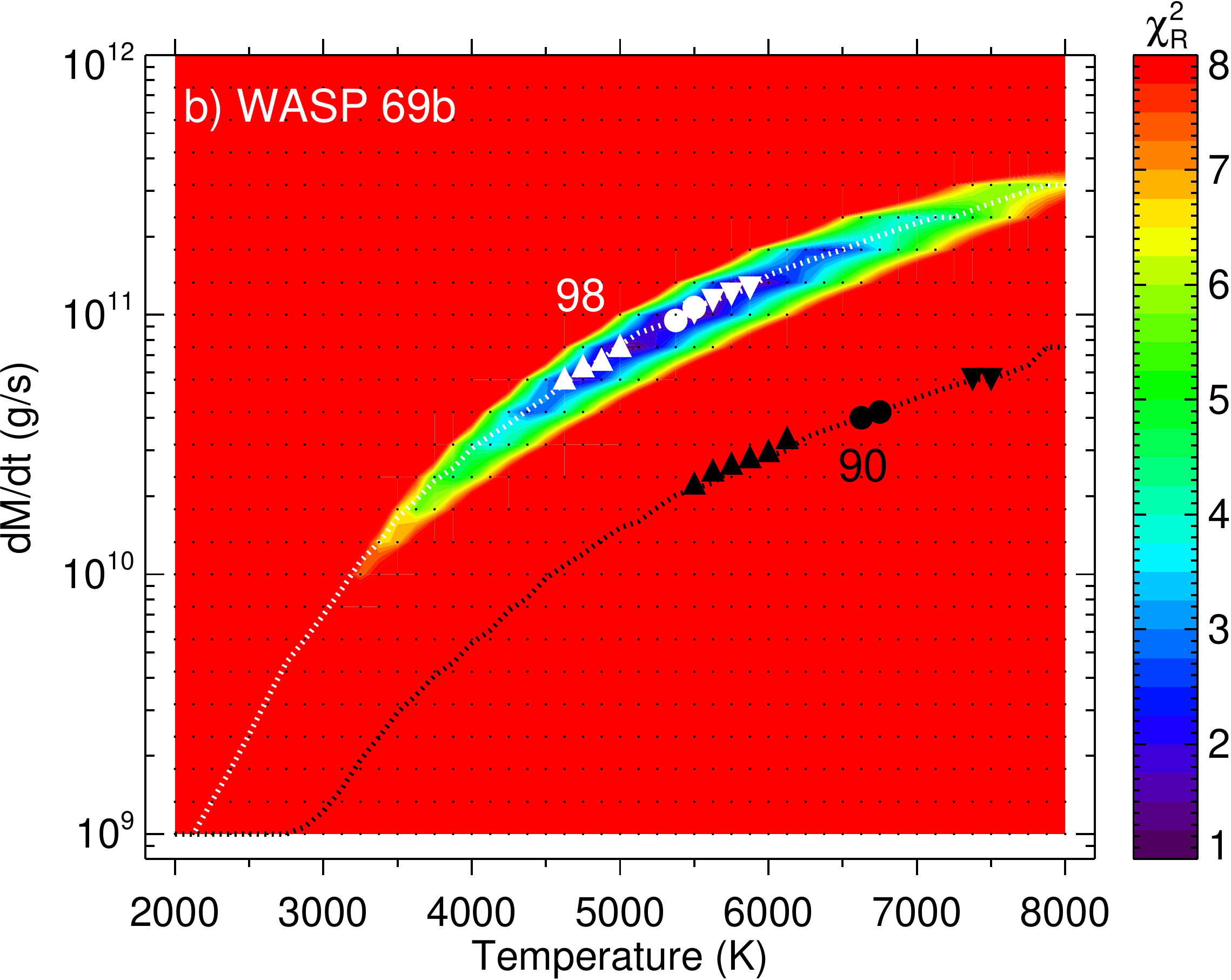} \\
\includegraphics[angle=90, width=1.\columnwidth]{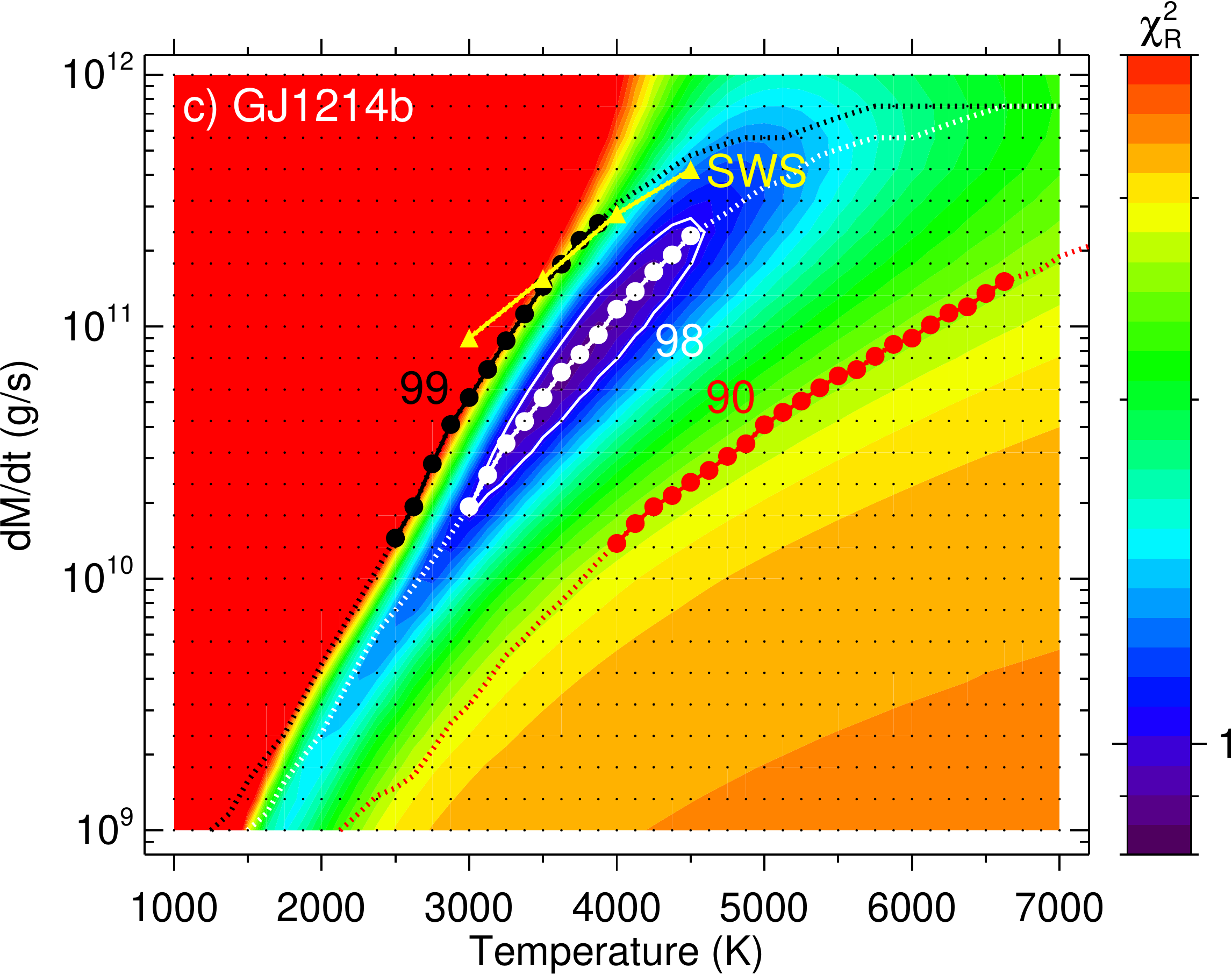}&
\includegraphics[angle=90, width=1.\columnwidth]{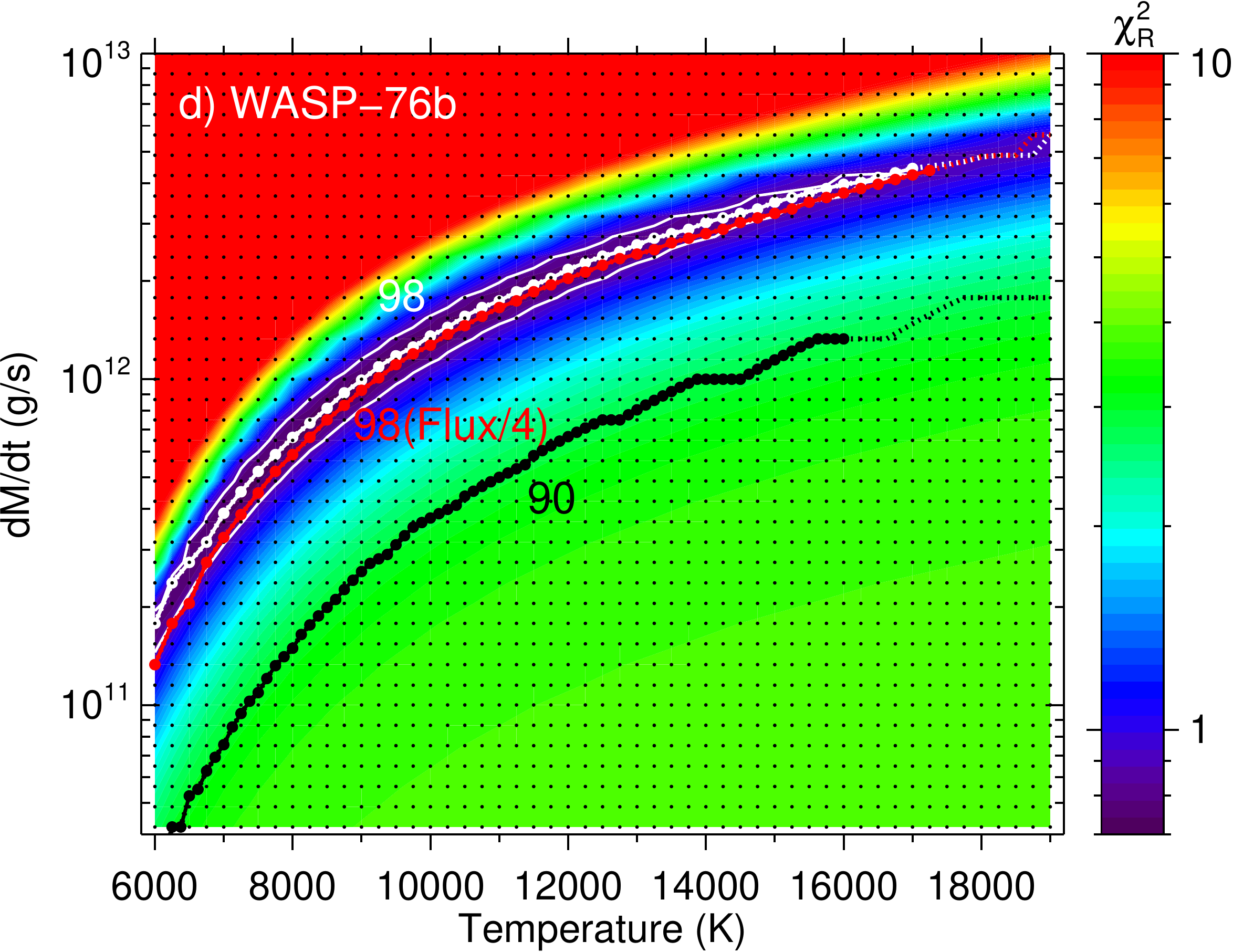} \\
\end{tabular}
\caption{
Contour maps of the reduced $\chi^2$ of the \het\ absorption for a) \hatp32, b) \w69, c) \gju12, and d) \wa76 (with different scales of temperature and \mlr). Dotted curves represent the best fits; the  large symbols denote the constrained ranges for a confidence level of 95\% (see Sect.\,\ref{grid}). 
In panels a) and b) the upward triangles represent the limits imposed by including blue and red components (see text). The downward triangles in panel a) are the $T$-\mlr\ obtained when assuming null radial velocities of the gas inside the ionisation front (IF) region and in panel b) when not including the turbulence broadening.  
The yellow upward triangles in panel c) correspond to the $T$-\mlr\ range obtained when the fast solar wind is considered. The red symbols in panel d) correspond to the flux density reduced by a factor of four and those in white for the nominal  upper limit. Overplotted are also the curves and symbols for several H/He ratios (e.g.  labelled  `90' for a H/He of 90/10, `98' for H/He=98/2). The black dots represent the $T$-\mlr\ grid of the simulations. The minima shown are those obtained from the solutions of the hydrodynamic and non-LTE models for a physically meaningful range of the parameter space. 
}
\label{chi2}
\end{figure*}

\begin{figure*}
\includegraphics[angle=90.0, width=1.\columnwidth]{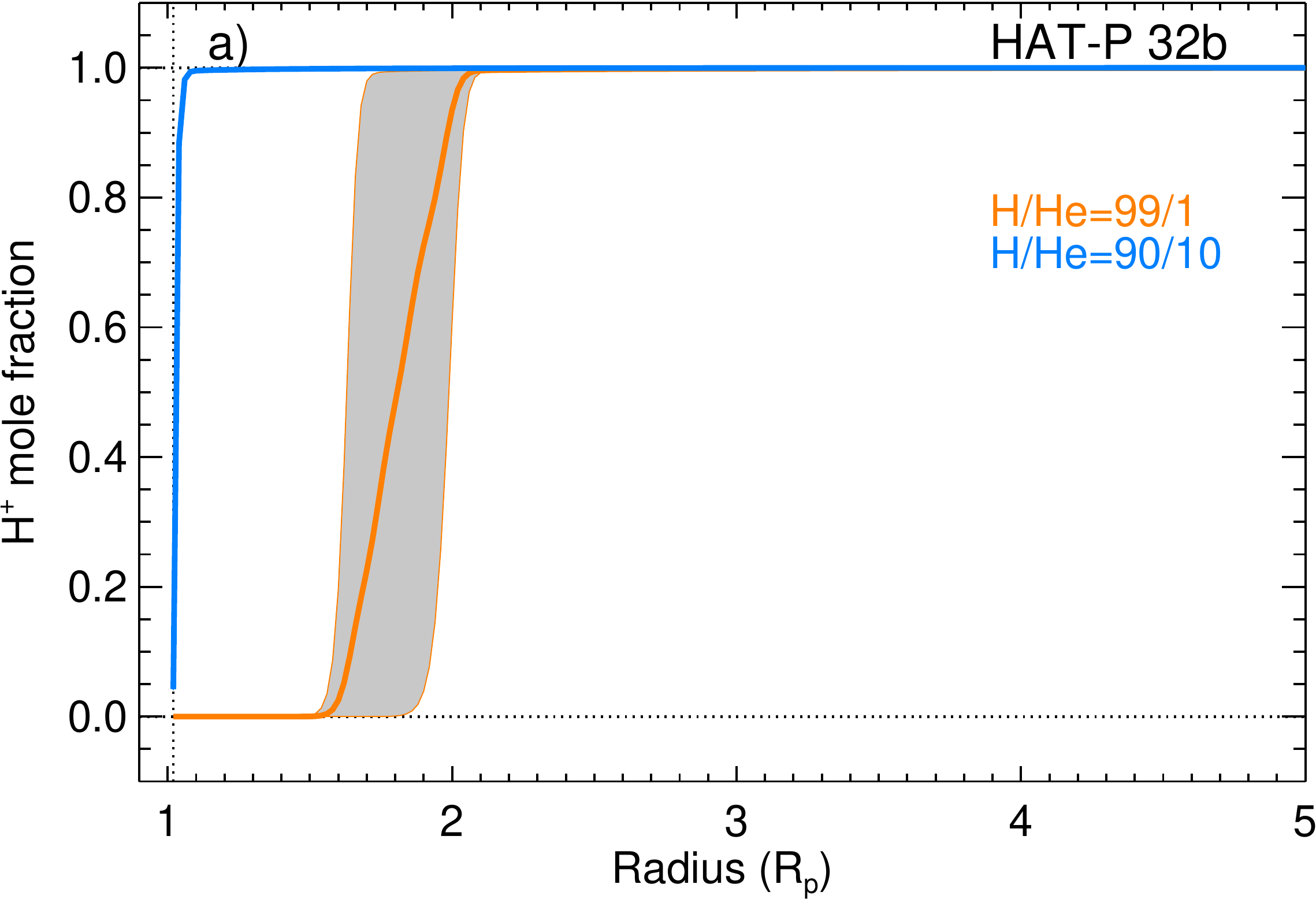}
\includegraphics[angle=90.0, width=1.\columnwidth]{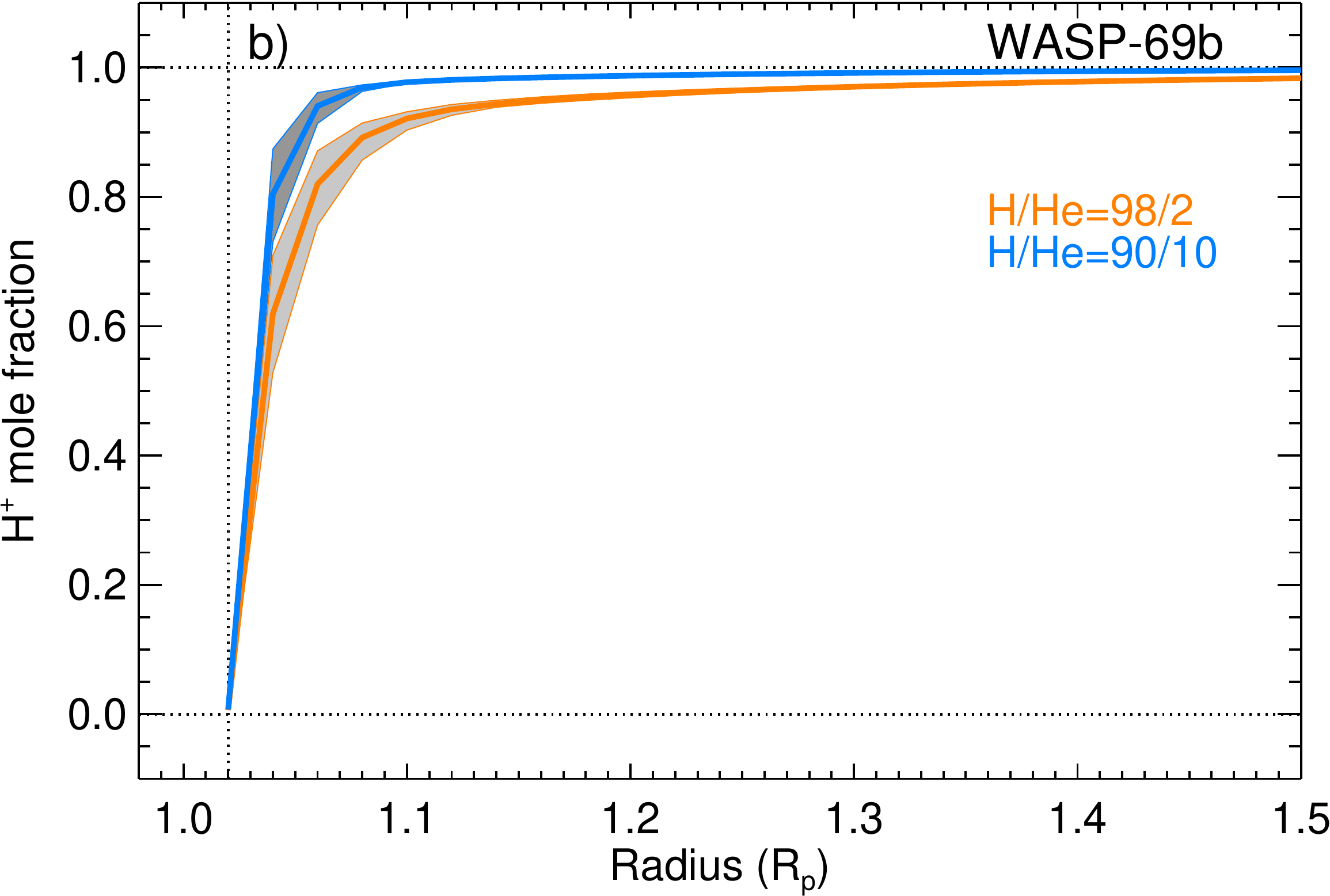}
\includegraphics[angle=90.0, width=1.\columnwidth]{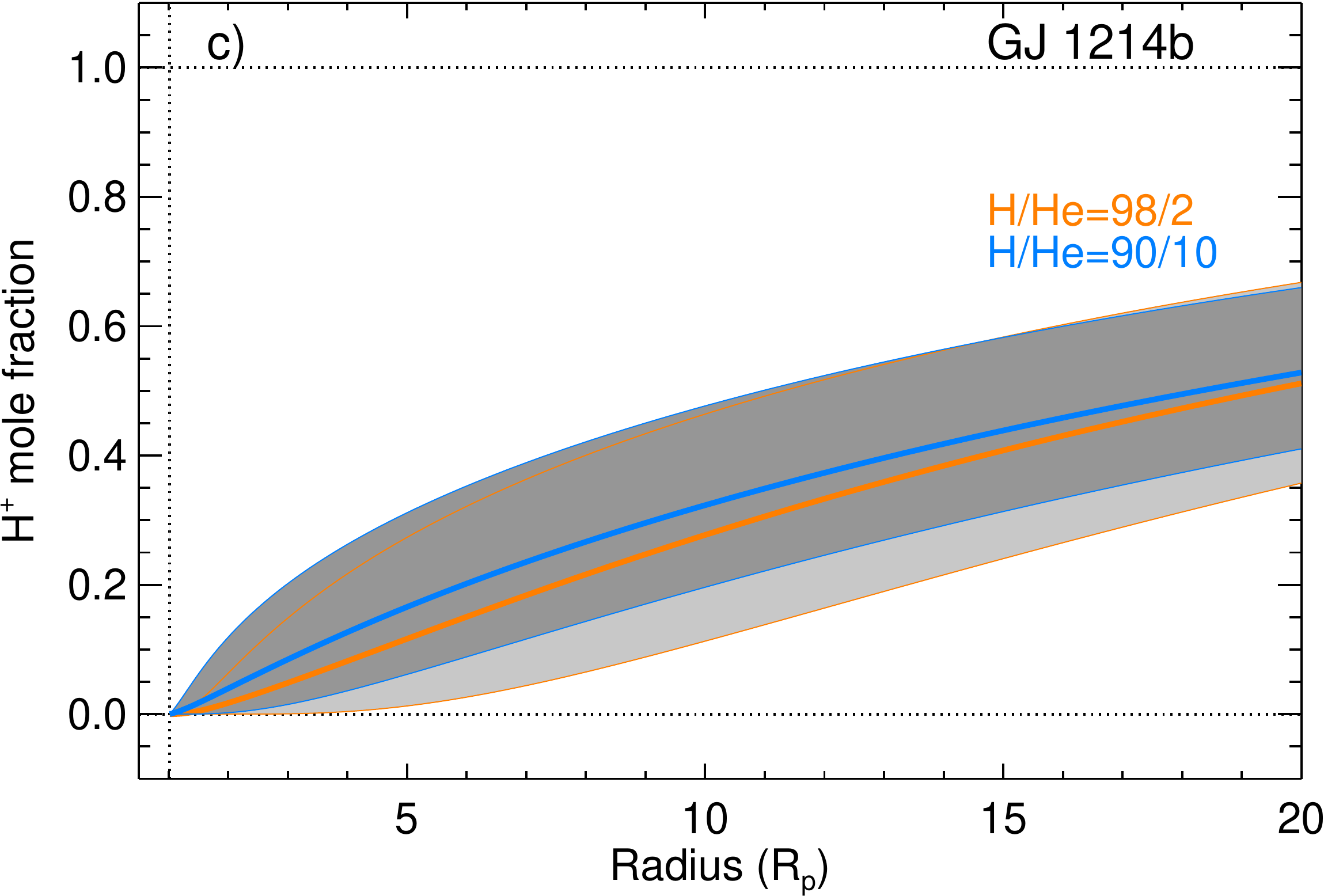}
\includegraphics[angle=90.0, width=1.\columnwidth]{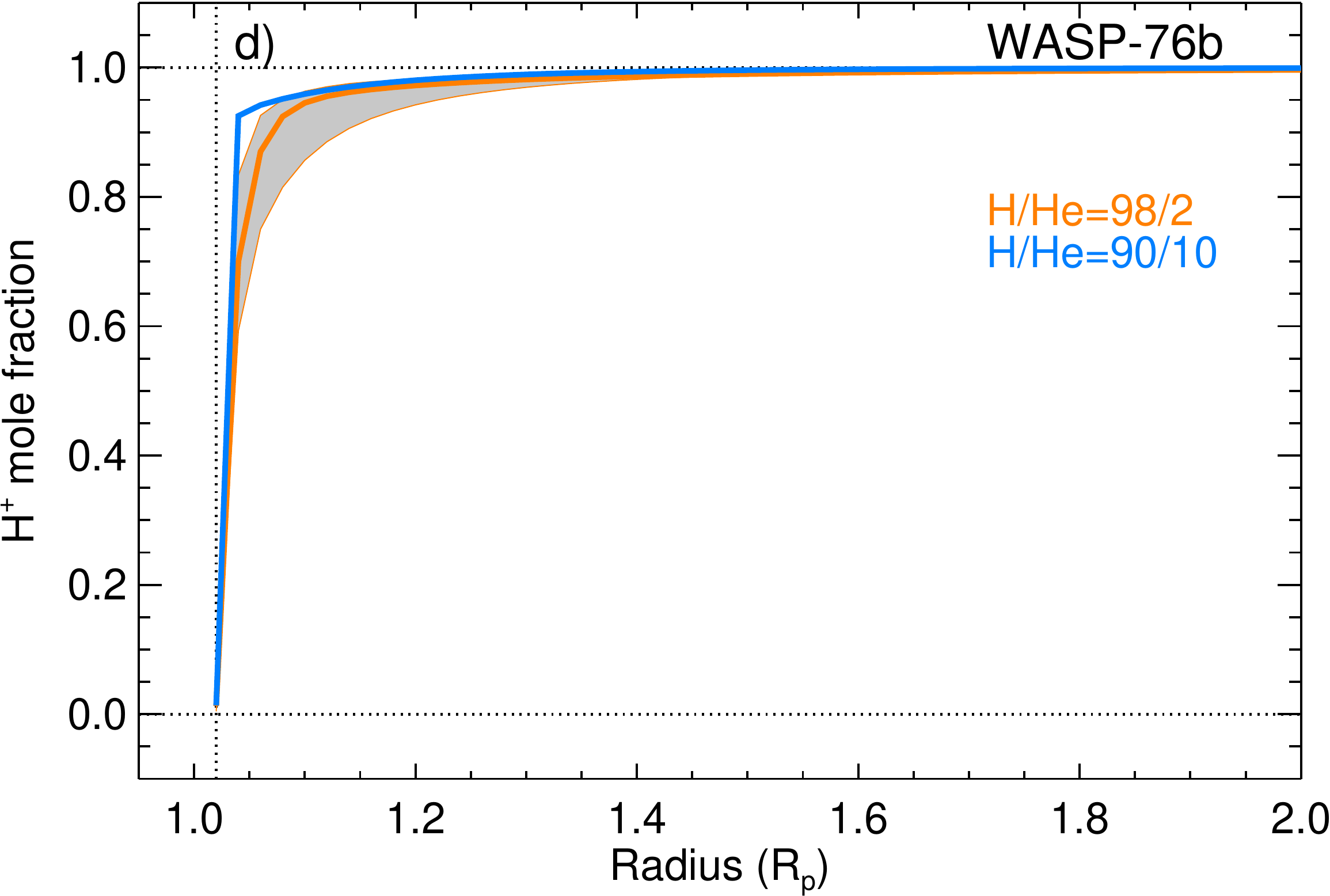}
\caption{H$^+$ mole fraction profiles resulting from the fit of the measured absorption (large circles in Fig.~\ref{chi2}) for \hatp32 (H/He=99/1, top left panel), WASP-69b (H/He=98/2, top right panel), GJ 1214b (H/He=98/2, bottom left panel), and WASP-76b (H/He=98/2, bottom right panel). For comparison,   also included are the results for H/He=90/10 for all planets. The  x-axis ranges are different.  The solid thicker lines are the mean profiles.}
\label{hvmr_candidates} 
\end{figure*}

\begin{figure}
\includegraphics[angle=90.0, width=\columnwidth]{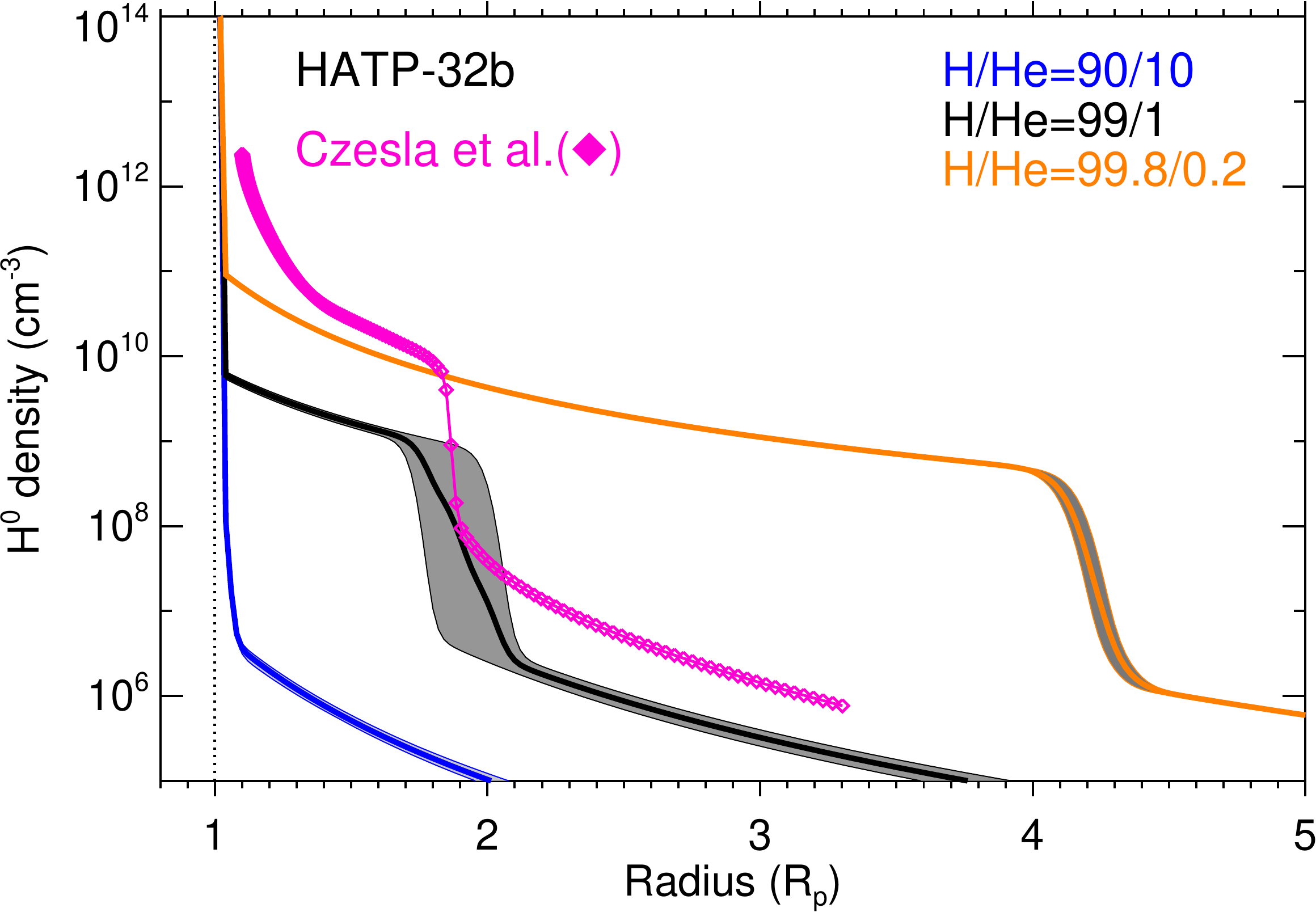}
\caption{Range of the neutral hydrogen concentration profiles (grey shaded areas) for \hatp32 resulting from the fit of the measured absorption for H/He ratios of 90/10, 99/1, and 99.8/0.2 (for H/He=90/10 and 99/1 see Fig.~\ref{chi2}a). The solid thicker curves are the mean profiles. The H$^0$ density derived from \halpha\ measurements by \cite{Czesla_2022} are also shown (magenta diamonds).}
\label{hden_candidates} 
\end{figure}

\begin{figure*}
\centering
\includegraphics[angle=90, width=0.9\columnwidth]{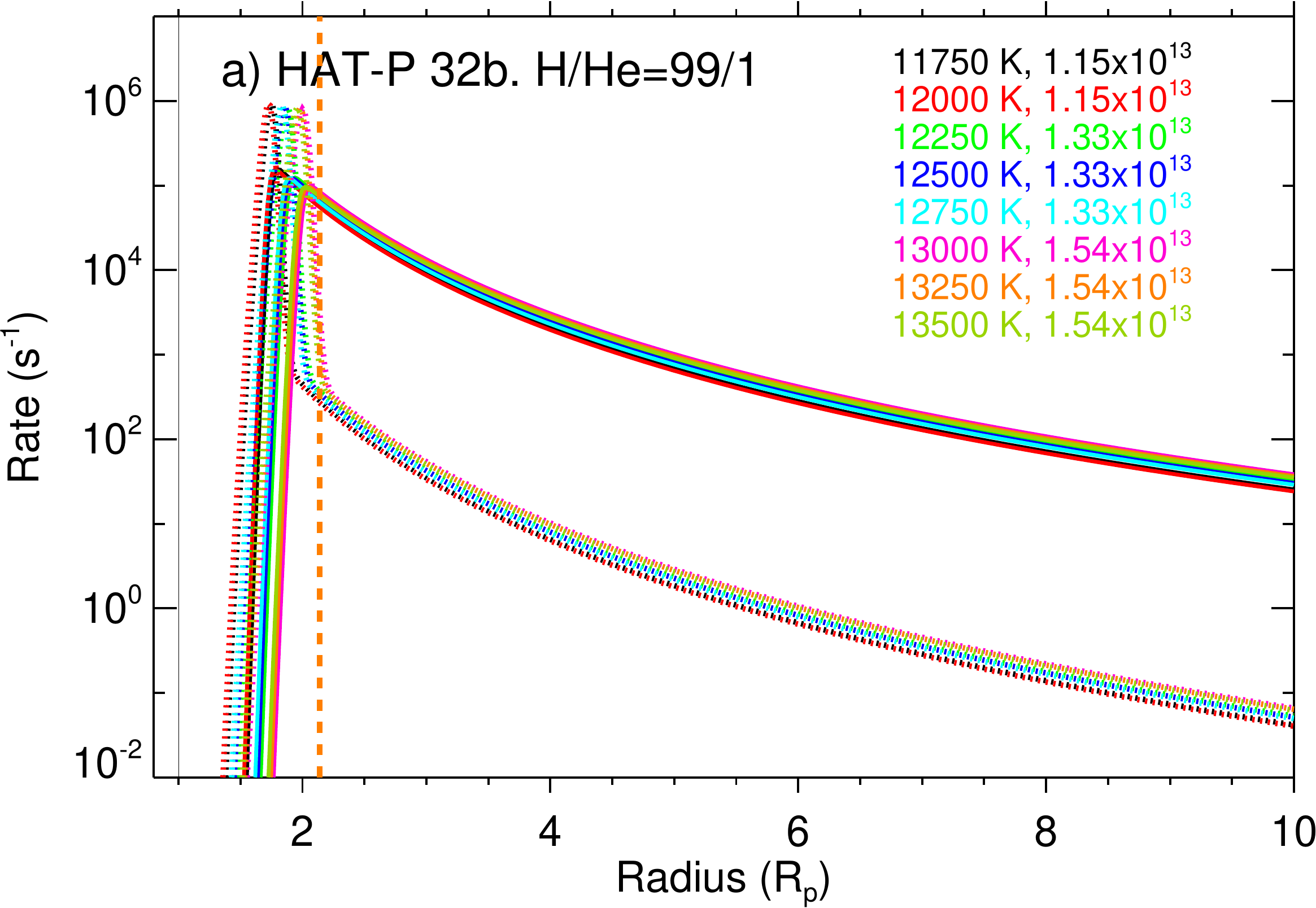}
\includegraphics[angle=90, width=0.9\columnwidth]{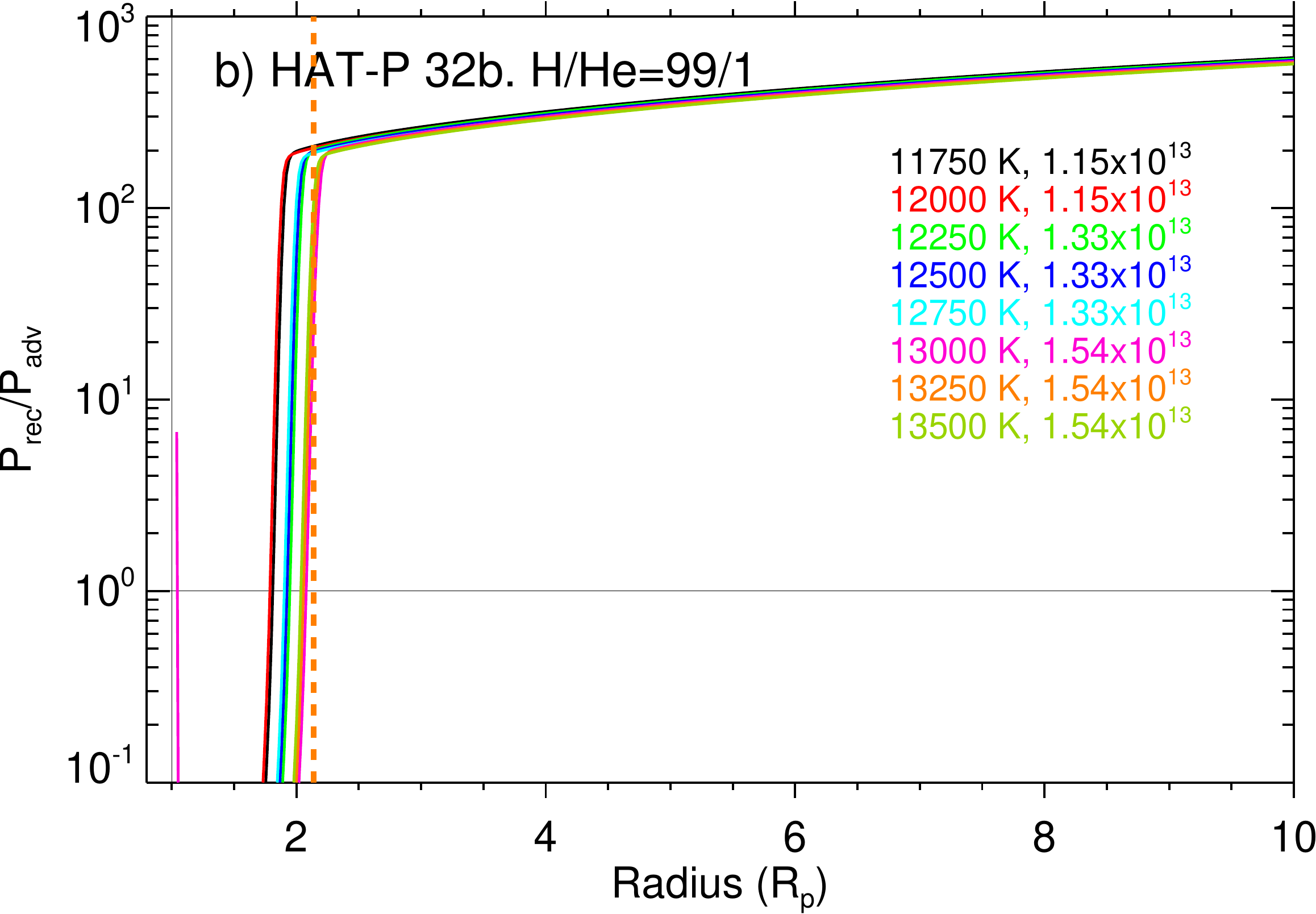}\\
\includegraphics[angle=90, width=0.9\columnwidth]{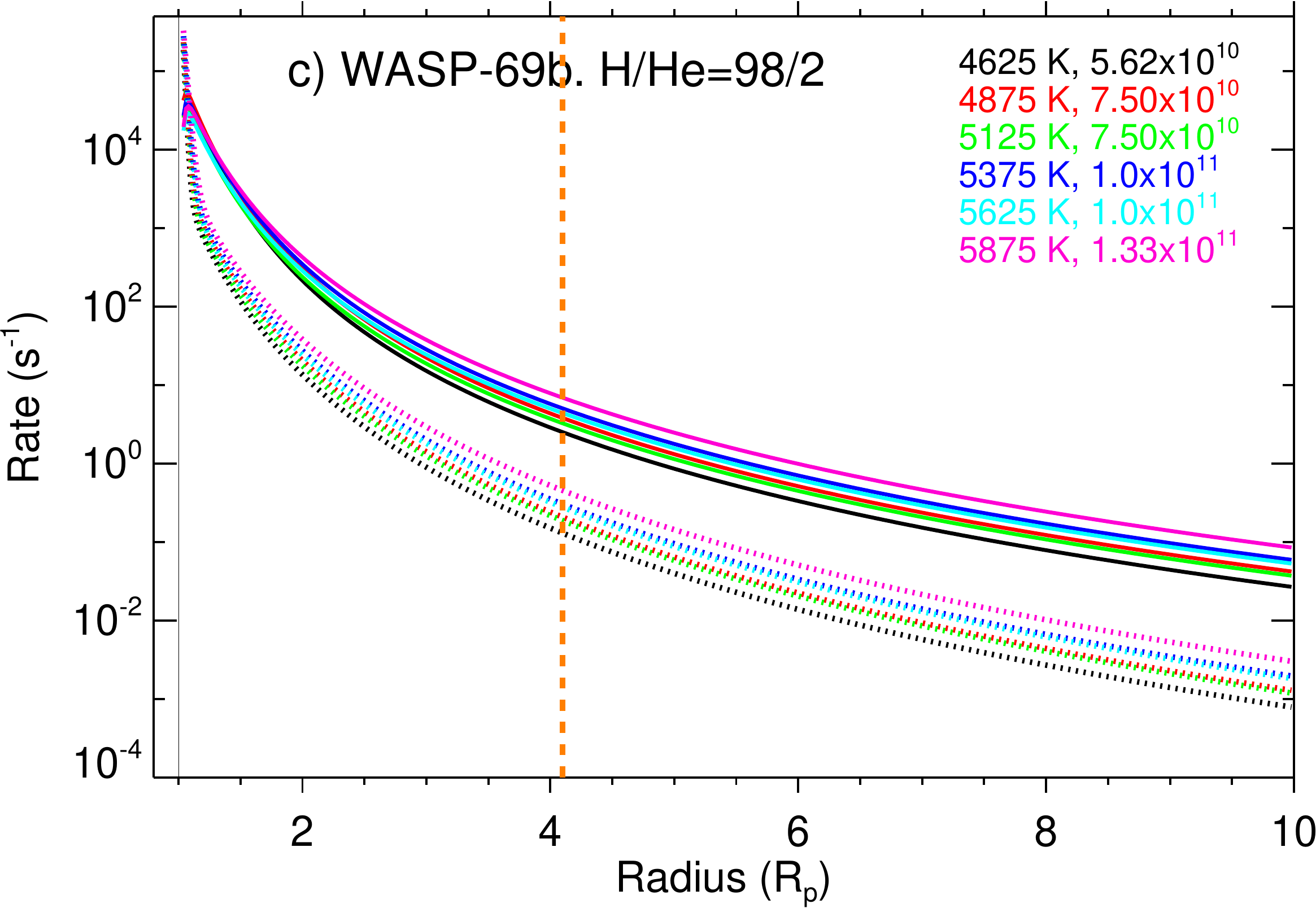}
\includegraphics[angle=90, width=0.9\columnwidth]{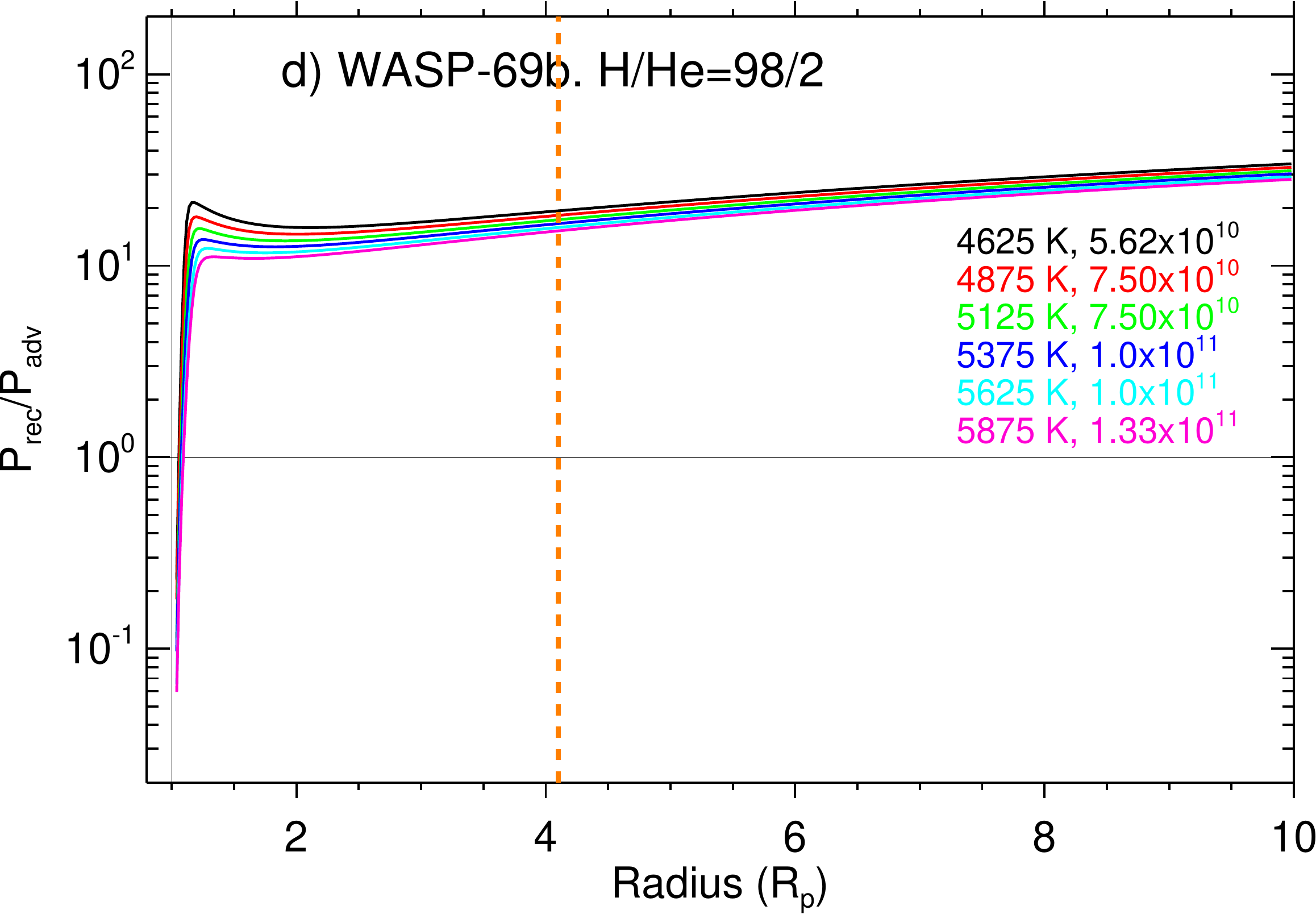}\\
\includegraphics[angle=90, width=0.9\columnwidth]{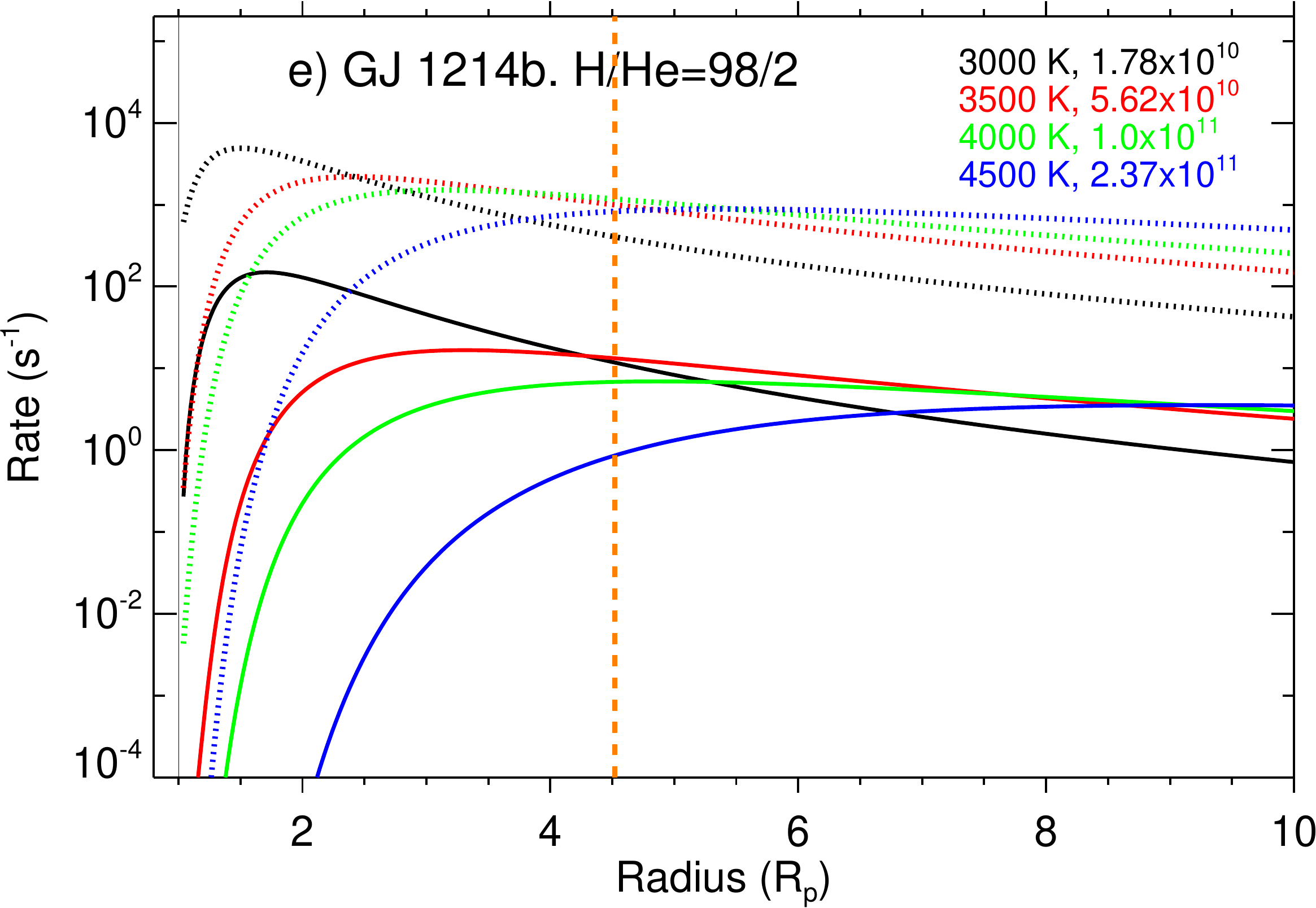}
\includegraphics[angle=90, width=0.9\columnwidth]{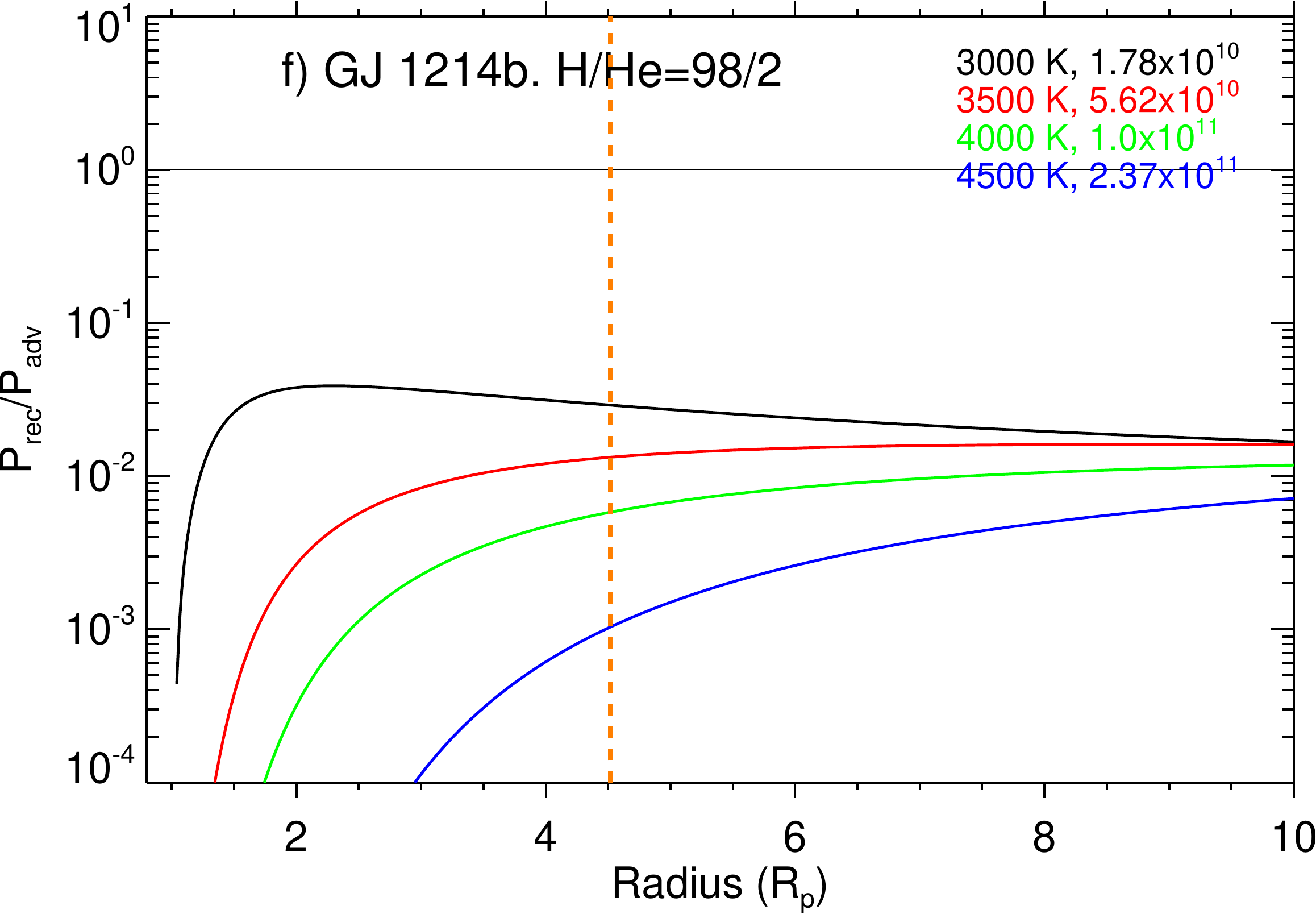}\\
\includegraphics[angle=90, width=0.9\columnwidth]{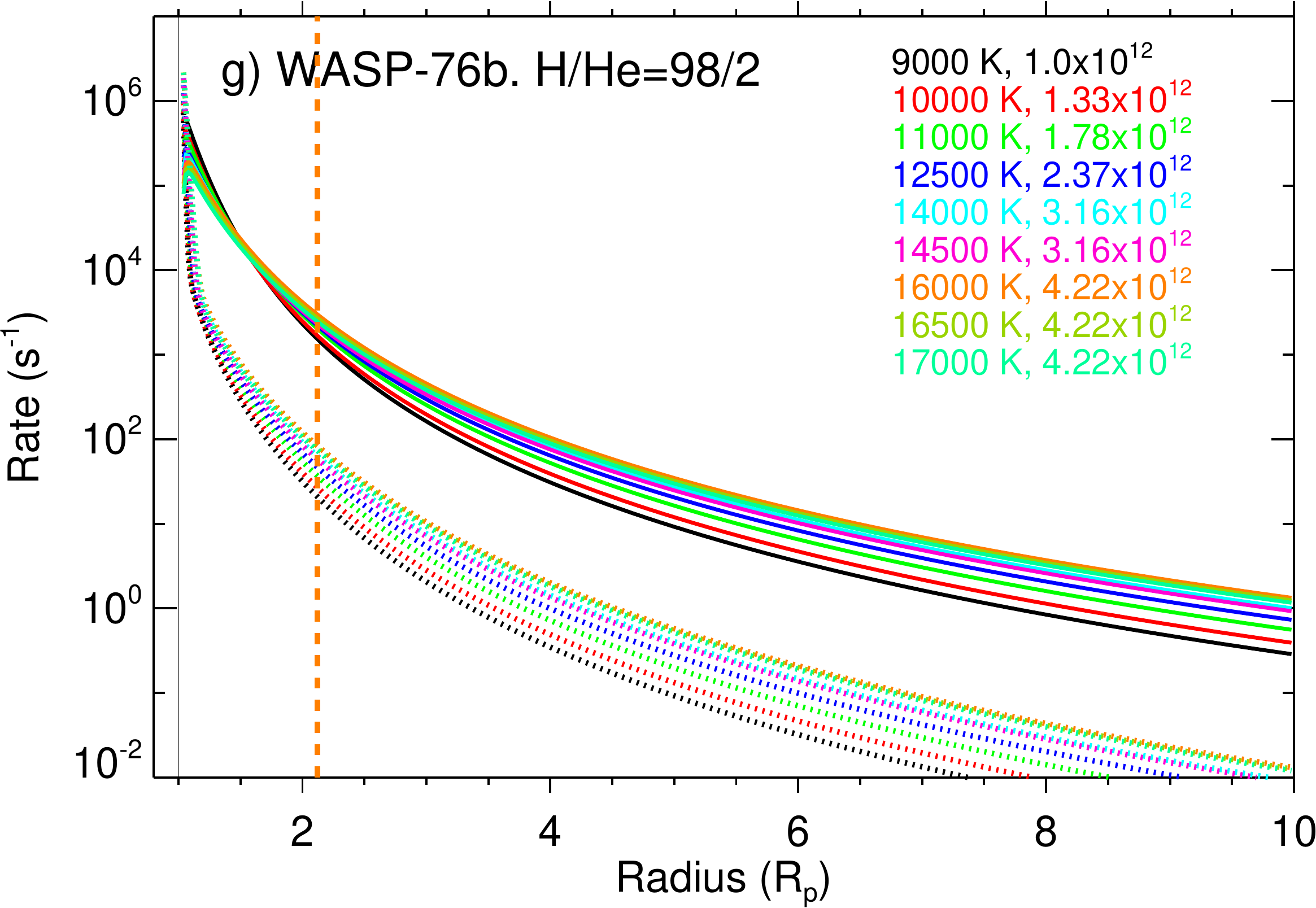}
\includegraphics[angle=90, width=0.9\columnwidth]{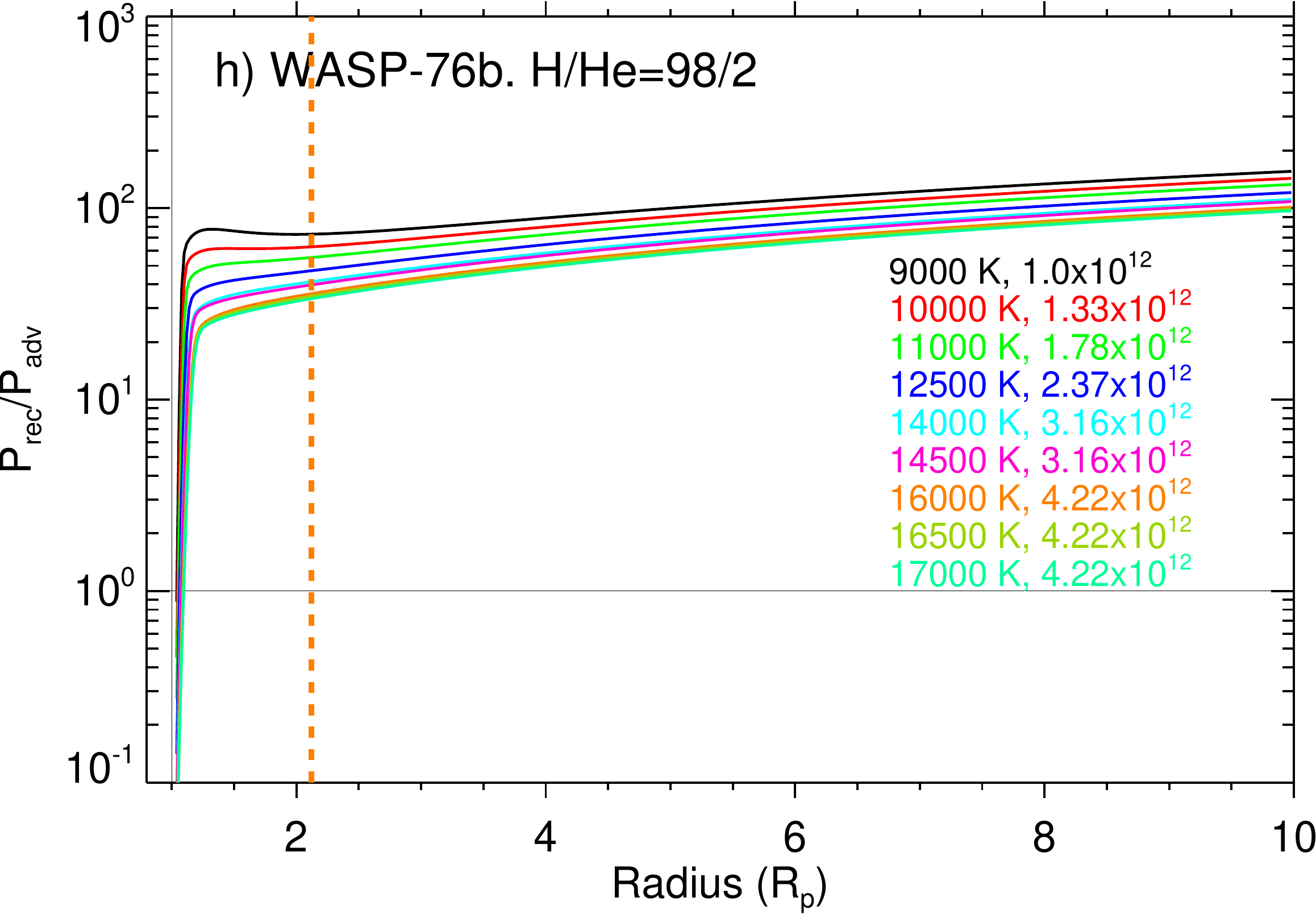}
\caption{Recombination and advection for HAT-P\,32b, WASP-69b, GJ\,1214b, and WASP-76b. Left column: Production rates of H by recombination  (solid lines) and advection (dashed lines) derived from the fits of the measured \het\ absorption spectra (see symbols in Fig.\,\ref{chi2}). Right column: Ratios of the recombination to the advection. Data, from top to bottom, for the atmospheres of  HAT-P\,32b, WASP-69b, GJ\,1214b, and WASP-76b. The vertical dashed orange lines indicate the mean Roche lobes. The different scales of the $x$- and $y$-axes for the different planets should be noted.}
\label{hterms} 
\end{figure*}

\subsection{\hatp32} \label{sec_hatp32}

The reduced $\chi^2$ contour map of \hatp32\ for our nominal case is shown in Fig.\,\ref{chi2}a.
In these calculations, we assumed a H/He ratio of 99/1 (see below) and included the turbulence broadening component at the corresponding temperature. The observed absorption line is well fitted by the simulations from the constrained $T$\,-\,\mlr\ range (white circles).
The red curve in Fig.\,\ref{absorption}a corresponds to the synthetic absorption of one of these simulations, $T$\,=\,12\,500\,K and \mlr\,=\,1.4\,$\times$10$^{13}$\,\gs.
At low (high) temperatures the gas radial outflow velocities are lower (higher) (see Fig.\,\ref{vel}a), which yields a narrower (broader) absorption line and consequently worse fits (higher $\chi^2$).

In the nominal calculations above we did not include any atmospheric component moving away from (red) or  towards (blue) the observer. However, \cite{Czesla_2022} showed  (see Sect.\,\ref{Obs}) that some of these components are present at other phases of the transit. Hence, they could also be present, although at different velocity shifts, in the observed mid-transit spectra. Thus, to explore these potential contributions to the derived mass-loss rates, we conducted further calculations following \cite{Czesla_2022}. These tests consider, in addition to the gas radial velocities from the hydrodynamic model,
a 20\% fraction of the atmosphere moving with a redshift of 20\,\kms. As this contribution broadens the synthetic absorption, the derived temperatures and mass-loss rates extend to lower values:  9500\,K and 
\mlr\,$\simeq$\,6\,$\times$10$^{12}$\,\gs\ (see Fig.\,\ref{chi2}a, upward white triangles). 

Similarly, to assess the effect of turbulence, we repeated the analysis by excluding its broadening. We found that this broadening does not significantly alter the derived temperatures and mass-loss rates because it is much smaller than the widening produced by the large gas radial outflow velocities in the upper atmosphere of this planet (see Fig.\,\ref{vel}a).  

As discussed in Sect.\,\ref{Obs}, the mid-transit spectrum also shows a significant absorption at  $\lambda$\,$\approx$\,10831.5\,\AA, which is appreciably blueshifted from the position of the weak \het\ line. However, as this feature is only significant at the mid- and end-transit phases, it is uncertain whether this absorption is produced by the bulk atmosphere. To determine the contribution of this possible absorption in our results, we reperformed the analysis by excluding this feature (i.e. considering   the fitting in the spectral range of 10831--10835\,\AA\ and the corresponding 32 spectral points).
However, we found no significant differences in the constrained $T$-\mlr\ range.

\hatp32\ shows a very narrow ionisation front (IF) located at high altitudes (1.5--2\,\rp; see Fig.\,\ref{hvmr_candidates}a),
which is a rather different scenario to those found for other planets (i.e. \hd20, \hdu18, \gj34, \w69, \gju12, and \wa76). 
The assumption of constant speed of sound that we apply in our hydrodynamic model could lead to the overestimation of the gas radial outflow velocities in the lower part of the IF region (e.g. compare  the velocity profiles in Fig.\,\ref{vel}a with those of Fig.\,16 in \citealt{Czesla_2022} obtained with the more comprehensive model of \citeauthor{Garcia_munoz_2019}, \citeyear{Garcia_munoz_2019}). This overestimation could produce a non-realistic broadening of the \het\ absorption if the IF region is narrow and located at high altitudes where the velocities and their gradients are significant. In order to analyse this possible degeneracy, we repeated the analysis for the nominal case but assuming zero radial velocities of the gas inside the IF region. We found that the \het\ absorption line inside the IF region is significantly narrowed, and that  the derived temperatures and mass-loss rates extend to higher values,  15250\,K and 
\mlr\,$\simeq$\,2\,$\times$10$^{13}$\,\gs, as shown in Fig.\,\ref{chi2}a (downward white triangles).

In addition to the \het\ absorption, \cite{Czesla_2022} also measured \ha\ absorption spectra in \hatp32. By modelling this absorption with a hydrodynamical model that includes only H (no He), they found that the \ha\ absorption (e.g. the concentration of H(2)) is mainly produced in the IF region, which is very narrow and located at $\sim$1.8 \rp\ \citep[see Fig.\,16 in][]{Czesla_2022}. Thus, we focus on reproducing with our model the concentrations of H$^0$ and H$^{+}$ derived from the \ha\ absorption just in that IF region.
In Fig.\,\ref{hden_candidates} we show the H$^0$ 
profiles that resulted from our models corresponding to the constrained $T$\,-\mlr\ ranges for the \het\ for H/He ratios of 90/10, 99/1, and 99.8/0.2 (blue, black, and orange curves, respectively), together with those derived by \cite{Czesla_2022} (magenta curve).
The H/He\,=\,99/1 case yields a   location of the IF region (at about 1.8\,\rp) similar to that obtained from the \ha\ measurements. 
We should note, however, that the H$^0$ and H$^{+}$ concentrations are lower than those derived by \cite{Czesla_2022}, so that our predicted \ha\ absorption underestimates the measurements. 
However, assuming lower gas radial outflow velocities in the lower part of the IF region (as discussed above), would yield higher H$^0$ and H$^{+}$ concentrations 
{that would reproduce the observed \ha\ absorption with no significant variation on our constrained $T$\,-\,\mlr\, ranges}.  
For H/He ratios of 90/10 and 99.8/0.2, the IF region is far away from that derived by \cite{Czesla_2022}.  
Therefore, we conclude that the H/He ratio in the upper atmosphere of \hatp32\ is (99.0/1.0)$^{+0.5}_{-1.0}$. For this H/He ratio and taking into account the complete constrained $T$\,-\,\mlr\ range (i.e. white circles, and upward and downward white triangles in Fig.\,\ref{chi2}a),
we derive a \mlr\,=\,(130\,$\pm$\,70)\,$\times$\,10$^{11}$\,\gs\ and $T$\,=\,12\,400\,$\pm$\,2900\,K.

Regarding the \het\ abundance, we find that its density distribution is confined to a narrow range, decreasing by about two orders of magnitude from $\sim$\,1.6\,\rp\ to $\sim$\,5.5\,\rp\ (see Fig.\,\ref{he3}a).
It peaks at the IF which, as already discussed, is very narrow and is located at relatively high altitudes, $\sim$\,1.6\,\rp\ (see Fig.\,\ref{hvmr_candidates}a).
The outflow is almost completely ionised in the whole upper atmosphere (as it begins where the IF starts). In this region, the H$^0$ production is dominated by recombination, as shown in the panels in the first row of Fig.\,\ref{hterms}, and the heating efficiency is very low, 0.017\,$\pm\,$0.003. Thus, we conclude that \hatp32\ is in the recombination-limited (RL) regime (see Sect.\,\ref{sec_RH}).

In summary, we find that \hatp32\ photo-evaporates at \mlr\,=\,(130\,$\pm$\,70)\,$\times$\,10$^{11}$\,\gs\ (thus refining the order of magnitude of 10$^{13}$\,\gs\ reported by \citealt{Czesla_2022}), has a maximum upper atmospheric temperature of 12\,400\,$\pm$\,2900\,K, has an upper atmospheric H/He ratio of $\sim$\, 99/1, is in the RL regime, and has a heating efficiency of 0.017\,$\pm$\,0.003.  
We also find that turbulence broadening is not significant in this planet, and the presence of the absorption at $\lambda$\,$\approx$\,10831.5\,\AA\ does not alter our results. Our derived $T$\,-\mlr\ range, accounts for plausible different wind configurations and different velocities in the IF zone, which have a significant effect on the absorption broadening.

\subsection{\w69} \label{sec_w69}

\subsubsection{General results} 
Figure\,\ref{chi2}b shows the reduced $\chi^2$ contour map 
of \w69\ for our nominal case. The results were obtained by including the hydrodynamical model outputs of the \het\ concentrations and the radial velocities of the gas. They were obtained for a H/He ratio of 98/2 (see discussion below). In the calculations of the line absorption, we included a net blueshift of $-$4.5\,\kms, as suggested by the observations (see Fig.\,\ref{absorption}b), and also the turbulence broadening at the appropriate temperature.
That panel shows two simulations for one $T$\,-\,\mlr\ pair, \mlr\,=\,10$^{11}$\,\gs, and $T$\,=\,5375\,K (i.e. for one of the white circles in Fig.\,\ref{chi2}b). We see, on the one hand, that the model reproduces  the observed absorption very well and, on the other, that the broadening of the lines produced by the radial outflow is significant (compare red and blue curves), and thus allows us to constrain the $T$\,-\,\mlr\ range.
As shown in Fig.\,\ref{chi2}b (white dots) the range of $T$\,-\,\mlr\ is very narrow, mainly caused by the dependence of the gas radial outflow velocities on temperature (see Fig. \ref{vel}b). 
At low (high) temperatures the radial velocities are lower (higher) and yield narrower (broader) synthetic profiles and consequently worse fits (higher $\chi^2$).

The narrow $T$\,-\,\mlr\ range described above was obtained under the assumption that we have an overall net blueshift of the whole upper atmosphere of $-$\,4.5\,\kms. However, the absorption of the combined T2-T3 transit spectra analysed here is also compatible with other combined atmospheric blue- and redshifted components, as the observations at different transit phases have shown the existence of such components \citep[see Sect. \ref{Obs} and][]{Nortmann2018}. 
Thus, we   further explored the $T$\,-\,\mlr\ range by replacing that entire blueshift by other components similar to those observed at ingress and egress. In particular, we considered the case of including two blueshifted components, at $-$3.6\,\kms\ and  $-$10.7\,\kms\ for respective fractions of 45\% and 30\% of the upper atmosphere 
\cite[sectors of 0.9$\pi$ and 0.6$\pi$\,rad, see Eq.\,17 in][]{Lampon2020}, together with 15\% (0.3$\pi$\,rad) of the atmosphere moving away from the observer at 1.4\,\kms. For this case the constrained $T$\,-\,\mlr\ range (upright white triangles) is extended to lower temperatures and lower mass-loss rates. The reason is that the blue and red components significantly broaden the absorption line, and thus fitting the line requires lower gas radial velocities, that is, lower temperatures.

We repeated the analysis, but excluded the turbulence broadening. This 
resulted in $T$\,-\,\mlr\ slightly larger (downward white triangles) 
since neglecting the turbulence broadening leads to narrower lines.
Thus, overall, accounting for the lack of knowledge of these effects, we find a slightly extended range of the constrained $T$\,-\,\mlr.

As we   discuss in Sect.\,\ref{Intro},   the H/He concentration has a large impact on the mass-loss rates derived from \het\ absorption measurements. 
Unfortunately, there are no available observations of the H$^0$ lines for \w69. \cite{Khalafinejad_2021} found a possible signature of \ha\ absorption, but they did not claim it as a significant detection. Thus, based on the fact that for most of the observed planets high H/He ratios have been derived (see Sect.\,\ref{Intro}), we assumed in our analysis a high H/He ratio of 98/2.
In any case, we complemented the analysis by assuming a low H/He concentration of 90/10. This results, as expected, in significantly warmer temperatures and lower mass-loss rates (see Fig.\,\ref{chi2}b, black dotted line and symbols). These temperatures and mass-loss rates can be considered as extreme values, although we think, based on other planets results, that they are unlikely, and we propose the range described above for the nominal model.

Our results show that the \het\ density distribution of \w69\ is rather spatially confined, peaking at the lower boundary of the model (i.e. at the base of the upper atmosphere) and decreasing by about two orders of magnitude near 4\rp\ (see Fig. \ref{he3}b).
The outflow is almost completely ionised in the upper atmosphere, with a narrow ionization front (see Fig.\,\ref{hvmr_candidates}b). The H$^0$ production is dominated by recombination in practically the entire upper atmosphere (see second row of panels in Fig.\,\ref{hterms}), 
and the heating efficiency is very low, 0.02\,$\pm$\,0.01, which indicates a significant radiative cooling of the gas. Thus, we conclude that \w69\ is in the recombination-limited (RL) regime.

\subsubsection{Discussion on \w69} 

In this section we compare our results with those derived from previous works.
\cite{Vissapragada_2020}
observed the \het\ absorption using ultra-narrow band photometry. From these observations, which agree well with those of \cite{Nortmann2018} analysed here, they constrained $T$ and \mlr\ using a 1D isothermal Parker wind model \citep{Oklopcic2018} and assuming a solar-like H/He ratio (90/10). For that H/He ratio, our mass-loss rates are about 30\% higher than those derived by \cite{Vissapragada_2020} for a given temperature 
in the range of 5000-10\,000 K. 
Therefore, despite analysing different \het\ measurements
from different observational techniques (spectroscopic and photometric) and using different stellar fluxes, our results are in reasonable agreement.
\cite{Vissapragada_2020} did not constrain the $T$\,-\mlr\, range;  more recently, \cite{Vissapragada_2022}, still maintaining a H/He=90/10, imposed constraints by studying its maximum mass-loss efficiency. Our constrained $T$\,-\mlr\, range is consistent with their estimations, as they found that the outflow must be cooler than 14\,000\,K and that \mlr\,$\lesssim$\,3$\times$10$^{11}$\,\gs.
Interestingly, in contrast to \cite{Nortmann2018}, \cite{Vissapragada_2022} found no evidence of a tail in the post-transit, which is probably due to the wide bandpass they used. 
Our derived range of $T$\,-\mlr, however, also accounts for this case of not having a tail.

\cite{Wang_2021} analysed the \het\ measurements reported by \cite{Nortmann2018} and \cite{Vissapragada_2020} using a 3D model. Assuming  a H/He ratio of
$\sim$\,91/9, they derived a mass-loss rate of $\sim$\,10$^{11}$\,\gs\ and a maximum temperature of  9000\,K. Our results for a similar H/He yield a mass-loss rate of about half of their value. 
However, when we consider a H/He of 98/2 which, as discussed above, seems to be more likely, we obtain a very similar mass-loss rate.
However, the temperature we obtain is lower. For a H/He ratio of 98/2 we found a temperature in the range   5000--6000\,K and  for H/He=90/10 we found 5500--7500\,K, which are lower than their maximum temperature.

\cite{Khalafinejad_2021}, by combining low- and high-resolution measurements in the visible including the Na lines,  obtained a thermospheric temperature of 6000$\pm$3000\,K for this planet. Their analysis suggests that the Na lines are probing pressures lower than the millibar region,  
which is at the lower boundary 
of our hydrodynamic model.
Their temperatures are thus consistent with the temperature range of 5250$\pm$750\,K obtained here.

In summary, we found that non-radial winds and turbulence significantly broaden the \het\ absorption line in this planet.
The H/He ratio cannot be constrained as H$^0$ observations are not available for \w69.
Nevertheless, we constrained   \mlr\ in the range  (0.9\,$\pm$\,0.5)$\times$10$^{11}$\,\gs\ and   temperature in the range  5250\,$\pm$\,750\,K, assuming a likely H/He ratio of 98/2.

\begin{figure}[t]
\includegraphics[angle=90.0, width=\columnwidth]{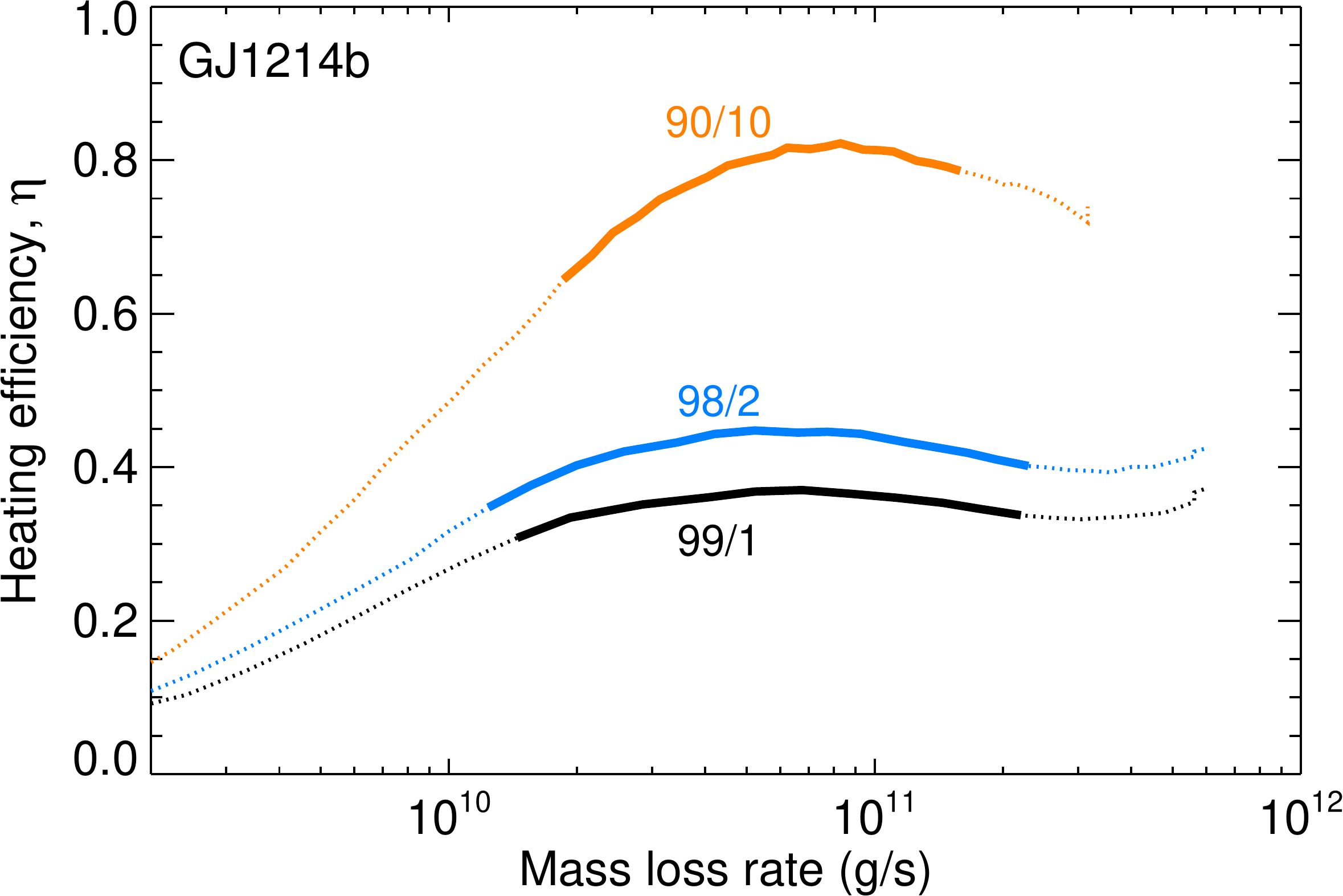}
\caption{Heating efficiencies for GJ\,1214b for different H/He ratios.}
\label{eta_gj1214} 
\end{figure}

\subsection{ \gju12} \label{sec_gj12}

For most of the planets that we have studied so far (\hd20, \hdu18, \gj34, and \hatp32), we determined the H/He ratio using the H$^0$ distribution derived from either \lya\ or \ha\ observations. For this planet, no such observations are available. In this case we therefore tried to put some constraints on that ratio based on arguments of heating efficiency. We found that the outflows derived from the \het\ absorption when considering a H/He ratio of 90/10 requires a very high heating efficiency, $\eta$, while those obtained with higher H/He ratios require much more moderate values of $\eta$.
Figure\,\ref{eta_gj1214} shows the heating efficiency of \gju12\ for H/He ratios of 90/10, 98/2, and 99/1 (see Sect.\,\ref{grid}, Eq.\,\ref{eq:energy_lim}), where it is worth   noting that $\eta$ decreases when the H/He ratio increases. Then, by calculating the upper limit of $\eta$ we constrained the H/He ratio.
The maximum heating efficiency, $\eta_{\textrm {max}}$, is given by (see Appendix\,\ref{ap:eta_max})
\begin{align}
    \eta_{\textrm {max}} = 1 - \frac{E_{0}\,\int_{v_{\textrm {0}}}^{\infty} I(v)\,dv}{F_{XUV}}~, 
    \label{eq:heating_max3}
\end{align}
where $v_0$ and E$_{0}$ are respectively the ionisation frequency and the ionisation energy of hydrogen in the ground state (i.e. E$_{0}$\,=\,13.6\,eV); $F_{XUV}$\,=\,$\int_{v_{\textrm 0}}^{\infty} E_{v}\,I(v)\,dv$; and I($v$) is the irradiation at the top of the upper atmosphere in photons\,s$^{-1}$\,cm$^{-2}$\,Hz$^{-1}$ (calculated from Fig.\,\ref{flux_xsec}). From Eq.\,\ref{eq:heating_max3} we obtained $\eta_{\textrm {max}}$\,$\sim$\,0.53 and, by comparison with the values obtained for different H/He ratios (see Fig.\,\ref{eta_gj1214}), we conclude that H/He\,$\gtrsim$\,98/2 for this planet. 
We note that including processes such as the He photo-ionisation energy loss and the radiative cooling and accounting for ionising photons not absorbed by the upper atmosphere, could reduce the value of $\eta_{\textrm {max}}$ \citep[see Appendix\,\ref{ap:eta_max} and e.g.][]{Shematovich_2014,Vissapragada_2022}.
However, to constrain the H/He ratio, we prefer to be conservative and use a higher upper limit, only dependent on $I$($v$). 

As the atmosphere of \gju12\ is very extended (see Fig.\,\ref{he3}), we conducted a test by extending the upper boundary of the model from the nominal value of 20\,\rp\ to 40\,\rp. We found, however, no significant changes in our constrained $T$\,-\,\mlr\ range, as the extra absorption produced is negligible.
Furthermore, for this planet, the turbulence does not significantly broaden the \het\ line, as the widening introduced by the high gas radial outflow velocities clearly dominates. Therefore, this parameter does not help to constrain the $T$-\mlr\ degeneracy in \gju12. 

The temperature and mass-loss rates obtained for \gju12\ are shown in Fig.\,\ref{chi2}c. We derived temperatures in the range of 3000--4500\,K and \mlr\ in the interval of (0.2--2.4)\,$\times$\,10$^{11}$\,\gs, which means that, due to the rather steep gradient of the derived $T$-\mlr\ curve, the uncertainty in the mass-loss rate is rather large. We note that some of the reduced $\chi^2_R$ values obtained for this planet (and for \wa76; see Sect. \ref{sec_wasp76} and Fig.\,\ref{chi2}d) are smaller than unity. 
This is likely caused by a too conservative estimation of the absorption errors for these planets. We   checked that rescaling the errors does not significantly change the derived temperatures and mass-loss rates. 
Regarding the potential effects of the SW for this planet, only   assuming the fast solar wind would yield a significant change (a factor between four and two) in the \mlr\ range (0.8--4.0)\,$\times$\,10$^{11}$\,\gs.

As mentioned above, \cite{Orell-Miquel_2022} reported a tentative detection of the \het\ signal, and later \cite{Spake_2022} could not confirm it, reporting an upper limit of 1.22\% for the He triplet absorption. If we consider this upper limit we would derive a mass-loss rate of about 0.2\,$\times$\,10$^{11}$\,\gs\ for a temperature of 3500\,K, which is about a factor of two smaller. The escape regime, however, is not affected.

Figure\,\ref{hvmr_candidates}c shows that the ionisation front of this planet is very extended, occupying the whole upper atmosphere. Advection dominates over recombination (see Fig.\,\ref{hterms}e and Fig.\,\ref{hterms}f) and the heating efficiency is 0.43\,$\pm$\,0.03 for a H/He range of 98/2 (see Fig.\,\ref{eta_gj1214}). We conclude that \gju12\, is in the photon-limited regime, which supports the result obtained by \cite{Orell-Miquel_2022}.

\begin{figure}[t]
\includegraphics[angle=90, width=\columnwidth]{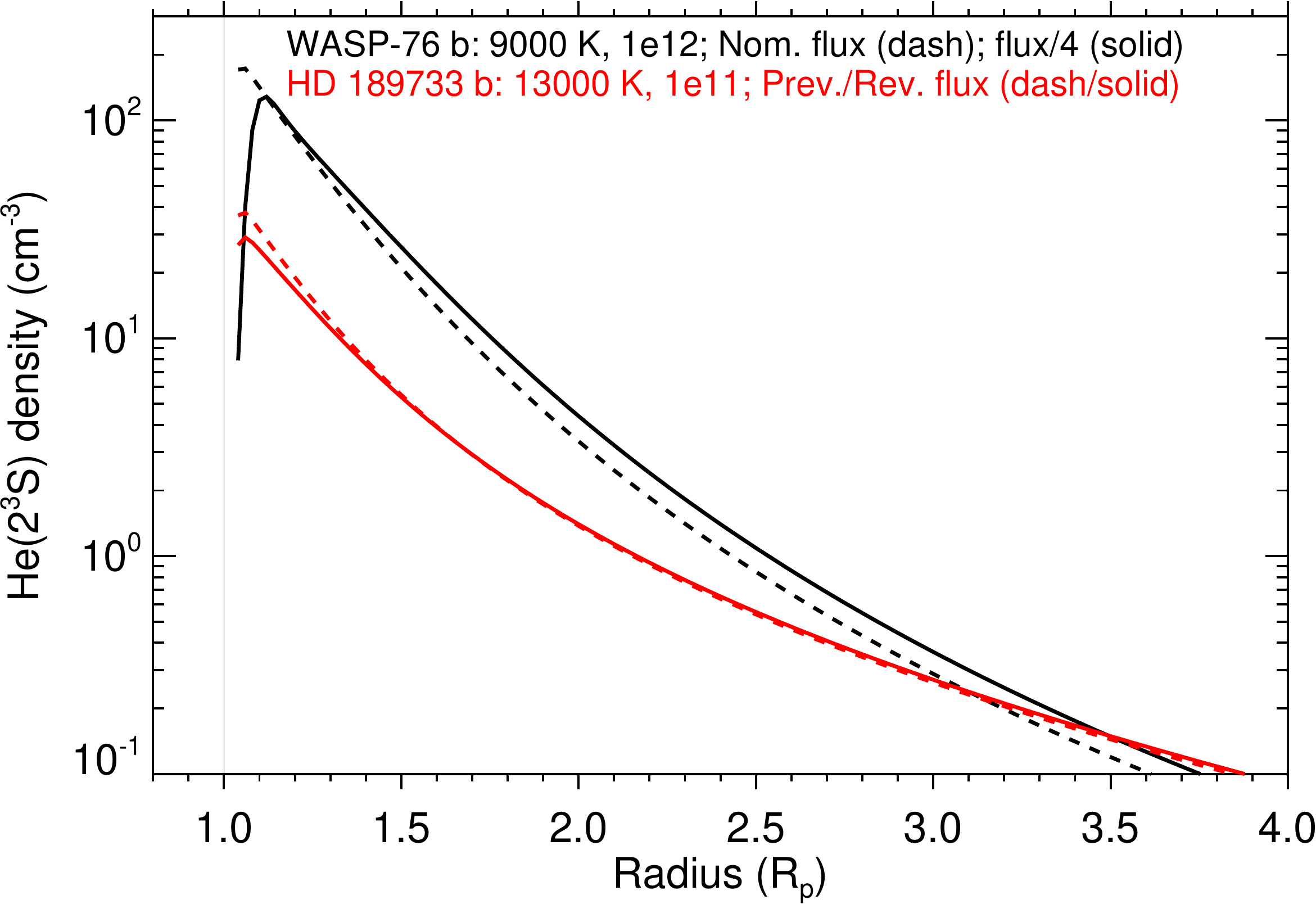} 
\caption{Effects of the XUV flux density upon the \het\ concentration for {WASP-76 b} and \hdu18. Dashed lines: nominal and previous fluxes; Solid lines: perturbed and revised (smaller) fluxes.
} 
\label{fluxdiv4}
\end{figure}

\subsection{WASP-76 b} \label{sec_wasp76}

The \het\ signal obtained by \cite{Casasayas_2021} is rather weak and affected by telluric contamination close to the stronger \het\ lines. The absorption of about 0.5\% is to be considered an upper limit. In addition, the \het\ signal was found to be redshifted, which is uncommon compared with the rest of \het\ observations. 
Further, this planet is an ultra-hot Jupiter, strongly irradiated in the XUV. Hence, for all these reasons the analysis of the \het\ signal is important in order to better understand the escape from gas giants.

As suggested from the observations, we included in the calculations of the line absorption a blueshifted component, at $-$33\,\kms\ and a redshifted component at 12\,\kms\ with respective fractions of 40\% and 30\% of the upper atmosphere \cite[sectors of 0.8$\pi$ and 0.6$\pi$\,rad, see Eq.\,17 in][]{Lampon2020}. The results on the temperature and mass-loss rates obtained for WASP-76 b are shown in Fig.\,\ref{chi2}d. Because of the larger errors of the absorption (see Fig.\,\ref{absorption}d),  
the temperature and mass-loss rate are poorly constrained.
Thus, the temperature is constrained from 6000\,K to  17000\,K, and the mass-loss rate spans from 2 to 45\,$\times$\,10$^{11}$\,\gs\ (see Fig.\,\ref{chi2}d). If we consider confidence levels in the $\chi^2$ of 90\% and 68\% instead of 95\%, the upper values are reduced to 15\,000\,K and 12\,000\,K for temperature and to 30\,$\times$\,10$^{11}$\,\gs\ and 20\,$\times$\,10$^{11}$\,\gs\ for \mlr,  respectively. 
The H/He ratio for its atmosphere has not been determined. 
The analysis of \ha\ by \cite{Tabernero2021} was inconclusive. 
Here we used a H/He ratio of 98/2, in line with the values obtained for several other planets (see discussion in Sect.\,\ref{grid}). Thus, because of the H/He uncertainty, the ranges of errors in the temperature and mass-loss rates reported above might be underestimated.

As described in Sect.\,\ref{fluxes}, the stellar flux density used for this planet is an upper limit, and hence we tested how the derived temperatures and mass-loss rates would be affected by uncertainties in the flux density. According to our analysis (see Sanz-Forcada et al., 2022, in prep. for more details), a reasonably lower limit is given by a reduction by a factor of 4 with respect to the nominal irradiation. The use of 
the reduced flux has minor effects on the derived temperatures and mass loss rates (see red dots in Fig.\,\ref{chi2}d).
The upper values have not changed and only the lower limit of the \mlr\ has been reduced from 2 to 1.5 $\times$\,10$^{11}$\,\gs. 
It is interesting though to understand why the mass-loss rates are lower for the reduced flux. By inspection of the \het\ concentration (Fig.\,\ref{fluxdiv4}) we observe that it is reduced at low altitudes, which is expected as for a lower flux density the electron and \hep\ densities are smaller. However, at larger radii the concentration is higher. This is a consequence of the higher density for the reduced flux. This means that for a fixed temperature and mass-loss rate, decreasing the irradiating flux leads to fewer ions and electrons, and hence to a higher mean molecular weight, which 
yields a lower velocity of the gas
giving rise to a higher density, and hence to higher He and \het\ concentrations.
Overall, this effect dominates over the decrease in the concentration at low altitudes in the absorption, leading to the counterintuitive result of having a larger absorption for the weaker flux. Further, to compensate for the stronger absorption we require a lower mass-loss rate in Fig.\,\ref{chi2}d. We note that for \hdu18 (see Sect.\,\ref{revisiting_gj34} below), a planet with a more compressed atmosphere, the opposite behaviour occurs, with the absorption in the inner layers dominating.

Although the mass-loss rate is poorly constrained, and temperature and H/He are essentially unconstrained, we can derive the hydrodynamic escape regime of \wa76. Figures\,\ref{hvmr_candidates}d, \ref{hterms}g and \ref{hterms}h, and \ref{eta} show that the upper atmosphere is strongly ionised; $P_{rec}/P_{adv}$\,$\gg$\,1 in almost the entire upper atmosphere; and the heating efficiency is very low, 0.039\,$\pm$\,0.036, which shows that this planet is in the RL 
regime (see Sect.\,\ref{sec_RH}).
We should note that for temperatures higher than 18\,500\,K the planet is between the RL and the EL regimes. For a H/He of 90/10, the results (not shown) are similar, and its hydrodynamic escape regime would not be altered.

\subsection{Reanalysis of \gj34 and \hdu18} 
\label{revisiting_gj34}

The \het\ transmission absorption of \gj34\ observed by \cite{Palle2020} was studied by \cite{Lampon_2021b} assuming that the mid-transit transmission spectrum corresponds to the average from T2 to T3 contacts. This approximation, however, might not be accurate \citep[see e.g.][]{Dos_santos_2021}.
This, together with the fact that the atmosphere of this planet is very extended, and that the XUV stellar flux has been revised (see Sect. \ref{fluxes} and Fig.\,\ref{flux_new}), 
might significantly alter the constrained temperatures and mass-loss rates.

Here, we reanalysed \gj34   using the same methods used by \cite{Lampon_2021b}, but considering that the measured spectrum is the average for the T2-T3 contacts and also by considering a more-extended atmosphere with an upper boundary of 30\,\rp. The latter was included in order to assure that it covers the whole stellar disc at any transit phase. Using higher upper boundaries does not significantly affect our results. 
The revision of the H/He concentration was also necessary in order to fit the H$^0$ concentration obtained in the new calculations with that derived from \lya\ measurements.

\begin{figure}[t] 
\includegraphics[angle=90, width=\columnwidth]{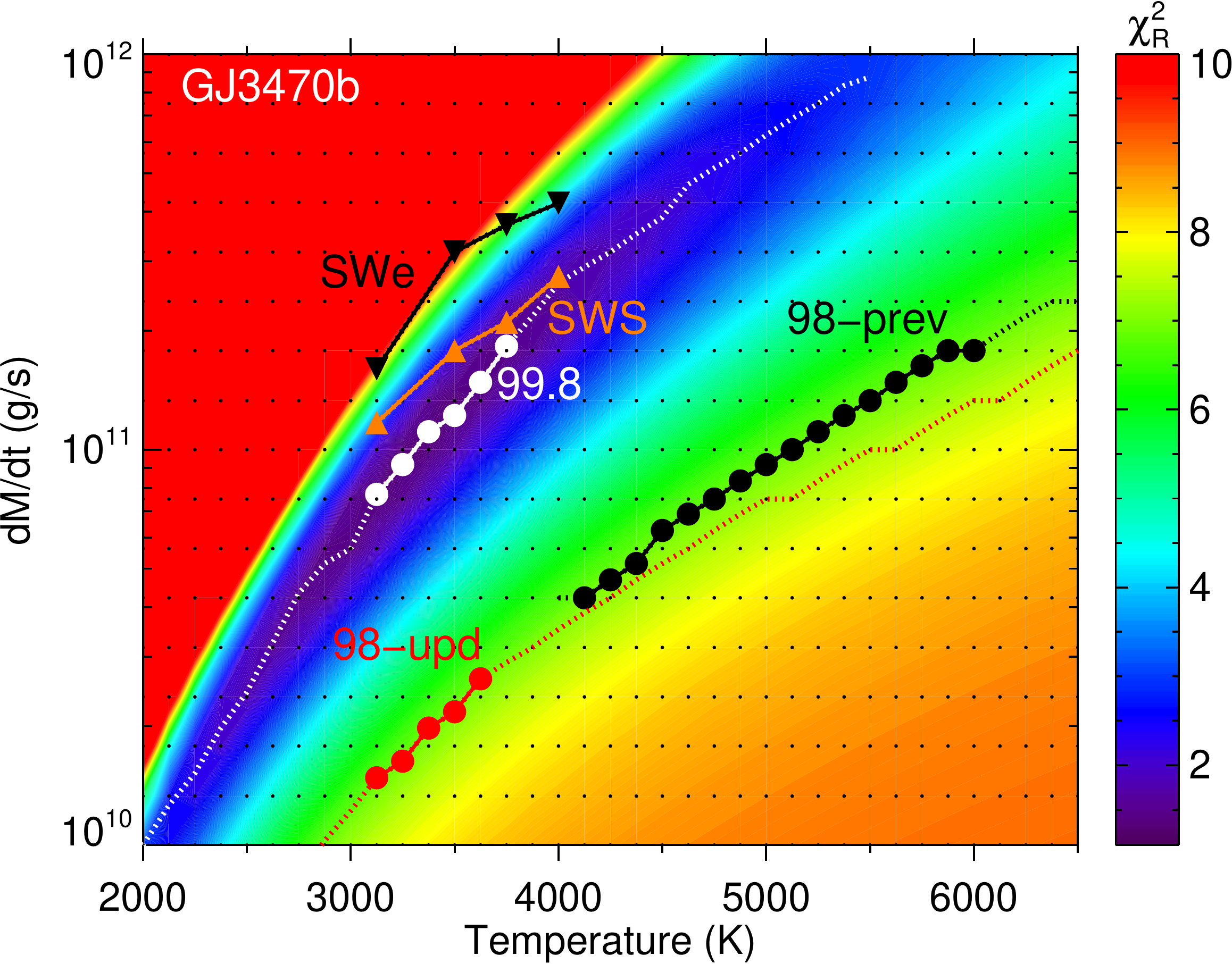} 
\caption{Contour maps of the reduced $\chi^2$ for \het\ absorption of \gj34. Dotted curves represent the best fits, with large circles denoting the constrained ranges for a confidence level of 95\%. 
Overplotted are also the curve and symbols for H/He\,=\,98/2 obtained previously by \cite{Lampon_2021a}, black solid circles, 
where the label `98' corresponds to the hydrogen percentage. The curves and triangles are the \mlr--$T$ ranges obtained for H/He\,=\,99.8 when including the effects of SW (see Sect.\,\ref{sw}) for a fast solar wind (upward triangles, orange) and a scaled stellar wind (downward triangles, black).
The black dots represent the grid of the simulations.} 
\label{chi2_gj34}
\end{figure}

Figure\,\ref{chi2_gj34} shows the derived $T$-\mlr\ range 
and the $T$-\mlr\ range constrained by \cite{Lampon_2021a}.
In the reanalysis, we found significantly colder temperatures, in the range of 3400$\pm$350\,K, 
although the mass-loss rate, in the range of  (1.3\,$\pm$\,0.6)$\times$10$^{11}$\,\gs, is practically the same. As mentioned before, in order to fit the H$^0$ distribution in the revision  we have to change the H/He concentration to $\approx$\,99.8/0.2, higher than that of 98/2 derived by \cite{Lampon_2021a}. 
In this revision we also evaluated the effects of the stellar winds (SW). Their impact on the \mlr--$T$ are small for the fast solar wind case; however, they are significant (\mlr\ $\sim$2 times  larger) when considering the scaled SW (see black and orange triangles in Fig.\,\ref{chi2_gj34}).

The heating efficiency, which is in the range of 0.11--0.12, is slightly lower. The hydrodynamic regime, however, did not change and our results still support that \gj34\ is in the photon-limited regime.

\begin{figure}[t] 
\includegraphics[angle=90, width=\columnwidth]{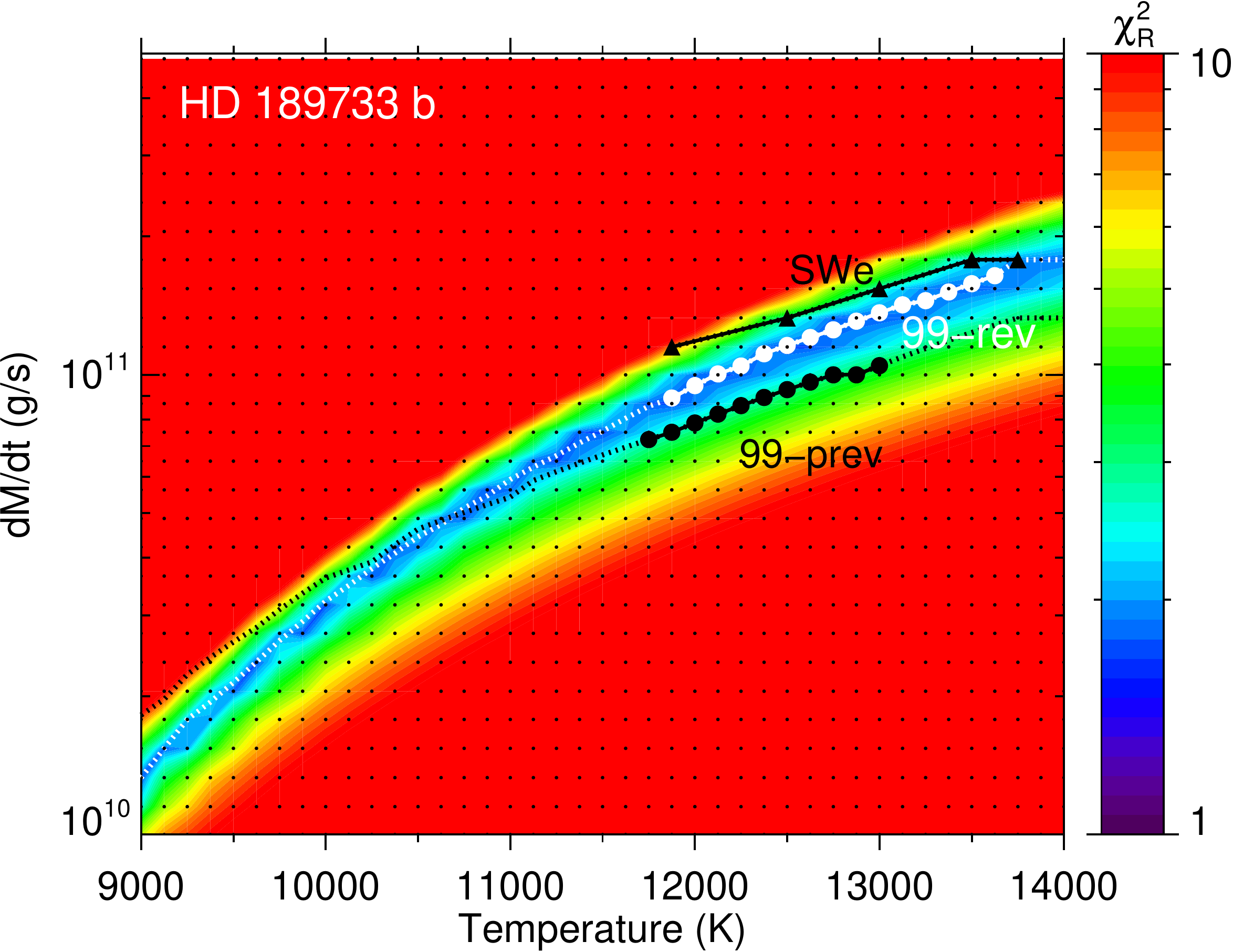} 
\caption{
Contour maps of the reduced $\chi^2$ for the \het\ absorption of \hdu18. Dotted curves represent the best fits with large symbols denoting the constrained ranges for a confidence level of 95\%. 
Overplotted are also the estimations when including the effects of the stellar winds with a scaled solar wind (see Sect.\,\ref{sw}, black triangles), and the curve and circles (in black) obtained previously by \cite{Lampon_2021a}.}
\label{chi2_hd18}
\end{figure}

\begin{table*}
\centering
\caption{\label{eqw}Planet parameters, the XUV flux, and  the equivalent width (EW) \het\ absorption.}
\begin{tabular}{l c c c c c c c  } 
\hline  \hline  \noalign{\smallskip}
Planet                  & \hd20 & \hdu18  & \gj34 & \gju12 & \w69 & \wa76 & \hatp32  \\
\hline \noalign{\smallskip}
Mass ($M_{\rm Jup}$)    & 0.685\,\,$^{+0.015}_{-0.014}$ & 1.162\,\,$^{+0.058}_{-0.039}$ & 0.036\,\,$^{+0.002}_{-0.002}$ &
0.026\,\,$^{+0.001}_{-0.001}$ &
0.260\,\,$^{+0.017}_{-0.017}$ &
0.894\,\,$^{+0.014}_{-0.013}$ &
0.585\,\,$^{+0.031}_{-0.031}$
 \\
\noalign{\smallskip}
Radius ($R_{\rm Jup}$)  & 1.359\,\,$^{+0.016}_{-0.019}$ & 1.230\,\,$^{+0.03}_{-0.03}$ &
0.36\,\,$^{+0.01}_{-0.01}$ &
0.245\,\,$^{+0.005}_{-0.005}$&
1.057\,\,$^{+0.047}_{-0.047}$&
1.854\,\,$^{+0.077}_{-0.076}$ &
1.789\,\,$^{+0.025}_{-0.025}$
\\
\noalign{\smallskip}
$\Phi$\,($\Phi_{\rm Jup}$)\,$^{(a)}$    & 
0.504 & 
0.944 &  
0.100 & 
0.105 & 
0.246 & 
0.482 &
0.327
\\
\noalign{\smallskip}
$R_{\textrm{lobe}}$ (\rp)\,$^{(b)}$     & 
4.2 &  
4.2 &  
5.8 & 
4.52 & 
4.1 &
2.12 &
2.14
\\
\noalign{\smallskip}
$F_{\rm XUV}$\,$^{(c)}$ & 
2.4     &  
25.4  &  
3.7  & 
0.64 & 
23.2 &
145.6 &
417.4
\\
\noalign{\smallskip}
EW (m\AA)\,$^{(d)}$     & 5.3$\pm$0.5   &  12.7$\pm$0.4 & 20.7$\pm$1.3  & 33.2$\pm$3.8 & 28.3$\pm$0.9  &
12.4$\pm$1.7 & 114$\pm$4 \\
\noalign{\smallskip}
H/He &
$\approx$98/2 &
(99.2/0.8)$\pm$0.1       &
$\approx$\,99.8/0.2 &
$\gtrsim$\,98/2 &
-- &
-- &
$\approx$\,99/1
\\
\noalign{\smallskip}
\mlr\,($\times$10$^{11}$\gs)\,$^{(e)}$  & 
0.7\,$\pm$\,0.3 &  
1.4\,$\pm$\,0.5 & 
1.3\,$\pm$\,0.6 & 
1.3\,$\pm$\,1.1 & 
0.9\,$\pm$\,0.5 & 
23.5\,$\pm$\,21.5 & 
130\,$\pm$\,70  \\
\noalign{\smallskip}
$T$\,(K)\,$^{(e)}$      & 
7600\,$\pm$\,500        &  
12\,700\,$\pm$\,900     & 
3400\,$\pm$\,350 & 
3750\,$\pm$\,750 & 
5250\,$\pm$\,750 & 
11\,500\,$\pm$\,5500 & 
12\,400\,$\pm$\,2900    \\
\noalign{\smallskip}
$\eta$\,$^{(e)}$        & 
0.15\,$\pm$\,0.05       &  
0.03\,$\pm$\,0.01       & 
0.115\,$\pm$\,0.005 & 
0.43\,$\pm$\,0.03 & 
0.02\,$\pm$\,0.01 & 
0.039\,$\pm\,$0.036 & 
0.017\,$\pm\,$0.003     \\
\noalign{\smallskip}
$\bar \eta_{90}/ \bar \eta_{h}$\,$^{(f)}$       & 
1.0     &  
0.04    & 
1.58 & 
1.80 & 
0.45 & 
0.33 & 
1.44    \\
\noalign{\smallskip}
HER\,$^{(g)}$ & 
EL      &  
RL      & 
PL & 
PL & 
RL & 
RL & 
RL      \\
\noalign{\smallskip}
\hline
\end{tabular}
\tablefoot{
Data of \hd20\  from \cite{Lampon2020} and references therein. Data for \hdu18\ and \gj34\ from this work, \cite{Lampon_2021a} and references therein.  
\tablefoottext{a}{Gravitational potential in units of Jupiter's potential.} 
\tablefoottext{b}{Roche lobe calculated following \cite{Eggleton_1983}. 
}
\tablefoottext{c}{XUV flux in units of 10$^{3}$ erg\,cm$^{-2}$\,s$^{-1}$ at $\lambda$ < 912 \AA\ at planetary distance, calculated from Fig.\,\ref{flux_xsec}.} 
\tablefoottext{d}{EW integrated in the range 10831.0$-$10834.5\,\AA.}
\tablefoottext{e}{\mlr, $T$, and $\eta$ are calculated assuming a H/He = 98/2 for \gju12, \w69,\ and \wa76. The range of $\eta$ for \hd20\ is assumed from \cite{Shematovich_2014}.}
\tablefoottext{f}{$\bar \eta_h$ is calculated for H/He = 98/2 in \hd20, \gju12, \w69, and \wa76; for H/He = 99/1 in \hdu18, and \hatp32; and  for H/He = 99.8/2 in \gj34.} 
\tablefoottext{g}{HER stands for hydrodynamic escape regime, where EL is energy-limited, RL recombination-limited, and PL photon-limited.} 
}
\label{list_planets}
\end{table*}

\begin{figure}
\includegraphics[angle=90, width=1\columnwidth]{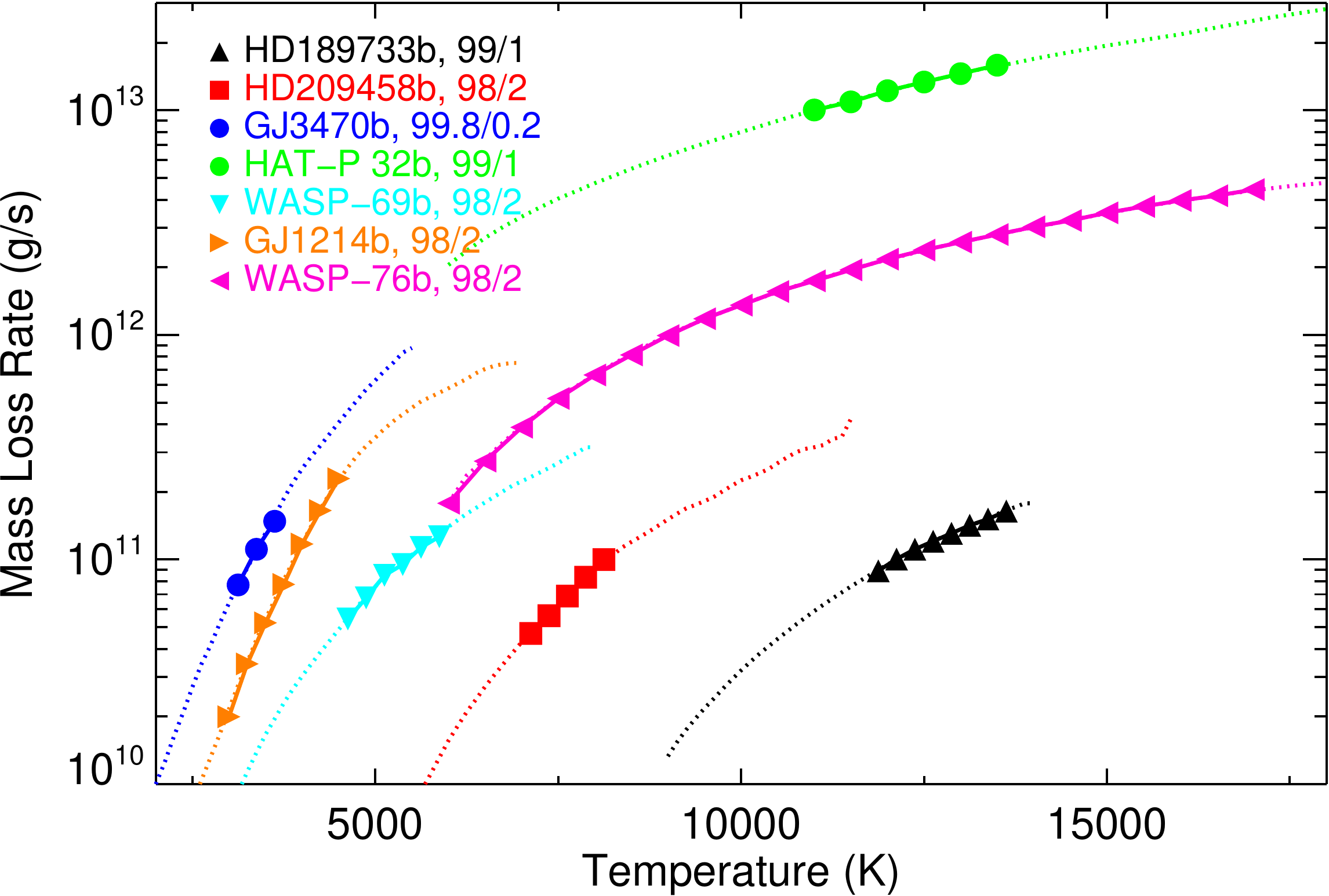} 
\caption{
Ranges of temperature and mass-loss rates for \hd20, \hdu18, \gj34, \hatp32, \w69, \gju12, and \wa76 for H/He ratios (as labelled).
Dotted lines show the extended $T$--$\dot M$ ranges and thick lines the derived $T$--$\dot M$ ranges (see dotted lines and symbols, respectively, in Figs.\,\ref{chi2} and \ref{chi2_gj34}). 
The values for \hd20\ were taken from \cite{Lampon2020}. 
}  
\label{mesos_chi2}
\end{figure}

\begin{figure}[t] 
\includegraphics[angle=90, width=1\columnwidth]{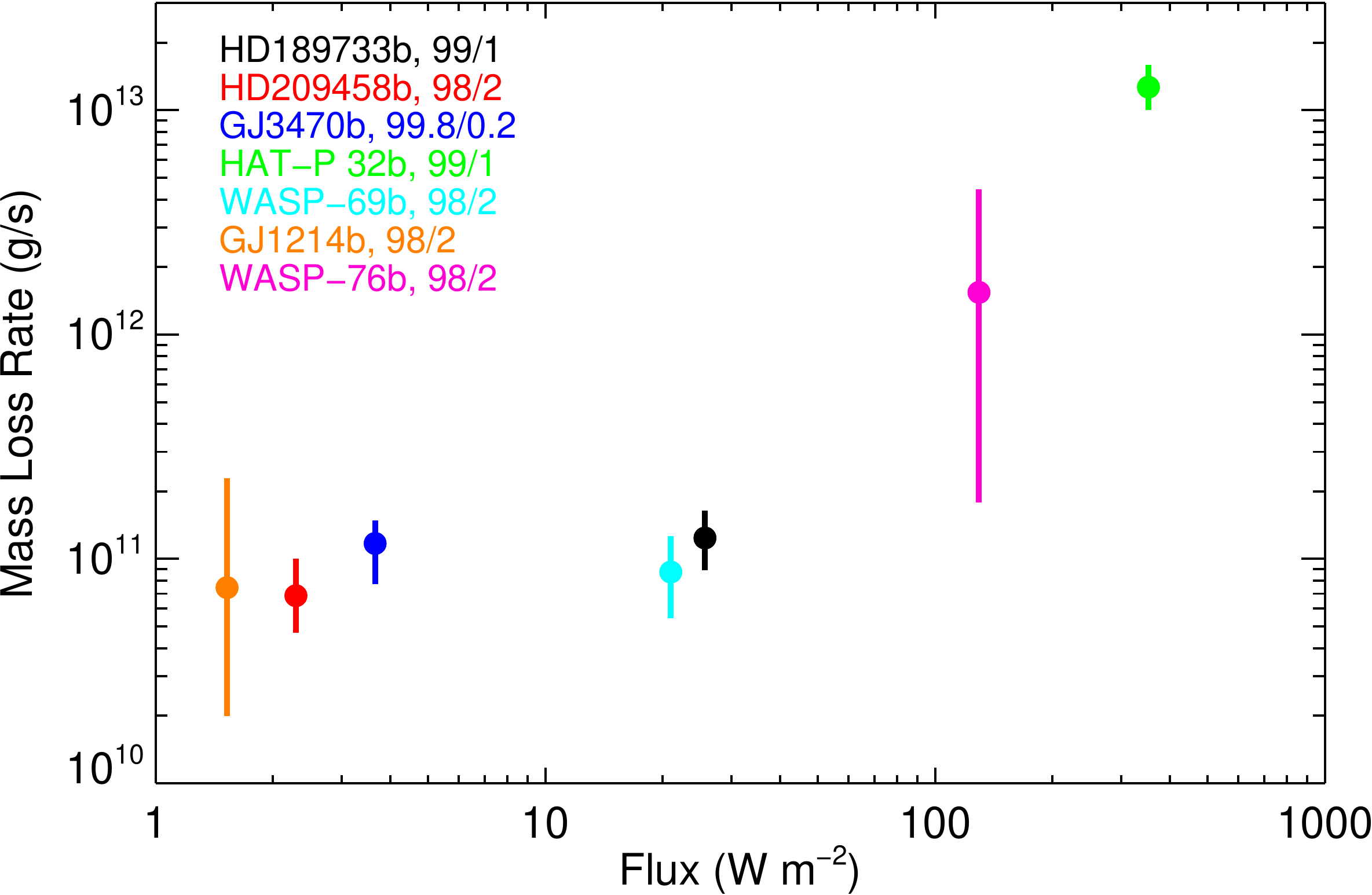} 
\caption{Mass-loss rates vs the XUV flux density for the seven studied planets.}  
\label{mlr_flux}
\end{figure}

We studied \hdu18 before \citep{Lampon_2021a}, but since the XUV flux density of its host star has been recently revised (see Sect.\,\ref{fluxes} and Fig.\,\ref{flux_new}), and we   perform a homogeneous analysis of seven planets in Sect.\,\ref{sec_comparison}, we   redid the analysis. The revised flux is significantly lower, by a factor from two to three at wavelengths between $\sim$250\,\AA\ and 1200\,\AA. However, the effect of this change in the derived \mlr\ and temperature is of a different sign than for the case of \wa76 discussed above. \hdu18 has a more compressed atmosphere and hence the effect of changing the flux on the absorption of the inner layers dominates over that in the outer layers (see Fig.\,\ref{fluxdiv4}). As a consequence, in this planet the derived mass-loss rate and temperature increase when considering the weaker XUV flux.  The effects, however, are not large (see Fig.\,\ref{chi2_hd18}): the mass-loss rate has increased by a factor of $\sim$1.5 and the upper value of the temperature by about 500\,K.
Regarding the impact of the SW, only the more strongly scaled SW slightly shifts the \mlr\ towards slightly larger values (no more than a factor of 1.3) (see black triangles in Fig.\,\ref{chi2_hd18}).

\section{Discussion} \label{sec_comparison}

\subsection{Mass-loss rates and temperatures}

Figure\,\ref{mesos_chi2} shows the \mlr--$T$ ranges derived for the planets we   analysed, \hd20, \hdu18, \gj34, \hatp32, \w69, \wa76, and \gju12.
On the one hand, five planets show very similar mass-loss rates around $\sim$10$^{11}$\,\gs
(\gj34, \gju12, \w69, \hd20, and \hdu18),  but with a wide range of temperatures; another two, \wa76 and \hatp32, have mass-loss rates about one and two orders of magnitude higher, respectively. The higher mass-loss rates of the latter seem to be caused by their high XUV irradiation (they received the highest fluxes, see Table\,\ref{list_planets} and Fig.\,\ref{mlr_flux}), in line with Eq.\,\ref{eq:energy_lim}.  
The former five planets received, however, significantly different XUV fluxes. \hdu18 and \w69 have the third and fourth highest XUV fluxes. Hence, it is reasonable that they also evaporate at a high \mlr, although not as high as \wa76 and \hatp32. \gj34, \gju12, and \hd20 seem to be the exceptions since they are irradiated at significantly lower XUV fluxes. However, \gj34 and \gju12 have the lowest gravitational potentials, which, following Eq.\,\ref{eq:energy_lim}, could well explain their relatively high evaporation rates. With a low gravitational potential for retaining their expanding upper atmospheres, they exhibit anomalously high mass-loss rates for the XUV flux at which they are irradiated \citep[see e.g.][]{Salz2018}.
The archetypal planet \hd20 does not fit, however, in that reasoning. Its mass-loss rate is rather high, although it is irradiated with a very low XUV flux 
and its gravitational potential is relatively large.
The reason seems to be that \hd20\ has a relatively large $R_{\rm XUV}$, larger than \gj34 and \gju12, which makes the absorption of the stellar radiation more efficient (the effective area is $\pi\,R_{\rm XUV}^2$), and thus compensates for its low XUV irradiation and stronger gravitational potential (see Eq.\,\ref{eq:energy_lim}).

In extremely irradiated planets (i.e. \hatp32\ and \wa76) the temperature is correlated with the flux density. However, for moderately irradiated planets (i.e. \gj34, \gju12, \w69, \hd20, and \hdu18) it better correlates with the gravitational potential:  the higher the gravitational potential, the higher the temperature maximum of the atmosphere (Figs.\,\ref{mesos_chi2} and \ref{mlr_flux}). 
We recall that the temperature we derived here is very close to the maximum of the thermospheric temperature obtained by more sophisticated hydrodynamic models that take into account the energy balance equation \citep[see Sect.\,3.1 in][]{Lampon2020}.
Our results agree very well with those of \cite{Salz2016}, who found that the level of irradiation affects the maximum temperature of the atmosphere only marginally. Instead, the temperature correlates well with the gravitational potential as a stronger adiabatic cooling is produced by a faster expansion of the outflow in a lower gravitational potential, and hence a lower maximum temperature.

\subsection{Heating efficiency and hydrodynamic regime}

\begin{figure}[t]
\includegraphics[angle=90, width=1.0\columnwidth]{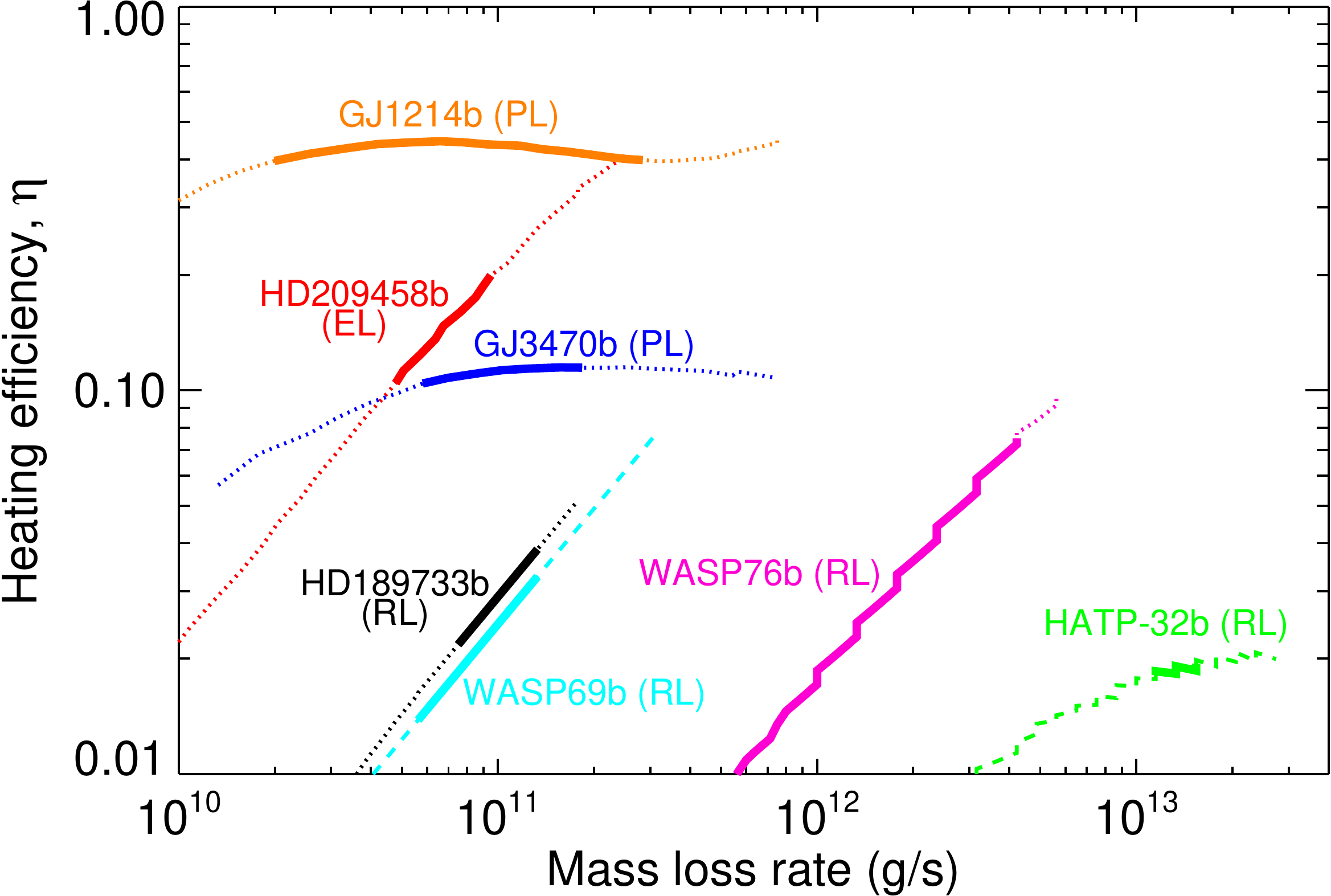} 
\caption{
Heating efficiency, $\eta$, vs mass-loss rates for the seven studied planets. Dotted lines show the extended $T$--$\dot M$ ranges and thick lines the derived $T$--$\dot M$ ranges (see  dotted lines and symbols, respectively, in Figs.\,\ref{chi2} and \ref{chi2_gj34}). The values for \hd20\ and \hdu18 are from \cite{Lampon_2021b}. 
} 
\label{eta}
\end{figure}

The heating efficiency is a very useful quantity for understanding the energy budget and the evaporation (including their hydrodynamic escape regime) of gaseous planets.
In Fig.\,\ref{eta} we show the heating efficiencies, $\eta$, for the seven studied planets 
using  Eq.\,\ref{eq:energy_lim},  
\begin{equation}
\eta = \frac{5}{4}\, \frac{K(\xi)}{4 \pi} \frac{\Phi}{R_{\rm XUV}^{2}}\, \frac{1}{F_{\rm XUV}}\, \dot M~, \label{eq:energy_lim2}
\end{equation}
with their derived mass-loss rates. 
Four planets (\hatp32, \wa76, \w69, and \hdu18) have low heating efficiencies, below $\sim$0.1, and all of them
received high XUV fluxes (see Fig.\,\ref{mlr_flux}). That is, planets that are strongly irradiated in the XUV tend to have low heating efficiencies (see Eq.\,\ref{eq:energy_lim2}). It may appear surprising that an external
quantity such as the XUV flux can alter the planet's heating efficiency. This can be understood by  noting that the XUV irradiance can actually change the density structure of the planet's upper atmosphere, and hence the location and slope of the ionisation fronts and, in consequence, the heating efficiency  \cite[see the discussion above for WASP-76\,b in Sect.\,\ref{sec_wasp76} and e.g.][and \citeauthor{Owen_2016} \citeyear{Owen_2016}]{Murray_Clay_2009}. 
The other three planets,
\gju12, \hd20, and \gj34, are more weakly irradiated (see Fig.\,\ref{mlr_flux}) and hence they are more efficient in thermalising the XUV flux.

\begin{figure}[]
\includegraphics[angle=90, width=1.0\columnwidth]{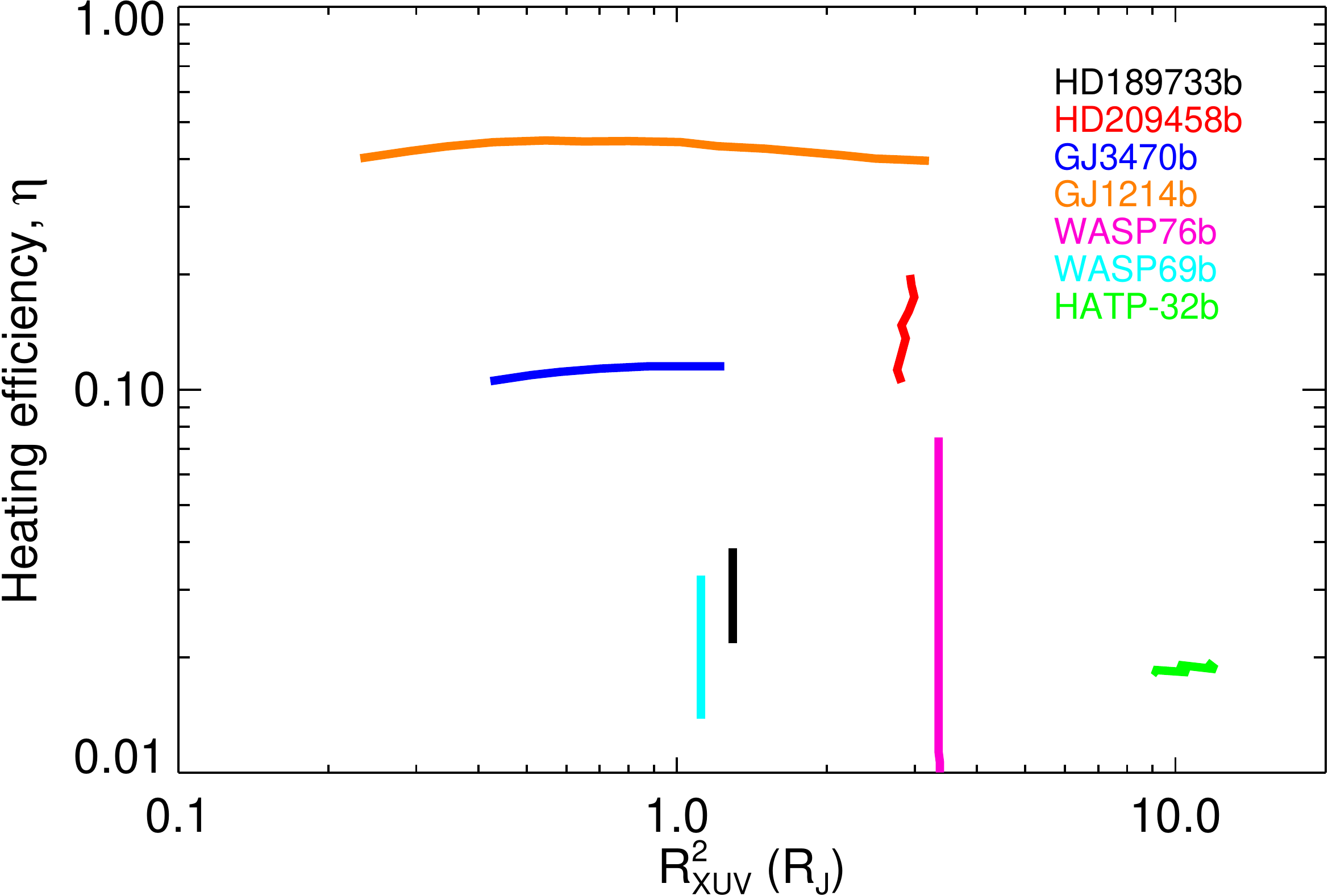} 
\caption{Heating efficiency vs $R_{XUV}^2$ for the seven studied planets.}
\label{eta_rxuv}
\end{figure}

Another factor affecting the heating efficiency is $R_{\rm XUV}$, as $\eta$ is inversely proportional to its square 
(see Eq.\,\ref{eq:energy_lim2}).  
Thus, the planets with large $R_{\rm XUV}$, \hatp32 and \wa76, have 
low heating efficiencies, 
and those with smaller $R_{\rm XUV}$  (\gj34, \gju12) have large efficiencies (see Fig.\,\ref{eta_rxuv}).  
\hdu18 and \w69 exhibit low heating efficiencies despite their smaller $R_{XUV}$ 
 because of their still high irradiation levels (compared with \hatp32 and \wa76, see Fig.\,\ref{mlr_flux}), just the opposite of \hd20, which shows a high heating efficiency due to a low irradiation flux, despite its high $R_{XUV}$.

The heating efficiency is also closely related to the hydrodynamic escape regime. As discussed above in Sect.\,\ref{sec_RH}, planets with very low heating efficiency, $\eta$\,$<$\,0.1, 
are in the recombination-limited (RL) regime, and planets with $\eta$\,$\gtrsim$\,0.1 are in either the energy-limited (EL) or the photon-limited (PL) regimes. Further, a high $\eta$, if nearly constant with \mlr, is indicative of the PL regime. Thus, \hatp32, \wa76, \w69, and \hdu18, all highly irradiated, are in the RL regime (see Fig.\,\ref{eta}). 
\hd20, on one hand, and \gj34 and \gju12, on the other, more moderately irradiated, are in the EL and PL regimes, respectively.
Our results are in agreement with the predictions of \cite{Owen_2016}.
Generally, highly irradiated planets would be in the RL regime, while planets with low irradiation would be in the EL regime if their gravitational potential is deep and in the PL regime if their gravitational potential is shallow.

\subsection{Relationship of the heating efficiency with H/He.}
\label{sec_eta-H/He}

We  also studied the dependence of the heating efficiency on the value of the H/He ratio.
We found that for \hdu18, \w69, and \wa76 the heating efficiency increases with the H/He ratio, while it decreases for \gj34, \gju12, and \hatp32 (see e.g. Fig.\,\ref{eta_gj1214} for GJ1214\,b).
This can be explained by looking at the relationship between \mlr, \rxuv,\ and  the H/He ratio. For a given temperature, as the H/He ratio increases, \mlr\ increases for all 
planets (see e.g. Fig.\,\ref{chi2}). In \hdu18, \w69, and \wa76, \rxuv\ does not significantly change with the H/He ratio
(see Fig. \ref{hvmr_candidates} for \w69 and \wa76, and Fig.\,8 from \cite{Lampon_2021a} for \hdu18). 
Thus, according to Eq.\,\ref{eq:energy_lim2}, where $\eta \propto \dot M/R^2_{\rm{XUV}}$, 
$\eta$ increases with the H/He ratio for these planets.
For \gj34, \gju12, and \hatp32, however, \rxuv\ significantly increases with the
H/He ratio (see Figs.\,\ref{hden_candidates} and \ref{hvmr_candidates} for \hatp32, and Fig.\,8 from \cite{Lampon_2021a} for \gj34), with $R^2_{\rm{XUV}}$ increasing in higher proportion than the mass-loss rate, and hence $\eta$ decreases. 

Thus, this $\eta$\,-\,H/He relationship is useful for constraining the H/He ratio. Assuming an upper limit for the heating efficiency allows us to  constrain the H/He ratio, as we did   for \gju12 (see Sect.\,\ref{sec_gj12}). Constraining the H/He ratio with this approach is mainly relevant
for planets in the PL and EL regimes as heating efficiencies are significantly higher than for those in the RL regime (see Fig.\,\ref{eta}). Actually, we could not apply this approach to \w69\ and \wa76\ (both in the RL regime).

\section{Summary} \label{summary}

In this work, we characterised the upper atmospheres of four exoplanets undergoing photo-evaporation, the hot Jupiters \hatp32\ and \w69, the warm sub-Neptune \gju12, and the ultra-hot Jupiter \wa76, through high-resolution measurements of their He {\sc i} triplet. In addition, we also reanalysed the warm Neptune \gj34 and the hot Jupiter \hdu18.

We used a 1D spherically symmetric hydrodynamic model, suitable for performing an extensive parameter study. The hydrodynamic model is coupled with a non-LTE code for deriving the He triplet distribution of the escaping outflow. Our 1D model does not account for some parameters that might affect the strength of the \het\ absorption signal, being  the stellar wind the most important. In order to gauge its potential impact, we  estimated its effect on the derived temperatures and mass-loss rate. Applying a radiative transfer model to the transit geometry, we computed the synthetic spectra. By comparing our calculations
with observations,
we constrained the main atmospheric parameters of the escape, namely the mass-loss rate, the thermospheric temperature, and the H/He ratio.
Additionally, by studying the processes dominating the production of the neutral H and the heating efficiency in the outflow, we classified these planets by their hydrodynamic escape regimes: energy-limited (EL), recombination-limited (RL), or photon-limited (PL).

For \hatp32 we analysed the \het\ absorption together with the \ha\ measurements reported by \cite{Czesla_2022}. The \ha\ absorption helps to constrain the H/He ratio in addition to the temperature and mass-loss rate. 
Our results show that \hatp32\ photo-evaporates at (130\,$\pm$\,70)$\times$10$^{11}$\,\gs\ under a very high XUV irradiation, which considerably heats its upper atmosphere to 12\,400\,$\pm$\,2900\,K. 
As recombination dominates in almost its entire upper atmosphere, the derived heating efficiency is low (0.017\,$\pm$\,0.003) and  the ionisation profile is very sharp, and hence we derived that \hatp32\ is in the recombination-limited (RL) regime. Further, we also constrained the H/He ratio of the outflow to the high value of (99.0/1.0)$^{+0.5}_{-1.0}$.

We found that \w69\ is losing its atmosphere with a mass-loss rate of (0.9\,$\pm$\,0.5)$\times$10$^{11}$\,\gs, at hot temperatures, 5250\,$\pm$\,750\,K, assuming a likely H/He ratio of 98/2. 
As there are no observations of H$^0$ lines for this planet, we could not constrain its H/He ratio. Therefore, we assumed such a high value based on the fact that most of the observed planets so far present large H/He ratios.  
In this planet recombination dominates over advection in practically the entire upper atmosphere, and hence it is in the RL regime. Further, we found a low heating efficiency of 0.02\,$\pm$\,0.01, in accordance with its regime.

The weak absorption of \wa76 prevents us from determining an accurate constraint of its temperature and mass-loss rate. 
Moreover, as for \w69, we could not constrain the H/He ratio.
By assuming a H/He of 98/2, we obtained a constrained narrow region of related temperatures and mass-loss rates, although with rather broad ranges of both parameters, 6000-17\,000\,K and (2-45)$\times$10$^{11}$\,\gs. 
As in \hatp32 and \w69, this planet undergoes hydrodynamic escape in the RL regime, with a heating efficiency of 0.039\,$\pm$\,0.036.

As for \w69\ and \wa76, there are no available measurements of H$^0$ lines in \gju12. However, due to the strong variation in the heating efficiency with respect to the H/He ratio for this planet, we could constrain the H/He ratio by calculating an upper limit for the heating efficiency. 
This method is relevant for planets in the PL and EL regimes, as their heating efficiencies are significantly higher than those in the RL regime, and it is especially valuable when we have no observations of the H$^0$ lines. 
When applied to this planet we found a H/He ratio in the range of 98/2--99/1.  
This planet has a light upper atmosphere 
that photo-evaporates at  (1.3\,$\pm$\,1.1)\,$\times$10$^{11}$\,\gs; a  relatively low temperature, 3750\,$\pm$\,750\,K; and high heating efficiency, in the range of 0.43\,$\pm$\,0.03.
In its outflow advection dominates over recombination in the entire upper atmosphere, and hence this planet is in the photon-limited (PL) regime.

We also reanalysed \gj34, for which a new stellar flux is available. 
We used the same methods as \cite{Lampon_2021a}, but took into account that the measured absorption spectrum is the average for the T2-T3 contacts, instead of the mid-transit spectrum, considering also a higher upper boundary and the new stellar flux. 
Comparing our results with \cite{Lampon_2021a}, we found lower temperatures, in the range of 3400\,$\pm$\,350\,K. 
Nevertheless, the new mass-loss rate, in the range of (1.3\,$\pm$\,0.6)$\times$10$^{11}$\,\gs, is practically the same. The H/He ratio of $\approx$\,99.8/0.2 is higher, and the heating efficiency, 0.115\,$\pm$\,0.005, is slightly lower.
The hydrodynamic regime did not change; that is, \gj34\ is in the PL regime.  

We also reanalysed \hdu18, as the XUV density flux has been updated. The effects are, however, relatively small (see Fig.\,\ref{chi2_hd18}). With respect to our previous analysis \citep{Lampon_2021a} its mass-loss rate   increased by a factor of $\sim$1.5 and the temperature by about 500\,K.

Our results from the homogeneous analysis of this sample of planets suggest that: (i) extremely irradiated planets,   \hatp32\ and \wa76, show very high mass-loss rates and temperatures; (ii) moderately irradiated planets,  \hd20, \hdu18, \gj34, \w69, and \gju12, show comparable mass-loss rates (for \gj34\ and \gju12 their low XUV irradiations are compensated by   their low gravitational potentials, while \hd20\ with its high $R_{\rm XUV}$); and  (iii) moderately irradiated planets have very different temperatures that correlate with their gravitational potential.
These findings are generally in line with the expected results of the current hydrodynamic escape models \citep[e.g.,][]{Salz2016, Salz2018,Owen_2020}.

Including the results of this work, four exoplanets were found to be in the recombination-limited regime (the hot Jupiters \hdu18, \hatp32, and \w69, and the ultra-hot Jupiter \wa76) \citep[][and this work]{Lampon_2021b,Czesla_2022},  
four in the PL regime (the Neptune \gj34, and the sub-Neptunes \gju12, HD\,63433\,b, and HD\,63433\,c) \citep[][]{Lampon_2021a,Orell-Miquel_2022,Zhang_2022a}, and only one in the energy-limited regime (the hot Jupiter \hd20) \citep{Lampon_2021a}.
Hot and ultra-hot Jupiters are mainly in the recombination-limited regime (with the exception of \hd20, which  is more weakly irradiated at XUV wavelengths), while warm Neptunes and sub-Neptunes, with XUV fluxes weaker than those of the hot and ultra-hot Jupiters,  
are in the photo-limited regime (see Table \ref{list_planets}).  
These results are in 
good agreement with the predictions of 
\cite{Owen_2016}. They predicted that the planets irradiated at low levels ($\sim$3\,W\,m$^{-2}$) in the XUV are mostly in either
the energy-limited (those with large gravitational potential) or in the photon-limited regime (those with weak gravitational potential). Here we found that of the three planets studied with low flux, one (\hd20) is in EL and two (the Neptune \gj34 and the sub-Neptune \gju12) are in PL; the former have a significantly larger gravitational potential than the latter. In addition,
\cite{Murray_Clay_2009} and
\cite{Owen_2016} found that those planets irradiated at high levels (e.g. $\sim$300\,W\,m$^{-2}$) are mostly in the recombination-limited scenario, as we have found for 
the hot Jupiters \hdu18, \hatp32, and \w69 and for  the ultra-hot Jupiter \wa76.

The H/He ratios derived in this work support the hypothesis that hydrodynamic escape outflows tend to have very low mean molecular weight, as suggested by \cite{Lampon_2021b}. The five planets for which we derived the H/He concentration, \hd20, \hdu18, \gj34,  \hatp32, and \gju12, all have a H/He ratio $\gtrsim$\,98/2. 
In addition to these, WASP-52\,b, HAT-P-11\,b, and WASP-80\,b, have also been reported with a high H/He ratio \citep[][respectively]{dongdong_2022,Dos_santos_2021,Fossati_2022}. Only WASP-107\,b has been reported with a solar-like value  \citep[H/He in the range of 87/13-93/7,][]{Khodachenko_2021}. 
There is no obvious reason (e.g. gravitational potential, XUV flux or age) why this planet should have such a different H/He ratio. Measurements of \lya, \halpha, or studies of its heating efficiency such as those performed here for \gju12, would be very interesting for corroborating its low H/He ratio.

Future confirmation of He triplet observations in \wa76\ and \gju12\ are crucial for supporting our conclusions about both exoplanets.
On the other hand, as the H/He ratio is key for testing the trend of outflows with very low mean molecular weight, and for reducing the degeneracy in the mass-loss rate and temperature, future observations of H$^0$ lines in \w69 and \wa76 would be very valuable in order to constrain this parameter.
Further, to fill the different scenarios of hydrodynamic escape regimes, it would be interesting to observe other massive planets ($>M_J$) weakly irradiated in the XUV, which might prove to be in the recombination-limited regime, and Neptunes or sub-Neptunes with strong XUV flux densities, which are predicted to be in the energy-limited or recombination-limited regime \citep{Owen_2016}.

\begin{acknowledgements}
We thank the referee for very useful comments.
We are grateful to Antonio García Muñoz for his very fruitful comments on the manuscript. CARMENES is an instrument for the Centro Astron\'omico Hispano-Alem\'an (CAHA) at Calar Alto (Almer\'{\i}a, Spain), operated jointly by the Junta de Andaluc\'ia and the Instituto de Astrof\'isica de Andaluc\'ia (CSIC).
CARMENES was funded by the Max-Planck-Gesellschaft (MPG), the Consejo Superior de Investigaciones Cient\'{\i}ficas (CSIC), the Ministerio de Econom\'ia y Competitividad (MINECO) and the European Regional Development Fund (ERDF) through projects FICTS-2011-02, ICTS-2017-07-CAHA-4, and CAHA16-CE-3978, and the members of the CARMENES Consortium (Max-Planck-Institut f\"ur Astronomie,
Instituto de Astrof\'{\i}sica de Andaluc\'{\i}a,
Landessternwarte K\"onigstuhl, 
Institut de Ci\`encies de l'Espai, 
Institut f\"ur Astrophysik G\"ottingen, 
Universidad Complutense de Madrid, 
Th\"uringer Landessternwarte Tautenburg, 
Instituto de Astrof\'{\i}sica de Canarias, 
Hamburger Sternwarte, 
Centro de Astrobiolog\'{\i}a and 
Centro Astron\'omico Hispano-Alem\'an), 
with additional contributions by the MINECO, the Deutsche Forschungsgemeinschaft through the Major Research Instrumentation Programme and Research Unit FOR2544 ``Blue Planets around Red Stars'', the Klaus Tschira Stiftung, the states of Baden-W\"urttemberg and Niedersachsen, and by the Junta de Andaluc\'{\i}a.
We acknowledge financial support from the State Agency for Research of the Spanish MCIU and the ERDF  through projects
PGC2018-099425--B--I00,
PID2019-109522GB-C51/2/3/4,
PGC2018-098153-B-C33,            
PID2019-110689RB-I00/AEI/10.13039/501100011033, 
and the Centre of Excellence ``Severo Ochoa'' and ``Mar\'ia de Maeztu'' awards to the
Instituto de Astrof\'isica de Andaluc\'ia (SEV-2017-0709),
Instituto de Astrof\'isica de Canarias (SEV-2015-0548), 
and Centro de Astrobiolog\'ia (MDM-2017-0737), and the Generalitat de Catalunya/CERCA programme. 
A.S.L. acknowledges funding from the European Research Council under the European Union's Horizon 2020 research and innovation program under grant agreement No 694513. K.M. acknowledges funding by the Excellence Cluster ORIGINS, funded by the Deutsche Forschungsgemeinschaft (DFG, German Research Foundation) under Germany's Excellence Strategy -EXC-2094-390783311. T.H. acknowledges support from the European Research Council under the Horizon 2020 Framework Program via the ERC Advanced Grant Origins 832428.
\end{acknowledgements}

\bibliographystyle{aa} 
\bibliography{ref.bib}

\begin{thebibliography}{101}
\expandafter\ifx\csname natexlab\endcsname\relax\def\natexlab#1{#1}\fi

\bibitem[{Allart {et~al.}(2019)Allart, Bourrier, Lovis, Ehrenreich, Aceituno,
  Guijarro, Pepe, Sing, Spake, \& Wyttenbach}]{Allart_2019}
Allart, R., Bourrier, V., Lovis, C., {et~al.} 2019, \aap, 623, A58

\bibitem[{{Allart} {et~al.}(2018){Allart}, {Bourrier}, {Lovis}, {Ehrenreich},
  {Spake}, {Wyttenbach}, {Pino}, {Pepe}, {Sing}, \& {Lecavelier des
  Etangs}}]{Allart_2018}
{Allart}, R., {Bourrier}, V., {Lovis}, C., {et~al.} 2018, Science, 362, 1384

\bibitem[{Alonso-Floriano {et~al.}(2019)Alonso-Floriano, Snellen, Czesla,
  Bauer, Salz, Lamp{\'o}n, Lara, Nagel, L{\'o}pez-Puertas, Nortmann, \&
  et~al.}]{Alonso2019}
Alonso-Floriano, F.~J., Snellen, I. A.~G., Czesla, S., {et~al.} 2019, \aap,
  629, A110

\bibitem[{Anderson {et~al.}(2014)Anderson, Collier~Cameron, Delrez, Doyle,
  Faedi, Fumel, Gillon, Gómez Maqueo~Chew, Hellier, Jehin, Lendl, Maxted,
  Pepe, Pollacco, Queloz, Ségransan, Skillen, Smalley, Smith, Southworth,
  Triaud, Turner, Udry, \& West}]{Anderson_2014}
Anderson, D.~R., Collier~Cameron, A., Delrez, L., {et~al.} 2014, MNRAS, 445,
  1114

\bibitem[{Ben-Jaffel \& Ballester(2013)}]{Ben_Jaffel_2013}
Ben-Jaffel, L. \& Ballester, G.~E. 2013, \aap, 553, A52

\bibitem[{Berta {et~al.}(2011)Berta, Charbonneau, Bean, Irwin, Burke,
  D{\'{e}}sert, Nutzman, \& Falco}]{Berta_2011}
Berta, Z.~K., Charbonneau, D., Bean, J., {et~al.} 2011, ApJ, 736, 12

\bibitem[{{Bourrier} {et~al.}(2018){Bourrier}, {Lecavelier des Etangs},
  {Ehrenreich}, {Sanz-Forcada}, {Allart}, {Ballester}, {Buchhave}, {Cohen},
  {Deming}, {Evans}, {Garc{\'\i}a Mu{\~n}oz}, {Henry}, {Kataria}, {Lavvas},
  {Lewis}, {L{\'o}pez-Morales}, {Marley}, {Sing}, \& {Wakeford}}]{Bourrier2018}
{Bourrier}, V., {Lecavelier des Etangs}, A., {Ehrenreich}, D., {et~al.} 2018,
  \aap, 620, A147

\bibitem[{{Casasayas-Barris} {et~al.}(2021){Casasayas-Barris}, {Orell-Miquel,
  J.}, {Stangret, M.}, {Nortmann, L.}, {Yan, F.}, {Oshagh, M.}, {Palle, E.},
  {Sanz-Forcada, J.}, {L\'opez-Puertas, M.}, {Nagel, E.}, {Luque, R.},
  {Morello, G.}, {Snellen, I. A. G.}, {Zechmeister, M.}, {Quirrenbach, A.},
  {Caballero, J. A.}, {Ribas, I.}, {Reiners, A.}, {Amado, P. J.}, {Bergond,
  G.}, {Czesla, S.}, {Henning, Th.}, {Khalafinejad, S.}, {Molaverdikhani, K.},
  {Montes, D.}, {Perger, M.}, {S\'anchez-L\'opez, A.}, \& {Sedaghati,
  E.}}]{Casasayas_2021}
{Casasayas-Barris}, N., {Orell-Miquel, J.}, {Stangret, M.}, {et~al.} 2021,
  A\&A, 654, A163

\bibitem[{{Casasayas-Barris} {et~al.}(2018){Casasayas-Barris}, {Pall\'e, E.},
  {Yan, F.}, {Chen, G.}, {Albrecht, S.}, {Nortmann, L.}, {Van Eylen, V.},
  {Snellen, I.}, {Talens, G. J. J.}, {Gonz\'alez Hern\'andez, J. I.}, {Rebolo,
  R.}, \& {Otten, G. P. P. L.}}]{Casasayas_2018}
{Casasayas-Barris}, N., {Pall\'e, E.}, {Yan, F.}, {et~al.} 2018, A\&A, 616,
  A151

\bibitem[{Castelli \& Kurucz(2003)}]{Castelli-Kurucz_2003}
Castelli, F. \& Kurucz, R. 2003, in IAU Symposium 210, ed. N.~Piskunov,
  W.~Weiss, \& D.~Gray, Vol. ASP-S210

\bibitem[{{Chadney} {et~al.}(2015){Chadney}, {Galand}, {Unruh}, {Koskinen}, \&
  {Sanz-Forcada}}]{chad15}
{Chadney}, J.~M., {Galand}, M., {Unruh}, Y.~C., {Koskinen}, T.~T., \&
  {Sanz-Forcada}, J. 2015, \icarus, 250, 357

\bibitem[{Cloutier {et~al.}(2021)Cloutier, Charbonneau, Deming, Bonfils, \&
  Astudillo-Defru}]{Cloutier_2021}
Cloutier, R., Charbonneau, D., Deming, D., Bonfils, X., \& Astudillo-Defru, N.
  2021, AJ, 162, 174

\bibitem[{Cubillos {et~al.}(2020)Cubillos, Fossati, Koskinen, Young, Salz,
  France, Sreejith, \& Haswell}]{Cubillos_2020}
Cubillos, P.~E., Fossati, L., Koskinen, T., {et~al.} 2020, The Astronomical
  Journal, 159, 111

\bibitem[{{Czesla} {et~al.}(2022){Czesla}, {Lamp\'on, M.}, {Sanz-Forcada, J.},
  {Garc\'{\i}a Mu\~noz, A.}, {L\'opez-Puertas, M.}, {Nortmann, L.}, {Yan, D.},
  {Nagel, E.}, {Yan, F.}, {Schmitt, J. H. M. M.}, {Aceituno, J.}, {Amado, P.
  J.}, {Caballero, J. A.}, {Casasayas-Barris, N.}, {Henning, Th.},
  {Khalafinejad, S.}, {Molaverdikhani, K.}, {Montes, D.}, {Pall\'e, E.},
  {Reiners, A.}, {Schneider, P. C.}, {Ribas, I.}, {Quirrenbach, A.}, {Zapatero
  Osorio, M. R.}, \& {Zechmeister, M.}}]{Czesla_2022}
{Czesla}, S., {Lamp\'on, M.}, {Sanz-Forcada, J.}, {et~al.} 2022, A\&A, 657, A6

\bibitem[{Dos~Santos {et~al.}(2022)Dos~Santos, Vidotto, Vissapragada, Alam,
  Allart, Bourrier, Kirk, Seidel, \& Ehrenreich}]{Dos_santos_2021}
Dos~Santos, L.~A., Vidotto, A.~A., Vissapragada, S., {et~al.} 2022, A\&A, 659,
  A62

\bibitem[{{Eggleton}(1983)}]{Eggleton_1983}
{Eggleton}, P.~P. 1983, \apj, 268, 368

\bibitem[{{Ehrenreich} {et~al.}(2011){Ehrenreich}, {Lecavelier Des Etangs}, \&
  {Delfosse}}]{Ehrenreich_2011}
{Ehrenreich}, D., {Lecavelier Des Etangs}, A., \& {Delfosse}, X. 2011, \aap,
  529, A80

\bibitem[{{Ehrenreich} {et~al.}(2020){Ehrenreich}, {Lovis}, {Allart}, {Zapatero
  Osorio}, {Pepe}, {Cristiani}, {Rebolo}, {Santos}, {Borsa}, {Demangeon},
  {Dumusque}, {Gonz{\'a}lez Hern{\'a}ndez}, {Casasayas-Barris},
  {S{\'e}gransan}, {Sousa}, {Abreu}, {Adibekyan}, {Affolter}, {Allende Prieto},
  {Alibert}, {Aliverti}, {Alves}, {Amate}, {Avila}, {Baldini}, {Bandy}, {Benz},
  {Bianco}, {Bolmont}, {Bouchy}, {Bourrier}, {Broeg}, {Cabral}, {Calderone},
  {Pall{\'e}}, {Cegla}, {Cirami}, {Coelho}, {Conconi}, {Coretti}, {Cumani},
  {Cupani}, {Dekker}, {Delabre}, {Deiries}, {D'Odorico}, {Di Marcantonio},
  {Figueira}, {Fragoso}, {Genolet}, {Genoni}, {G{\'e}nova Santos}, {Hara},
  {Hughes}, {Iwert}, {Kerber}, {Knudstrup}, {Landoni}, {Lavie}, {Lizon},
  {Lendl}, {Lo Curto}, {Maire}, {Manescau}, {Martins}, {M{\'e}gevand},
  {Mehner}, {Micela}, {Modigliani}, {Molaro}, {Monteiro}, {Monteiro},
  {Moschetti}, {M{\"u}ller}, {Nunes}, {Oggioni}, {Oliveira}, {Pariani},
  {Pasquini}, {Poretti}, {Rasilla}, {Redaelli}, {Riva}, {Santana Tschudi},
  {Santin}, {Santos}, {Segovia Milla}, {Seidel}, {Sosnowska}, {Sozzetti},
  {Span{\`o}}, {Su{\'a}rez Mascare{\~n}o}, {Tabernero}, {Tenegi}, {Udry},
  {Zanutta}, \& {Zerbi}}]{Ehrenreich_2020}
{Ehrenreich}, D., {Lovis}, C., {Allart}, R., {et~al.} 2020, \nat, 580, 597

\bibitem[{Erkaev {et~al.}(2007)Erkaev, Kulikov, Lammer, Selsis, Langmayr,
  Jaritz, \& Biernat}]{Erkaev_2007}
Erkaev, N.~V., Kulikov, Y.~N., Lammer, H., {et~al.} 2007, \aap, 472, 329

\bibitem[{{Fossati} {et~al.}(2022){Fossati}, {Guilluy, G.}, {Shaikhislamov, I.
  F.}, {Carleo, I.}, {Borsa, F.}, {Bonomo, A. S.}, {Giacobbe, P.}, {Rainer,
  M.}, {Cecchi-Pestellini, C.}, {Khodachenko, M. L.}, {Efimov, M. A.},
  {Rumenskikh, M. S.}, {Miroshnichenko, I. B.}, {Berezutsky, A. G.},
  {Nascimbeni, V.}, {Brogi, M.}, {Lanza, A. F.}, {Mancini, L.}, {Affer, L.},
  {Benatti, S.}, {Biazzo, K.}, {Bignamini, A.}, {Carosati, D.}, {Claudi, R.},
  {Cosentino, R.}, {Covino, E.}, {Desidera, S.}, {Fiorenzano, A.},
  {Harutyunyan, A.}, {Maggio, A.}, {Malavolta, L.}, {Maldonado, J.}, {Micela,
  G.}, {Molinari, E.}, {Pagano, I.}, {Pedani, M.}, {Piotto, G.}, {Poretti, E.},
  {Scandariato, G.}, {Sozzetti, A.}, \& {Stoev, H.}}]{Fossati_2022}
{Fossati}, L., {Guilluy, G.}, {Shaikhislamov, I. F.}, {et~al.} 2022, A\&A, 658,
  A136

\bibitem[{Fossati {et~al.}(2010)Fossati, Haswell, Froning, Hebb, Holmes, Kolb,
  Helling, Carter, Wheatley, Cameron, Loeillet, Pollacco, Street, Stempels,
  Simpson, Udry, Joshi, West, Skillen, \& Wilson}]{Fossati_2010}
Fossati, L., Haswell, C.~A., Froning, C.~S., {et~al.} 2010, The Astrophysical
  Journal Letters, 714, L222

\bibitem[{{Gaia Collaboration} {et~al.}(2021){Gaia Collaboration}, {Brown},
  {Vallenari}, {Prusti}, {de Bruijne}, {Babusiaux}, {Biermann}, {Creevey},
  {Evans}, {Eyer}, {Hutton}, {Jansen}, {Jordi}, {Klioner}, {Lammers},
  {Lindegren}, {Luri}, {Mignard}, {Panem}, {Pourbaix}, {Randich}, {Sartoretti},
  {Soubiran}, {Walton}, {Arenou}, {Bailer-Jones}, {Bastian}, {Cropper},
  {Drimmel}, {Katz}, {Lattanzi}, {van Leeuwen}, {Bakker}, {Cacciari},
  {Casta{\~n}eda}, {De Angeli}, {Ducourant}, {Fabricius}, {Fouesneau},
  {Fr{\'e}mat}, {Guerra}, {Guerrier}, {Guiraud}, {Jean-Antoine Piccolo},
  {Masana}, {Messineo}, {Mowlavi}, {Nicolas}, {Nienartowicz}, {Pailler},
  {Panuzzo}, {Riclet}, {Roux}, {Seabroke}, {Sordo}, {Tanga}, {Th{\'e}venin},
  {Gracia-Abril}, {Portell}, {Teyssier}, {Altmann}, {Andrae}, {Bellas-Velidis},
  {Benson}, {Berthier}, {Blomme}, {Brugaletta}, {Burgess}, {Busso}, {Carry},
  {Cellino}, {Cheek}, {Clementini}, {Damerdji}, {Davidson}, {Delchambre},
  {Dell'Oro}, {Fern{\'a}ndez-Hern{\'a}ndez}, {Galluccio}, {Garc{\'\i}a-Lario},
  {Garcia-Reinaldos}, {Gonz{\'a}lez-N{\'u}{\~n}ez}, {Gosset}, {Haigron},
  {Halbwachs}, {Hambly}, {Harrison}, {Hatzidimitriou}, {Heiter},
  {Hern{\'a}ndez}, {Hestroffer}, {Hodgkin}, {Holl}, {Jan{\ss}en}, {Jevardat de
  Fombelle}, {Jordan}, {Krone-Martins}, {Lanzafame}, {L{\"o}ffler}, {Lorca},
  {Manteiga}, {Marchal}, {Marrese}, {Moitinho}, {Mora}, {Muinonen}, {Osborne},
  {Pancino}, {Pauwels}, {Petit}, {Recio-Blanco}, {Richards}, {Riello},
  {Rimoldini}, {Robin}, {Roegiers}, {Rybizki}, {Sarro}, {Siopis}, {Smith},
  {Sozzetti}, {Ulla}, {Utrilla}, {van Leeuwen}, {van Reeven}, {Abbas}, {Abreu
  Aramburu}, {Accart}, {Aerts}, {Aguado}, {Ajaj}, {Altavilla}, {{\'A}lvarez},
  {{\'A}lvarez Cid-Fuentes}, {Alves}, {Anderson}, {Anglada Varela}, {Antoja},
  {Audard}, {Baines}, {Baker}, {Balaguer-N{\'u}{\~n}ez}, {Balbinot}, {Balog},
  {Barache}, {Barbato}, {Barros}, {Barstow}, {Bartolom{\'e}}, {Bassilana},
  {Bauchet}, {Baudesson-Stella}, {Becciani}, {Bellazzini}, {Bernet}, {Bertone},
  {Bianchi}, {Blanco-Cuaresma}, {Boch}, {Bombrun}, {Bossini}, {Bouquillon},
  {Bragaglia}, {Bramante}, {Breedt}, {Bressan}, {Brouillet}, {Bucciarelli},
  {Burlacu}, {Busonero}, {Butkevich}, {Buzzi}, {Caffau}, {Cancelliere},
  {C{\'a}novas}, {Cantat-Gaudin}, {Carballo}, {Carlucci}, {Carnerero},
  {Carrasco}, {Casamiquela}, {Castellani}, {Castro-Ginard}, {Castro Sampol},
  {Chaoul}, {Charlot}, {Chemin}, {Chiavassa}, {Cioni}, {Comoretto}, {Cooper},
  {Cornez}, {Cowell}, {Crifo}, {Crosta}, {Crowley}, {Dafonte}, {Dapergolas},
  {David}, {David}, {de Laverny}, {De Luise}, {De March}, {De Ridder}, {de
  Souza}, {de Teodoro}, {de Torres}, {del Peloso}, {del Pozo}, {Delbo},
  {Delgado}, {Delgado}, {Delisle}, {Di Matteo}, {Diakite}, {Diener},
  {Distefano}, {Dolding}, {Eappachen}, {Edvardsson}, {Enke}, {Esquej}, {Fabre},
  {Fabrizio}, {Faigler}, {Fedorets}, {Fernique}, {Fienga}, {Figueras},
  {Fouron}, {Fragkoudi}, {Fraile}, {Franke}, {Gai}, {Garabato},
  {Garcia-Gutierrez}, {Garc{\'\i}a-Torres}, {Garofalo}, {Gavras}, {Gerlach},
  {Geyer}, {Giacobbe}, {Gilmore}, {Girona}, {Giuffrida}, {Gomel}, {Gomez},
  {Gonzalez-Santamaria}, {Gonz{\'a}lez-Vidal}, {Granvik},
  {Guti{\'e}rrez-S{\'a}nchez}, {Guy}, {Hauser}, {Haywood}, {Helmi}, {Hidalgo},
  {Hilger}, {H{\l}adczuk}, {Hobbs}, {Holland}, {Huckle}, {Jasniewicz},
  {Jonker}, {Juaristi Campillo}, {Julbe}, {Karbevska}, {Kervella}, {Khanna},
  {Kochoska}, {Kontizas}, {Kordopatis}, {Korn}, {Kostrzewa-Rutkowska},
  {Kruszy{\'n}ska}, {Lambert}, {Lanza}, {Lasne}, {Le Campion}, {Le Fustec},
  {Lebreton}, {Lebzelter}, {Leccia}, {Leclerc}, {Lecoeur-Taibi}, {Liao},
  {Licata}, {Lindstr{\o}m}, {Lister}, {Livanou}, {Lobel}, {Madrero Pardo},
  {Managau}, {Mann}, {Marchant}, {Marconi}, {Marcos Santos}, {Marinoni},
  {Marocco}, {Marshall}, {Martin Polo}, {Mart{\'\i}n-Fleitas}, {Masip},
  {Massari}, {Mastrobuono-Battisti}, {Mazeh}, {McMillan}, {Messina},
  {Michalik}, {Millar}, {Mints}, {Molina}, {Molinaro}, {Moln{\'a}r},
  {Montegriffo}, {Mor}, {Morbidelli}, {Morel}, {Morris}, {Mulone}, {Munoz},
  {Muraveva}, {Murphy}, {Musella}, {Noval}, {Ord{\'e}novic}, {Orr{\`u}},
  {Osinde}, {Pagani}, {Pagano}, {Palaversa}, {Palicio}, {Panahi}, {Pawlak},
  {Pe{\~n}alosa Esteller}, {Penttil{\"a}}, {Piersimoni}, {Pineau}, {Plachy},
  {Plum}, {Poggio}, {Poretti}, {Poujoulet}, {Pr{\v{s}}a}, {Pulone}, {Racero},
  {Ragaini}, {Rainer}, {Raiteri}, {Rambaux}, {Ramos}, {Ramos-Lerate}, {Re
  Fiorentin}, {Regibo}, {Reyl{\'e}}, {Ripepi}, {Riva}, {Rixon}, {Robichon},
  {Robin}, {Roelens}, {Rohrbasser}, {Romero-G{\'o}mez}, {Rowell}, {Royer},
  {Rybicki}, {Sadowski}, {Sagrist{\`a} Sell{\'e}s}, {Sahlmann}, {Salgado},
  {Salguero}, {Samaras}, {Sanchez Gimenez}, {Sanna}, {Santove{\~n}a},
  {Sarasso}, {Schultheis}, {Sciacca}, {Segol}, {Segovia}, {S{\'e}gransan},
  {Semeux}, {Shahaf}, {Siddiqui}, {Siebert}, {Siltala}, {Slezak}, {Smart},
  {Solano}, {Solitro}, {Souami}, {Souchay}, {Spagna}, {Spoto}, {Steele},
  {Steidelm{\"u}ller}, {Stephenson}, {S{\"u}veges}, {Szabados}, {Szegedi-Elek},
  {Taris}, {Tauran}, {Taylor}, {Teixeira}, {Thuillot}, {Tonello}, {Torra},
  {Torra}, {Turon}, {Unger}, {Vaillant}, {van Dillen}, {Vanel}, {Vecchiato},
  {Viala}, {Vicente}, {Voutsinas}, {Weiler}, {Wevers}, {Wyrzykowski}, {Yoldas},
  {Yvard}, {Zhao}, {Zorec}, {Zucker}, {Zurbach}, \& {Zwitter}}]{Gaia_EDR3}
{Gaia Collaboration}, {Brown}, A.~G.~A., {Vallenari}, A., {et~al.} 2021, \aap,
  649, A1

\bibitem[{{Garc{\'\i}a Mu{\~n}oz} {et~al.}(2021){Garc{\'\i}a Mu{\~n}oz},
  {Fossati}, {Youngblood}, {Nettelmann}, {Gandolfi}, {Cabrera}, \&
  {Rauer}}]{Garcia_Munoz2021}
{Garc{\'\i}a Mu{\~n}oz}, A., {Fossati}, L., {Youngblood}, A., {et~al.} 2021,
  \apjl, 907, L36

\bibitem[{Garc{\'\i}a-Mu{\~n}oz(2007)}]{Garcia_munoz_2007}
Garc{\'\i}a-Mu{\~n}oz, A. 2007, Planetary and Space Science, 55, 1426

\bibitem[{{Guilluy} {et~al.}(2020){Guilluy}, {Andretta, V.}, {Borsa, F.},
  {Giacobbe, P.}, {Sozzetti, A.}, {Covino, E.}, {Bourrier, V.}, {Fossati, L.},
  {Bonomo, A. S.}, {Esposito, M.}, {Giampapa, M. S.}, {Harutyunyan, A.},
  {Rainer, M.}, {Brogi, M.}, {Bruno, G.}, {Claudi, R.}, {Frustagli, G.},
  {Lanza, A. F.}, {Mancini, L.}, {Pino, L.}, {Poretti, E.}, {Scandariato, G.},
  {Affer, L.}, {Baffa, C.}, {Baruffolo, A.}, {Benatti, S.}, {Biazzo, K.},
  {Bignamini, A.}, {Boschin, W.}, {Carleo, I.}, {Cecconi, M.}, {Cosentino, R.},
  {Damasso, M.}, {Desidera, S.}, {Falcini, G.}, {Martinez Fiorenzano, A. F.},
  {Ghedina, A.}, {Gonz\'alez-\'Alvarez, E.}, {Guerra, J.}, {Hernandez, N.},
  {Leto, G.}, {Maggio, A.}, {Malavolta, L.}, {Maldonado, J.}, {Micela, G.},
  {Molinari, E.}, {Nascimbeni, V.}, {Pagano, I.}, {Pedani, M.}, {Piotto, G.},
  \& {Reiners, A.}}]{guilluy_2020}
{Guilluy}, G., {Andretta, V.}, {Borsa, F.}, {et~al.} 2020, \aap, 639, A49

\bibitem[{{Harps\o{}e} {et~al.}(2013){Harps\o{}e}, {Hardis, S.}, {Hinse, T.
  C.}, {J\o{}rgensen, U. G.}, {Mancini, L.}, {Southworth, J.}, {Alsubai, K.
  A.}, {Bozza, V.}, {Browne, P.}, {Burgdorf, M. J.}, {Calchi Novati, S.},
  {Dodds, P.}, {Dominik, M.}, {Fang, X.-S.}, {Finet, F.}, {Gerner, T.}, {Gu,
  S.-H.}, {Hundertmark, M.}, {Jessen-Hansen, J.}, {Kains, N.}, {Kerins, E.},
  {Kjeldsen, H.}, {Liebig, C.}, {Lund, M. N.}, {Lundkvist, M.}, {Mathiasen,
  M.}, {Nesvorn\'y, D.}, {Nikolov, N.}, {Penny, M. T.}, {Proft, S.}, {Rahvar,
  S.}, {Ricci, D.}, {Sahu, K. C.}, {Scarpetta, G.}, {Sch\"afer, S.},
  {Sch\"onebeck, F.}, {Snodgrass, C.}, {Skottfelt, J.}, {Surdej, J.},
  {Tregloan-Reed, J.}, \& {Wertz, O.}}]{Harpsoe_2013}
{Harps\o{}e}, K. B.~W., {Hardis, S.}, {Hinse, T. C.}, {et~al.} 2013, A\&A, 549,
  A10

\bibitem[{Hartman {et~al.}(2011)Hartman, Bakos, Torres, Latham, Kov{\'{a}}cs,
  B{\'{e}}ky, Quinn, Mazeh, Shporer, Marcy, Howard, Fischer, Johnson, Esquerdo,
  Noyes, Sasselov, Stefanik, Fernandez, Szklen{\'{a}}r, L{\'{a}}z{\'{a}}r,
  Papp, \& S{\'{a}}ri}]{Hartman_2011}
Hartman, J.~D., Bakos, G.~{\'{A}}., Torres, G., {et~al.} 2011, ApJ, 742, 59

\bibitem[{Haswell {et~al.}(2012)Haswell, Fossati, Ayres, France, Froning,
  Holmes, Kolb, Busuttil, Street, Hebb, Cameron, Enoch, Burwitz, Rodriguez,
  West, Pollacco, Wheatley, \& Carter}]{Haswell_2012}
Haswell, C.~A., Fossati, L., Ayres, T., {et~al.} 2012, The Astrophysical
  Journal, 760, 79

\bibitem[{Hu {et~al.}(2015)Hu, Seager, \& Yung}]{Hu_2015}
Hu, R., Seager, S., \& Yung, Y.~L. 2015, \apj, 807, 8

\bibitem[{{Jackson} {et~al.}(2012){Jackson}, {Davis}, \&
  {Wheatley}}]{jackson_2012}
{Jackson}, A.~P., {Davis}, T.~A., \& {Wheatley}, P.~J. 2012, \mnras, 422, 2024

\bibitem[{Jin \& Mordasini(2018)}]{Jin_2018}
Jin, S. \& Mordasini, C. 2018, \apj, 853, 163

\bibitem[{{Johnstone} {et~al.}(2015{\natexlab{a}}){Johnstone}, {G\"udel, M.},
  {Brott, I.}, \& {L\"uftinger, T.}}]{Johnstone_2015b}
{Johnstone}, C.~P., {G\"udel, M.}, {Brott, I.}, \& {L\"uftinger, T.}
  2015{\natexlab{a}}, A\&A, 577, A28

\bibitem[{{Johnstone} {et~al.}(2015{\natexlab{b}}){Johnstone}, {G\"udel, M.},
  {L\"uftinger, T.}, {Toth, G.}, \& {Brott, I.}}]{Johnstone_2015}
{Johnstone}, C.~P., {G\"udel, M.}, {L\"uftinger, T.}, {Toth, G.}, \& {Brott,
  I.} 2015{\natexlab{b}}, A\&A, 577, A27

\bibitem[{{Khalafinejad} {et~al.}(2021){Khalafinejad}, {Molaverdikhani, K.},
  {Blecic, J.}, {Mallonn, M.}, {Nortmann, L.}, {Caballero, J. A.}, {Rahmati,
  H.}, {Kaminski, A.}, {Sadegi, S.}, {Nagel, E.}, {Carone, L.}, {Amado, P. J.},
  {Azzaro, M.}, {Bauer, F. F.}, {Casasayas-Barris, N.}, {Czesla, S.}, {von
  Essen, C.}, {Fossati, L.}, {G\"udel, M.}, {Henning, Th.}, {L\'opez-Puertas,
  M.}, {Lendl, M.}, {L\"uftinger, T.}, {Montes, D.}, {Oshagh, M.}, {Pall\'e,
  E.}, {Quirrenbach, A.}, {Reffert, S.}, {Reiners, A.}, {Ribas, I.}, {Stock,
  S.}, {Yan, F.}, {Zapatero Osorio, M. R.}, \& {Zechmeister,
  M.}}]{Khalafinejad_2021}
{Khalafinejad}, S., {Molaverdikhani, K.}, {Blecic, J.}, {et~al.} 2021, A\&A,
  656, A142

\bibitem[{{Khodachenko} {et~al.}(2021){Khodachenko}, {Shaikhislamov},
  {Fossati}, {Lammer}, {Rumenskikh}, {Berezutsky}, {Miroshnichenko}, \&
  {Efimof}}]{Khodachenko_2021}
{Khodachenko}, M.~L., {Shaikhislamov}, I.~F., {Fossati}, L., {et~al.} 2021,
  \mnras, 503, L23

\bibitem[{Khodachenko {et~al.}(2019)Khodachenko, Shaikhislamov, Lammer,
  Berezutsky, Miroshnichenko, Rumenskikh, Kislyakova, \&
  Dwivedi}]{Khodachenko_2019}
Khodachenko, M.~L., Shaikhislamov, I.~F., Lammer, H., {et~al.} 2019, The
  Astrophysical Journal, 885, 67

\bibitem[{Khodachenko {et~al.}(2021)Khodachenko, Shaikhislamov, Lammer,
  Miroshnichenko, Rumenskikh, Berezutsky, \& Fossati}]{Khodachenko_2021b}
Khodachenko, M.~L., Shaikhislamov, I.~F., Lammer, H., {et~al.} 2021, MNRAS,
  507, 3626

\bibitem[{{Kirk} {et~al.}(2020){Kirk}, {Alam}, {L{\'o}pez-Morales}, \&
  {Zeng}}]{Kirk_2020}
{Kirk}, J., {Alam}, M.~K., {L{\'o}pez-Morales}, M., \& {Zeng}, L. 2020, \aj,
  159, 115

\bibitem[{Kirk {et~al.}(2022)Kirk, Santos, L{\'{o}}pez-Morales, Alam,
  Oklop{\v{c}}i{\'{c}}, MacLeod, Zeng, \& Zhou}]{Kirk_2022}
Kirk, J., Santos, L. A.~D., L{\'{o}}pez-Morales, M., {et~al.} 2022, AJ, 164, 24

\bibitem[{Kulow {et~al.}(2014)Kulow, France, Linsky, \&
  Parke~Loyd}]{Kulow_2014}
Kulow, J.~R., France, K., Linsky, J., \& Parke~Loyd, R.~O. 2014, \apj, 786, 132

\bibitem[{{Lammer} {et~al.}(2020{\natexlab{a}}){Lammer}, {Leitzinger},
  {Scherf}, {Odert}, {Burger}, {Kubyshkina}, {Johnstone}, {Maindl},
  {Sch{\"a}fer}, {G{\"u}del}, {Tosi}, {Nikolaou}, {Marcq}, {Erkaev}, {Noack},
  {Kislyakova}, {Fossati}, {Pilat-Lohinger}, {Ragossnig}, \&
  {Dorfi}}]{Lammer_2020b}
{Lammer}, H., {Leitzinger}, M., {Scherf}, M., {et~al.} 2020{\natexlab{a}},
  \icarus, 339, 113551

\bibitem[{{Lammer} {et~al.}(2020{\natexlab{b}}){Lammer}, {Scherf}, {Kurokawa},
  {Ueno}, {Burger}, {Maindl}, {Johnstone}, {Leizinger}, {Benedikt}, {Fossati},
  {Kislyakova}, {Marty}, {Avice}, {Fegley}, \& {Odert}}]{Lammer_2020a}
{Lammer}, H., {Scherf}, M., {Kurokawa}, H., {et~al.} 2020{\natexlab{b}}, \ssr,
  216, 74

\bibitem[{{Lamp\'on} {et~al.}(2020){Lamp\'on}, {L\'opez-Puertas}, {Lara},
  S\'anchez-L\'opez, Salz, {Czesla, S.}, {Sanz-Forcada, J.}, {Molaverdikhani,
  K.}, {Alonso-Floriano, F. J.}, {Nortmann, L.}, {Caballero, J. A.}, {Bauer, F.
  F.}, {Pall\'e, E.}, {Montes, D.}, {Quirrenbach, A.}, {Nagel, E.}, {Ribas,
  I.}, {Reiners, A.}, \& {Amado, P. J.}}]{Lampon2020}
{Lamp\'on}, M., {L\'opez-Puertas}, M., {Lara}, L.~M., {et~al.} 2020, A\&A, 636,
  A13

\bibitem[{{Lamp\'on} {et~al.}(2021{\natexlab{a}}){Lamp\'on}, {L\'opez-Puertas,
  M.}, {Czesla, S.}, {S\'anchez-L\'opez, A.}, {Lara, L. M.}, {Salz, M.},
  {Sanz-Forcada, J.}, {Molaverdikhani, K.}, {Quirrenbach, A.}, {Pall\'e, E.},
  {Caballero, J. A.}, {Henning, Th.}, {Nortmann, L.}, {Amado, P. J.}, {Montes,
  D.}, {Reiners, A.}, \& {Ribas, I.}}]{Lampon_2021b}
{Lamp\'on}, M., {L\'opez-Puertas, M.}, {Czesla, S.}, {et~al.}
  2021{\natexlab{a}}, A\&A, 648, L7

\bibitem[{{Lamp\'on} {et~al.}(2021{\natexlab{b}}){Lamp\'on}, {L\'opez-Puertas,
  M.}, {Sanz-Forcada, J.}, {S\'anchez-L\'opez, A.}, {Molaverdikhani, K.},
  {Czesla, S.}, {Quirrenbach, A.}, {Pall\'e, E.}, {Caballero, J. A.}, {Henning,
  T.}, {Salz, M.}, {Nortmann, L.}, {Aceituno, J.}, {Amado, P. J.}, {Bauer, F.
  F.}, {Montes, D.}, {Nagel, E.}, {Reiners, A.}, \& {Ribas, I.}}]{Lampon_2021a}
{Lamp\'on}, M., {L\'opez-Puertas, M.}, {Sanz-Forcada, J.}, {et~al.}
  2021{\natexlab{b}}, A\&A, 647, A129

\bibitem[{Lecavelier~des Etangs {et~al.}(2012)Lecavelier~des Etangs, Bourrier,
  Wheatley, Dupuy, Ehrenreich, Vidal-Madjar, H{\'e}brard, Ballester,
  D{\'e}sert, Ferlet, \& et~al.}]{Lecavelier_des_Etangs_2012}
Lecavelier~des Etangs, A., Bourrier, V., Wheatley, P.~J., {et~al.} 2012, \aap,
  543, L4

\bibitem[{Lopez \& Fortney(2013)}]{Lopez_2013}
Lopez, E.~D. \& Fortney, J.~J. 2013, \apj, 776, 2

\bibitem[{Malsky \& Rogers(2020)}]{Malsky_2020}
Malsky, I. \& Rogers, L.~A. 2020, \apj, 896, 48

\bibitem[{Mansfield {et~al.}(2018)Mansfield, Bean, Oklop{\v c}i{\'c},
  Kreidberg, D{\'e}sert, Kempton, Line, Fortney, Henry, Mallonn, \&
  et~al.}]{Mansfield_2018}
Mansfield, M., Bean, J.~L., Oklop{\v c}i{\'c}, A., {et~al.} 2018, \apj, 868,
  L34

\bibitem[{Mordasini(2020)}]{Mordasini2020}
Mordasini, C. 2020, A\&A, 638, A52

\bibitem[{Mu{\~{n}}oz \& Schneider(2019)}]{Garcia_munoz_2019}
Mu{\~{n}}oz, A.~G. \& Schneider, P.~C. 2019, \apj, 884, L43

\bibitem[{Murray-Clay {et~al.}(2009)Murray-Clay, Chiang, \&
  Murray}]{Murray_Clay_2009}
Murray-Clay, R.~A., Chiang, E.~I., \& Murray, N. 2009, \apj, 693, 23

\bibitem[{Ninan {et~al.}(2020)Ninan, Stefansson, Mahadevan, Bender, Robertson,
  Ramsey, Terrien, Wright, Diddams, Kanodia, Cochran, Endl, Ford, Fredrick,
  Halverson, Hearty, Jennings, Kaplan, Lubar, Metcalf, Monson, Nitroy, Roy, \&
  Schwab}]{Ninan_2020}
Ninan, J.~P., Stefansson, G., Mahadevan, S., {et~al.} 2020, \apj, 894, 97

\bibitem[{Nortmann {et~al.}(2018)Nortmann, Palle, Salz, Sanz-Forcada, Nagel,
  Alonso-Floriano, Czesla, Yan, Chen, Snellen, Zechmeister, Schmitt,
  L{\'o}pez-Puertas, Casasayas-Barris, Bauer, Amado, Caballero, Dreizler,
  Henning, Lamp{\'o}n, Montes, Molaverdikhani, Quirrenbach, Reiners, Ribas,
  S{\'a}nchez-L{\'o}pez, Schneider, \& Zapatero~Osorio}]{Nortmann2018}
Nortmann, L., Palle, E., Salz, M., {et~al.} 2018, Science, 362, 1388

\bibitem[{Oklop{\v c}i{\'c} \& Hirata(2018)}]{Oklopcic2018}
Oklop{\v c}i{\'c}, A. \& Hirata, C.~M. 2018, \apj, 855, L11

\bibitem[{{Orell-Miquel} {et~al.}(2022){Orell-Miquel}, {Murgas, F.}, {Pall\'e,
  E.}, {Lamp\'on, M.}, {L\'opez-Puertas, M.}, {Sanz-Forcada, J.}, {Nagel, E.},
  {Kaminski, A.}, {Casasayas-Barris, N.}, {Nortmann, L.}, {Luque, R.},
  {Molaverdikhani, K.}, {Sedaghati, E.}, {Caballero, J. A.}, {Amado, P. J.},
  {Bergond, G.}, {Czesla, S.}, {Hatzes, A. P.}, {Henning, Th.}, {Khalafinejad,
  S.}, {Montes, D.}, {Morello, G.}, {Quirrenbach, A.}, {Reiners, A.}, {Ribas,
  I.}, {S\'anchez-L\'opez, A.}, {Schweitzer, A.}, {Stangret, M.}, {Yan, F.}, \&
  {Zapatero Osorio, M. R.}}]{Orell-Miquel_2022}
{Orell-Miquel}, J., {Murgas, F.}, {Pall\'e, E.}, {et~al.} 2022, A\&A, 659, A55

\bibitem[{Owen \& Alvarez(2016)}]{Owen_2016}
Owen, J.~E. \& Alvarez, M.~A. 2016, \apj, 816, 34

\bibitem[{{Owen} {et~al.}(2020){Owen}, {Shaikhislamov}, {Lammer}, {Fossati}, \&
  {Khodachenko}}]{Owen_2020}
{Owen}, J.~E., {Shaikhislamov}, I.~F., {Lammer}, H., {Fossati}, L., \&
  {Khodachenko}, M.~L. 2020, \ssr, 216, 129

\bibitem[{Owen \& Wu(2013)}]{Owen_2013}
Owen, J.~E. \& Wu, Y. 2013, \apj, 775, 105

\bibitem[{Owen \& Wu(2017)}]{Owen_2017}
Owen, J.~E. \& Wu, Y. 2017, \apj, 847, 29

\bibitem[{Pallé {et~al.}(2020)Pallé, Nortmann, {Casasayas-Barris},
  Lamp{\'o}n, {L{\'o}pez-Puertas}, Caballero, {Sanz-Forcada}, Lara, Nagel, Yan,
  {Alonso-Floriano}, Amado, Chen, Cifuentes, {Cort{\'e}s-Contreras}, Czesla,
  Molaverdikhani, Montes, Passegger, Quirrenbach, Reiners, Ribas,
  {S{\'a}nchez-L{\'o}pez}, Schweitzer, Stangret, Osorio, \&
  Zechmeister}]{Palle2020}
Pallé, E., Nortmann, L., {Casasayas-Barris}, N., {et~al.} 2020, \aap, 638, A61

\bibitem[{Paragas {et~al.}(2021)Paragas, Vissapragada, Knutson,
  Oklop{\v{c}}i{\'{c}}, Chachan, Greklek-McKeon, Dai, Tinyanont, \&
  Vasisht}]{Paragas_2021}
Paragas, K., Vissapragada, S., Knutson, H.~A., {et~al.} 2021, ApJ Letters, 909,
  L10

\bibitem[{{Quirrenbach} {et~al.}(2014){Quirrenbach}, {Amado}, {Caballero},
  {Mundt}, {Reiners}, {Ribas}, {Seifert}, {Abril}, {Aceituno},
  {Alonso-Floriano}, {Ammler-von Eiff}, {Antona Jim{\'e}nez},
  {Anwand-Heerwart}, {Azzaro}, {Bauer}, {Barrado}, {Becerril}, {B{\'e}jar},
  {Ben{\'\i}tez}, {Berdi{\~n}as}, {C{\'a}rdenas}, {Casal}, {Claret},
  {Colom{\'e}}, {Cort{\'e}s-Contreras}, {Czesla}, {Doellinger}, {Dreizler},
  {Feiz}, {Fern{\'a}ndez}, {Galad{\'\i}}, {G{\'a}lvez-Ortiz},
  {Garc{\'\i}a-Piquer}, {Garc{\'\i}a-Vargas}, {Garrido}, {Gesa}, {G{\'o}mez
  Galera}, {Gonz{\'a}lez {\'A}lvarez}, {Gonz{\'a}lez Hern{\'a}ndez},
  {Gr{\"o}zinger}, {Gu{\`a}rdia}, {Guenther}, {de Guindos},
  {Guti{\'e}rrez-Soto}, {Hagen}, {Hatzes}, {Hauschildt}, {Helmling}, {Henning},
  {Hermann}, {Hern{\'a}ndez Casta{\~n}o}, {Herrero}, {Hidalgo}, {Holgado},
  {Huber}, {Huber}, {Jeffers}, {Joergens}, {de Juan}, {Kehr}, {Klein},
  {K{\"u}rster}, {Lamert}, {Lalitha}, {Laun}, {Lemke}, {Lenzen}, {L{\'o}pez del
  Fresno}, {L{\'o}pez Mart{\'\i}}, {L{\'o}pez-Santiago}, {Mall}, {Mandel},
  {Mart{\'\i}n}, {Mart{\'\i}n-Ruiz}, {Mart{\'\i}nez-Rodr{\'\i}guez}, {Marvin},
  {Mathar}, {Mirabet}, {Montes}, {Morales Mu{\~n}oz}, {Moya}, {Naranjo},
  {Ofir}, {Oreiro}, {Pall{\'e}}, {Panduro}, {Passegger}, {P{\'e}rez-Calpena},
  {P{\'e}rez Medialdea}, {Perger}, {Pluto}, {Ram{\'o}n}, {Rebolo}, {Redondo},
  {Reffert}, {Reinhardt}, {Rhode}, {Rix}, {Rodler}, {Rodr{\'\i}guez},
  {Rodr{\'\i}guez-L{\'o}pez}, {Rodr{\'\i}guez-P{\'e}rez}, {Rohloff}, {Rosich},
  {S{\'a}nchez-Blanco}, {S{\'a}nchez Carrasco}, {Sanz-Forcada}, {Sarmiento},
  {Sch{\"a}fer}, {Schiller}, {Schmidt}, {Schmitt}, {Solano}, {Stahl}, {Storz},
  {St{\"u}rmer}, {Su{\'a}rez}, {Ulbrich}, {Veredas}, {Wagner}, {Winkler},
  {Zapatero Osorio}, {Zechmeister}, {Abell{\'a}n de Paco},
  {Anglada-Escud{\'e}}, {del Burgo}, {Klutsch}, {Lizon}, {L{\'o}pez-Morales},
  {Morales}, {Perryman}, {Tulloch}, \& {Xu}}]{Quirrenbach14}
{Quirrenbach}, A., {Amado}, P.~J., {Caballero}, J.~A., {et~al.} 2014, in
  Society of Photo-Optical Instrumentation Engineers (SPIE) Conference Series,
  Vol. 9147, Ground-based and Airborne Instrumentation for Astronomy V, ed.
  S.~K. {Ramsay}, I.~S. {McLean}, \& H.~{Takami}, 91471F

\bibitem[{Rumenskikh {et~al.}(2022)Rumenskikh, Shaikhislamov, Khodachenko,
  Lammer, Miroshnichenko, Berezutsky, \& Fossati}]{Rumenskikh_2022}
Rumenskikh, M.~S., Shaikhislamov, I.~F., Khodachenko, M.~L., {et~al.} 2022,
  ApJ, 927, 238

\bibitem[{{Salpeter}(1973)}]{Salpeter1973}
{Salpeter}, E.~E. 1973, \apjl, 181, L83

\bibitem[{Salz {et~al.}(2018)Salz, Czesla, Schneider, Nagel, Schmitt, Nortmann,
  Alonso~Floriano, L{\'o}pez-Puertas, Lamp{\'o}n, Bauer, Snellen, Palle,
  Caballero, Yan, Chen, Sanz~Forcada, Amado, Quirrenbach, Ribas, Reiners,
  Bejar, Casasayas-Barris, Cortes-Contreras, Dreizler, Guenther, Henning,
  Jeffers, Kaminski, K{\"u}rster, Lafarga, Lara, Molaverdikhani, Montes,
  Morales, S{\'a}nchez-L{\'o}pez, Seifert, Zapatero-Osorio, \&
  Zechmeister}]{Salz2018}
Salz, M., Czesla, S., Schneider, P.~C., {et~al.} 2018, \aap, 620, A97

\bibitem[{Salz {et~al.}(2016)Salz, Czesla, Schneider, \& Schmitt}]{Salz2016}
Salz, M., Czesla, S., Schneider, P.~C., \& Schmitt, J. H. M.~M. 2016, \aap,
  586, A75

\bibitem[{Salz {et~al.}(2015)Salz, Schneider, Czesla, \& Schmitt}]{Salz_2015}
Salz, M., Schneider, P.~C., Czesla, S., \& Schmitt, J. H. M.~M. 2015, \aap,
  585, L2

\bibitem[{{Sanz-Forcada} {et~al.}(2011){Sanz-Forcada}, {Micela}, {Ribas},
  {Pollock}, {Eiroa}, {Velasco}, {Solano}, \&
  {Garc{\'\i}a-{\'A}lvarez}}]{san11}
{Sanz-Forcada}, J., {Micela}, G., {Ribas}, I., {et~al.} 2011, \aap, 532, A6

\bibitem[{Seidel {et~al.}(2020)Seidel, Ehrenreich, Pino, Bourrier, Lavie,
  Allart, Wyttenbach, \& Lovis}]{Seidel2020}
Seidel, J.~V., Ehrenreich, D., Pino, L., {et~al.} 2020, \aap, 633, A86

\bibitem[{{Shaikhislamov} {et~al.}(2021){Shaikhislamov}, {Khodachenko},
  {Lammer}, {Berezutsky}, {Miroshnichenko}, \&
  {Rumenskikh}}]{Shaikislamov_2021}
{Shaikhislamov}, I.~F., {Khodachenko}, M.~L., {Lammer}, H., {et~al.} 2021,
  \mnras, 500, 1404

\bibitem[{Shematovich {et~al.}(2014)Shematovich, Ionov, \&
  Lammer}]{Shematovich_2014}
Shematovich, V.~I., Ionov, D.~E., \& Lammer, H. 2014, \aap, 571, A94

\bibitem[{Sing {et~al.}(2019)Sing, Lavvas, Ballester, des Etangs, Marley,
  Nikolov, Ben-Jaffel, Bourrier, Buchhave, Deming, Ehrenreich, Mikal-Evans,
  Kataria, Lewis, López-Morales, Muñoz, Henry, Sanz-Forcada, Spake, Wakeford,
  \& collaboration)}]{Sing_2019}
Sing, D.~K., Lavvas, P., Ballester, G.~E., {et~al.} 2019, The Astronomical
  Journal, 158, 91

\bibitem[{{Smith} {et~al.}(2001){Smith}, {Brickhouse}, {Liedahl}, \&
  {Raymond}}]{smith}
{Smith}, R.~K., {Brickhouse}, N.~S., {Liedahl}, D.~A., \& {Raymond}, J.~C.
  2001, \apjl, 556, L91

\bibitem[{{Spake} {et~al.}(2021){Spake}, {Oklop{\v{c}}i{\'c}}, \&
  {Hillenbrand}}]{Spake_2021}
{Spake}, J.~J., {Oklop{\v{c}}i{\'c}}, A., \& {Hillenbrand}, L.~A. 2021, \aj,
  162, 284

\bibitem[{Spake {et~al.}(2022)Spake, Oklopčić, Hillenbrand, Knutson, Kasper,
  Dai, Orell-Miquel, Vissapragada, Zhang, \& Bean}]{Spake_2022}
Spake, J.~J., Oklopčić, A., Hillenbrand, L.~A., {et~al.} 2022, ApJL, 939, L11

\bibitem[{Spake {et~al.}(2018)Spake, Sing, Evans, Oklop{\v c}i{\'c}, Bourrier,
  Kreidberg, Rackham, Irwin, Ehrenreich, Wyttenbach, \& et~al.}]{Spake_2018}
Spake, J.~J., Sing, D.~K., Evans, T.~M., {et~al.} 2018, Nature, 557, 68

\bibitem[{Stevenson(1975)}]{Stevenson1975}
Stevenson, D.~J. 1975, Phys. Rev. B, 12, 3999

\bibitem[{Stevenson(1980)}]{Stevenson1980}
Stevenson, D.~J. 1980, Science, 208, 746

\bibitem[{Stone \& Proga(2009)}]{Stone_2009}
Stone, J.~M. \& Proga, D. 2009, \apj, 694, 205

\bibitem[{Tabernero {et~al.}(2021)Tabernero, Osorio, Allart, Borsa,
  {Casasayas-Barris}, Demangeon, Ehrenreich, {Lillo-Box}, Lovis, Pall{\'e},
  Sousa, Rebolo, Santos, Pepe, Cristiani, Adibekyan, Prieto, Alibert, Barros,
  Bouchy, Bourrier, D'Odorico, Dumusque, Faria, Figueira, Santos,
  Hern{\'a}ndez, Hojjatpanah, Curto, Lavie, Martins, Martins, Mehner, Micela,
  Molaro, Nunes, Poretti, Seidel, Sozzetti, Mascare{\~n}o, Udry, Aliverti,
  Affolter, Alves, Amate, Avila, Bandy, Benz, Bianco, Broeg, Cabral, Conconi,
  Coelho, Cumani, Deiries, Dekker, Delabre, Fragoso, Genoni, Genolet, Hughes,
  Knudstrup, Kerber, Landoni, Lizon, Maire, Manescau, Marcantonio,
  M{\'e}gevand, Monteiro, Monteiro, Moschetti, Mueller, Modigliani, Oggioni,
  Oliveira, Pariani, Pasquini, Rasilla, Redaelli, Riva, {Santana-Tschudi},
  Santin, Santos, Segovia, Sosnowska, Span{\`o}, Tenegi, Iwert, Zanutta, \&
  Zerbi}]{Tabernero2021}
Tabernero, H.~M., Osorio, M. R.~Z., Allart, R., {et~al.} 2021, A\&A, 646, A158

\bibitem[{Tian {et~al.}(2005)Tian, Toon, Pavlov, \& De~Sterck}]{Tian2005}
Tian, F., Toon, O.~B., Pavlov, A.~A., \& De~Sterck, H. 2005, \apj, 621, 1049

\bibitem[{Tripathi {et~al.}(2015)Tripathi, Kratter, Murray-Clay, \&
  Krumholz}]{Tripathi_2015}
Tripathi, A., Kratter, K.~M., Murray-Clay, R.~A., \& Krumholz, M.~R. 2015,
  \apj, 808, 173

\bibitem[{Vidal-Madjar {et~al.}(2003)Vidal-Madjar, des Etangs, D{\'e}sert,
  Ballester, Ferlet, H{\'e}brard, \& Mayor}]{VidalMadjar2003}
Vidal-Madjar, A., des Etangs, A.~L., D{\'e}sert, J.-M., {et~al.} 2003, Nature,
  422, 143

\bibitem[{Vidal-Madjar {et~al.}(2004)Vidal-Madjar, D{\'e}sert, des Etangs,
  H{\'e}brard, Ballester, Ehrenreich, Ferlet, McConnell, Mayor, \&
  Parkinson}]{VidalMadjar2004}
Vidal-Madjar, A., D{\'e}sert, J.-M., des Etangs, A.~L., {et~al.} 2004, \apj,
  604, 69

\bibitem[{Vidotto \& Cleary(2020)}]{Vidotto2020}
Vidotto, A.~A. \& Cleary, A. 2020, Monthly Notices of the Royal Astronomical
  Society, 494, 2417

\bibitem[{Vissapragada {et~al.}(2022)Vissapragada, Knutson, dos Santos, Wang,
  \& Dai}]{Vissapragada_2022}
Vissapragada, S., Knutson, H.~A., dos Santos, L.~A., Wang, L., \& Dai, F. 2022,
  ApJ, 927, 96

\bibitem[{Vissapragada {et~al.}(2020)Vissapragada, Knutson, Jovanovic, Harada,
  Oklop{\v{c}}i{\'{c}}, Eriksen, Mawet, Millar-Blanchaer, Tinyanont, \&
  Vasisht}]{Vissapragada_2020}
Vissapragada, S., Knutson, H.~A., Jovanovic, N., {et~al.} 2020, AJ, 159, 278

\bibitem[{Wang \& Dai(2021)}]{Wang_2021}
Wang, L. \& Dai, F. 2021, ApJ, 914, 98

\bibitem[{Watson {et~al.}(1981{\natexlab{a}})Watson, Donahue, \&
  Walker}]{Watson1981}
Watson, A.~J., Donahue, T.~M., \& Walker, J.~C. 1981{\natexlab{a}}, Icarus, 48,
  150

\bibitem[{Watson {et~al.}(1981{\natexlab{b}})Watson, Donahue, \&
  Walker}]{Watson_1981}
Watson, A.~J., Donahue, T.~M., \& Walker, J.~C. 1981{\natexlab{b}}, Icarus, 48,
  150

\bibitem[{{West} {et~al.}(2016){West}, {Hellier}, {Almenara}, {Anderson},
  {Barros}, {Bouchy}, {Brown}, {Collier Cameron}, {Deleuil}, {Delrez}, {Doyle},
  {Faedi}, {Fumel}, {Gillon}, {G{\'o}mez Maqueo Chew}, {H{\'e}brard}, {Jehin},
  {Lendl}, {Maxted}, {Pepe}, {Pollacco}, {Queloz}, {S{\'e}gransan}, {Smalley},
  {Smith}, {Southworth}, {Triaud}, \& {Udry}}]{West_2016}
{West}, R.~G., {Hellier}, C., {Almenara}, J.~M., {et~al.} 2016, \aap, 585, A126

\bibitem[{Wilson \& Militzer(2010)}]{Wilson2010}
Wilson, H.~F. \& Militzer, B. 2010, Phys. Rev. Lett., 104, 121101

\bibitem[{{Wyttenbach} {et~al.}(2020){Wyttenbach}, {Molli\`ere, P.},
  {Ehrenreich, D.}, {Cegla, H. M.}, {Bourrier, V.}, {Lovis, C.}, {Pino, L.},
  {Allart, R.}, {Seidel, J. V.}, {Hoeijmakers, H. J.}, {Nielsen, L. D.},
  {Lavie, B.}, {Pepe, F.}, {Bonfils, X.}, \& {Snellen, I. A.
  G.}}]{Wyttenbach_2020}
{Wyttenbach}, A., {Molli\`ere, P.}, {Ehrenreich, D.}, {et~al.} 2020, \aap, 638,
  A87

\bibitem[{Yan {et~al.}(2022)Yan, Seon, Guo, Chen, \& Li}]{dongdong_2022}
Yan, D., Seon, K.-i., Guo, J., Chen, G., \& Li, L. 2022, ApJ, 936, 177

\bibitem[{Yan \& Henning(2018)}]{Yan_2018}
Yan, F. \& Henning, T. 2018, Nature Astronomy, 2, 714

\bibitem[{Yelle(2004)}]{Yelle_2004}
Yelle, R.~V. 2004, Icarus, 170, 167

\bibitem[{Zhang {et~al.}(2023)Zhang, Knutson, Dai, Wang, Ricker, Schwarz, Mann,
  \& Collins}]{Zhang_2022c}
Zhang, M., Knutson, H.~A., Dai, F., {et~al.} 2023, The Astronomical Journal,
  165, 62

\bibitem[{Zhang {et~al.}(2022{\natexlab{a}})Zhang, Knutson, Wang, Dai, \&
  Barrag{\'{a}}n}]{Zhang_2022b}
Zhang, M., Knutson, H.~A., Wang, L., Dai, F., \& Barrag{\'{a}}n, O.
  2022{\natexlab{a}}, AJ, 163, 67

\bibitem[{Zhang {et~al.}(2022{\natexlab{b}})Zhang, Knutson, Wang, Dai, dos
  Santos, Fossati, Henry, Ehrenreich, Alibert, Hoyer, Wilson, \&
  Bonfanti}]{Zhang_2022a}
Zhang, M., Knutson, H.~A., Wang, L., {et~al.} 2022{\natexlab{b}}, AJ, 163, 68

\bibitem[{Zhao {et~al.}(2014)Zhao, O{\textquotesingle}Rourke, Wright, Knutson,
  Burrows, Fortney, Ngo, Fulton, Baranec, Riddle, Law, Muirhead, Hinkley,
  Showman, Curtis, \& Burruss}]{Zhao_2014}
Zhao, M., O{\textquotesingle}Rourke, J.~G., Wright, J.~T., {et~al.} 2014, ApJ,
  796, 115

\end{thebibliography}

\begin{appendix}

\section{\het\ concentration profiles and gas radial velocities for the different planets} \label{ap:concentrations_winds}

\begin{figure}[htbp!]
\includegraphics[angle=90.0, width=0.88\columnwidth]{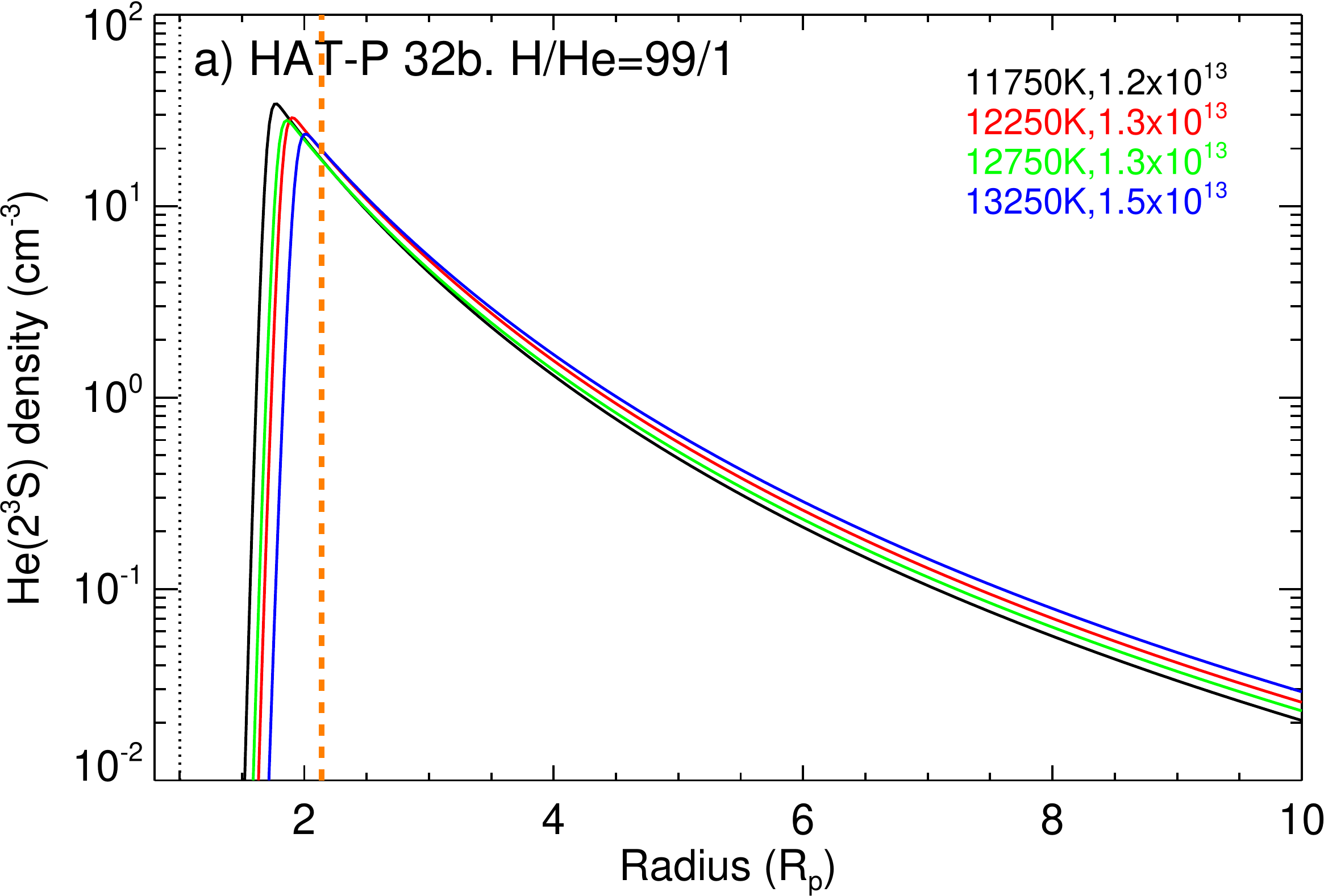} 
\includegraphics[angle=90.0, width=0.88\columnwidth]{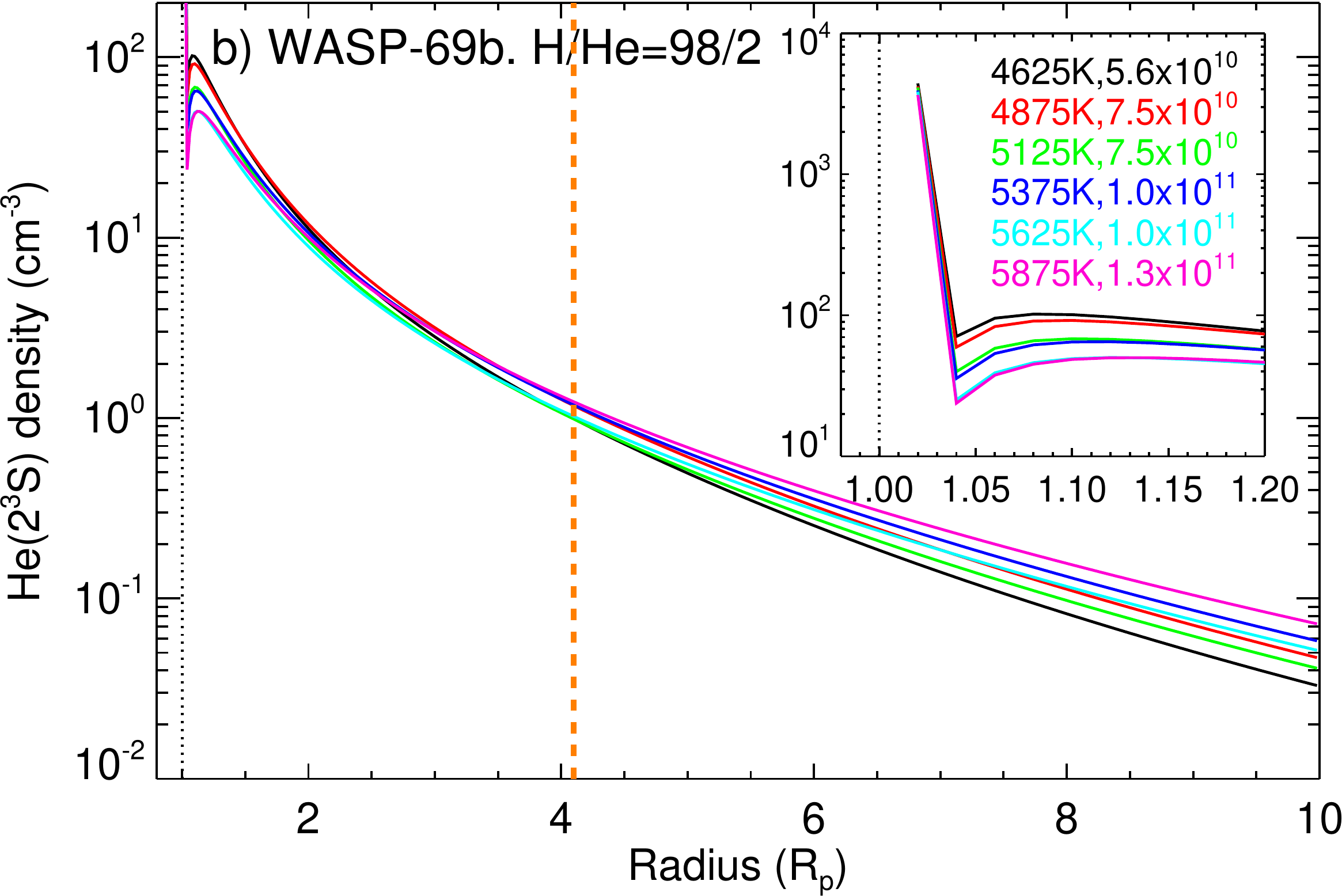} 
\includegraphics[angle=90.0, width=0.88\columnwidth]{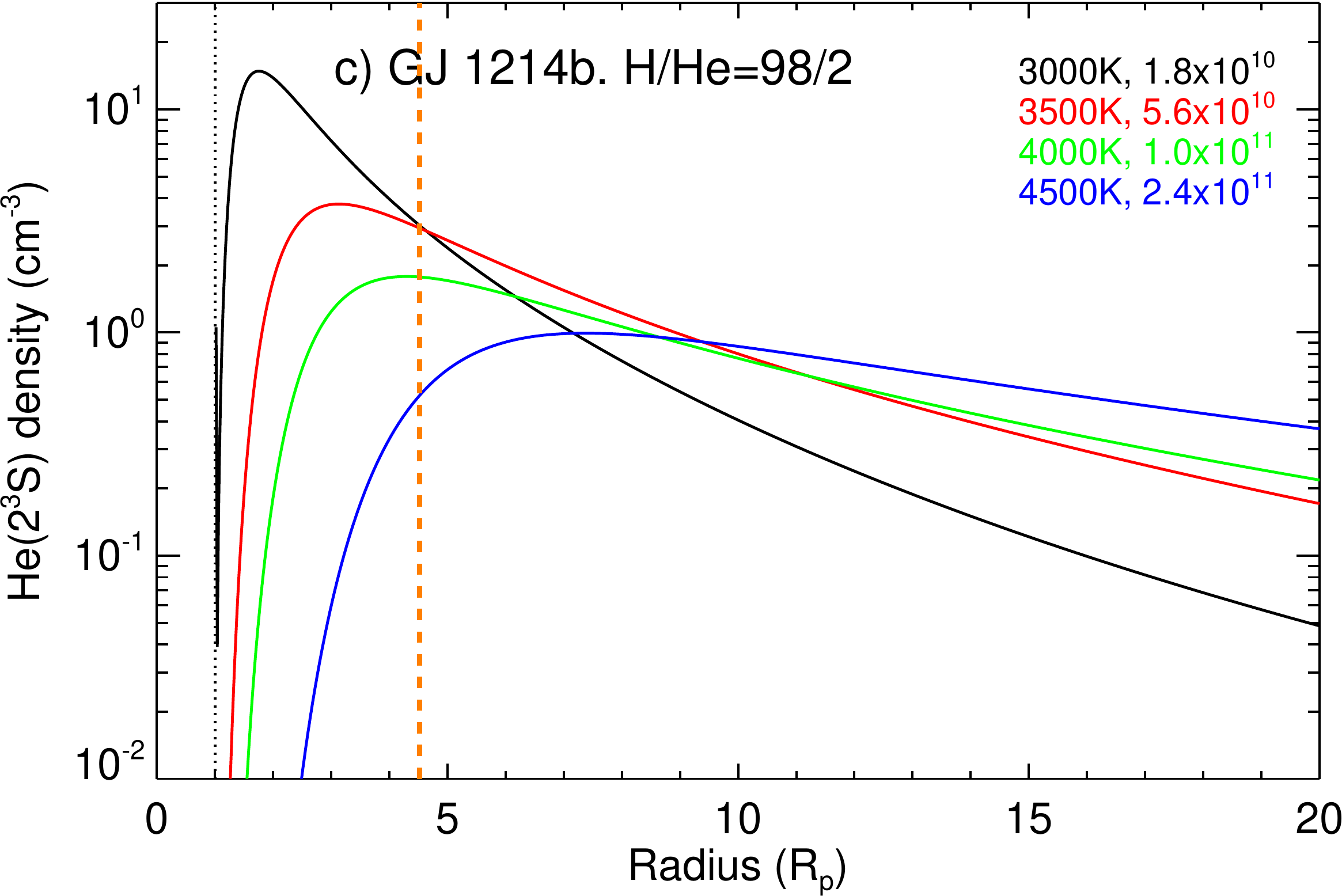} 
\includegraphics[angle=90.0, width=0.88\columnwidth]{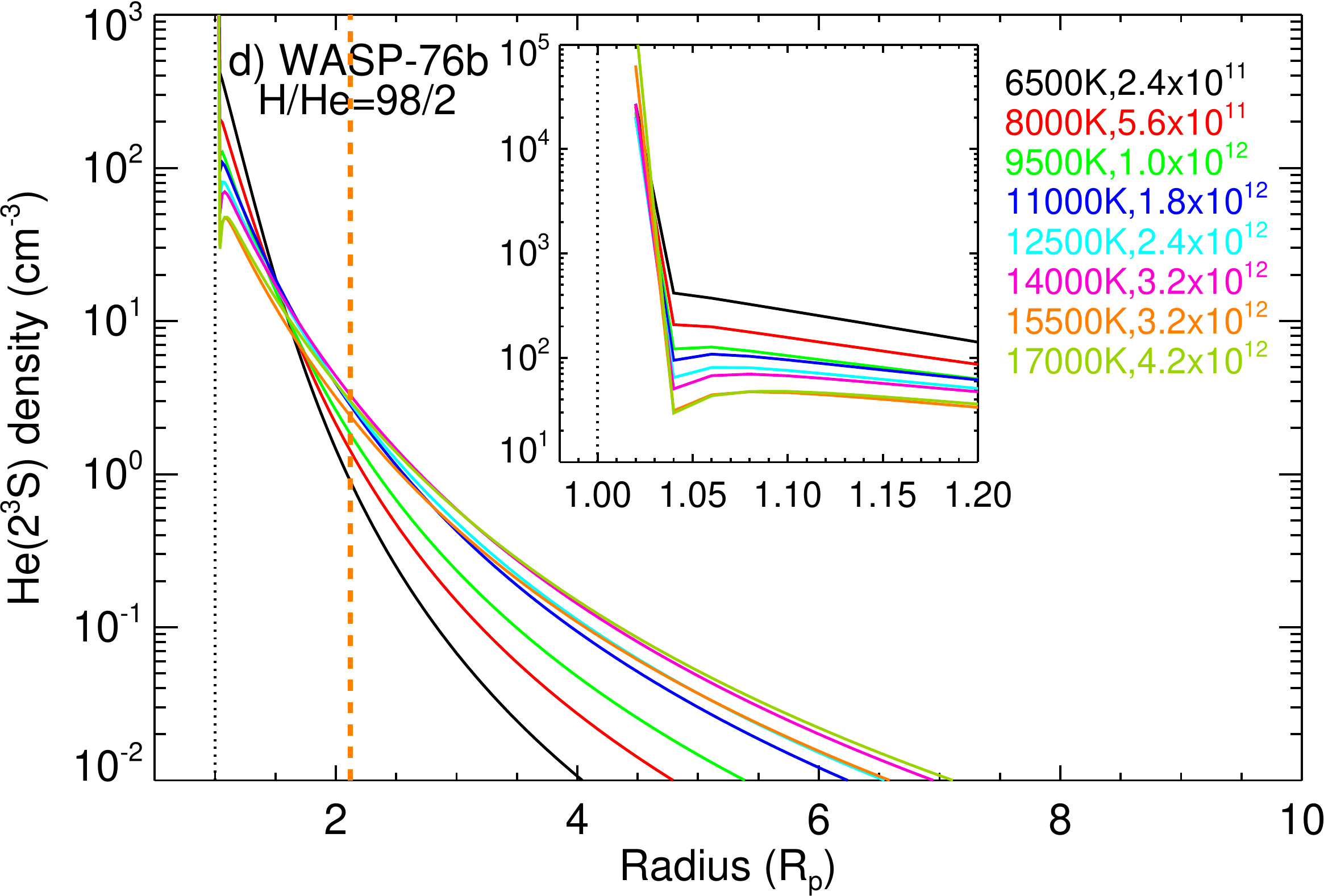}
\caption{\het\ concentration profiles that best fit the measured absorption (for the white symbols in Fig.~\ref{chi2}). The  scales of the x-axis are different. The vertical dashed orange lines indicate the mean Roche lobes. The insets in the right panels show   zoomed-in images at low radii.} 
\label{he3} 
\end{figure}

\begin{figure}[htbp!]
\includegraphics[angle=90.0, width=0.88\columnwidth]{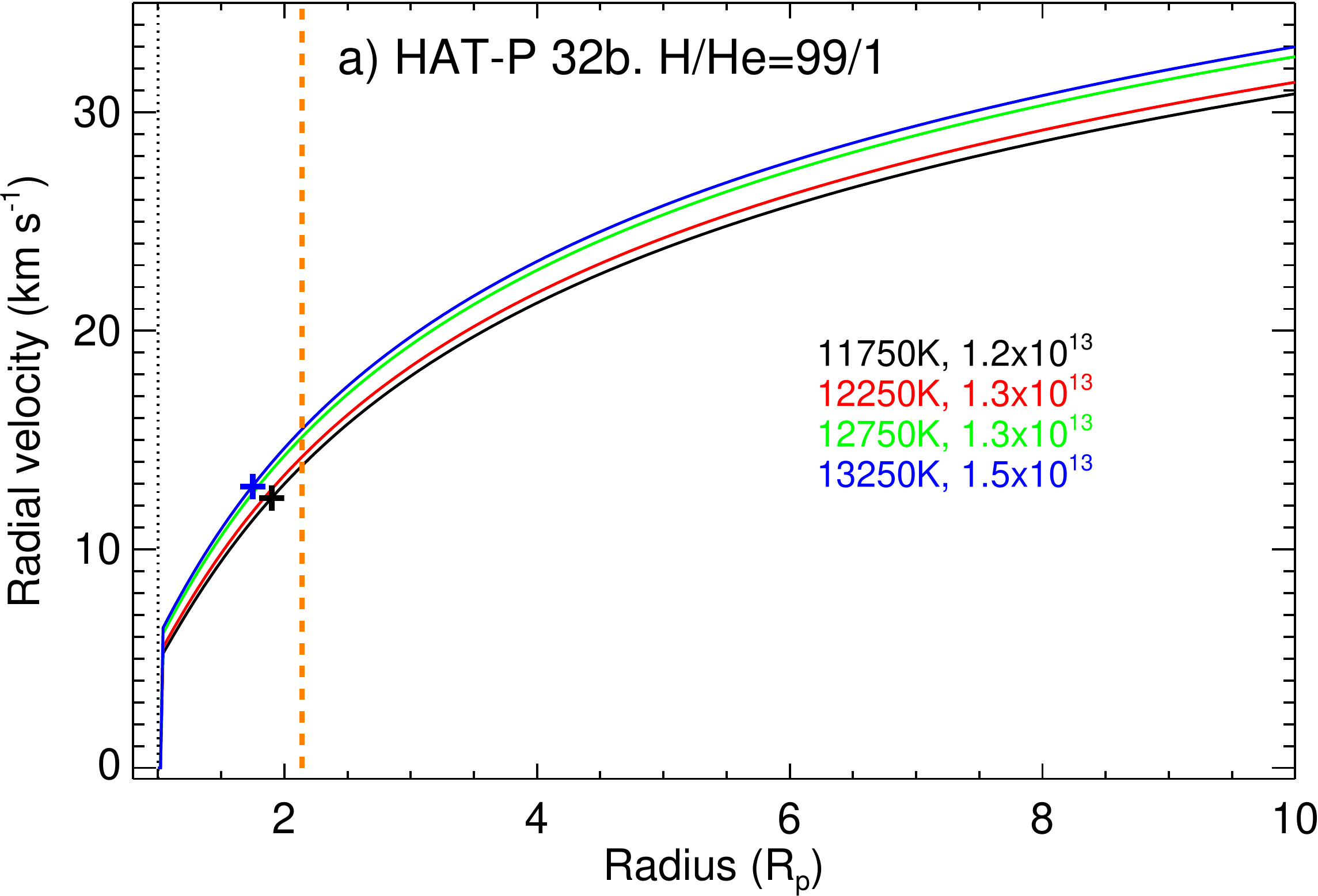}
\includegraphics[angle=90.0, width=0.88\columnwidth]{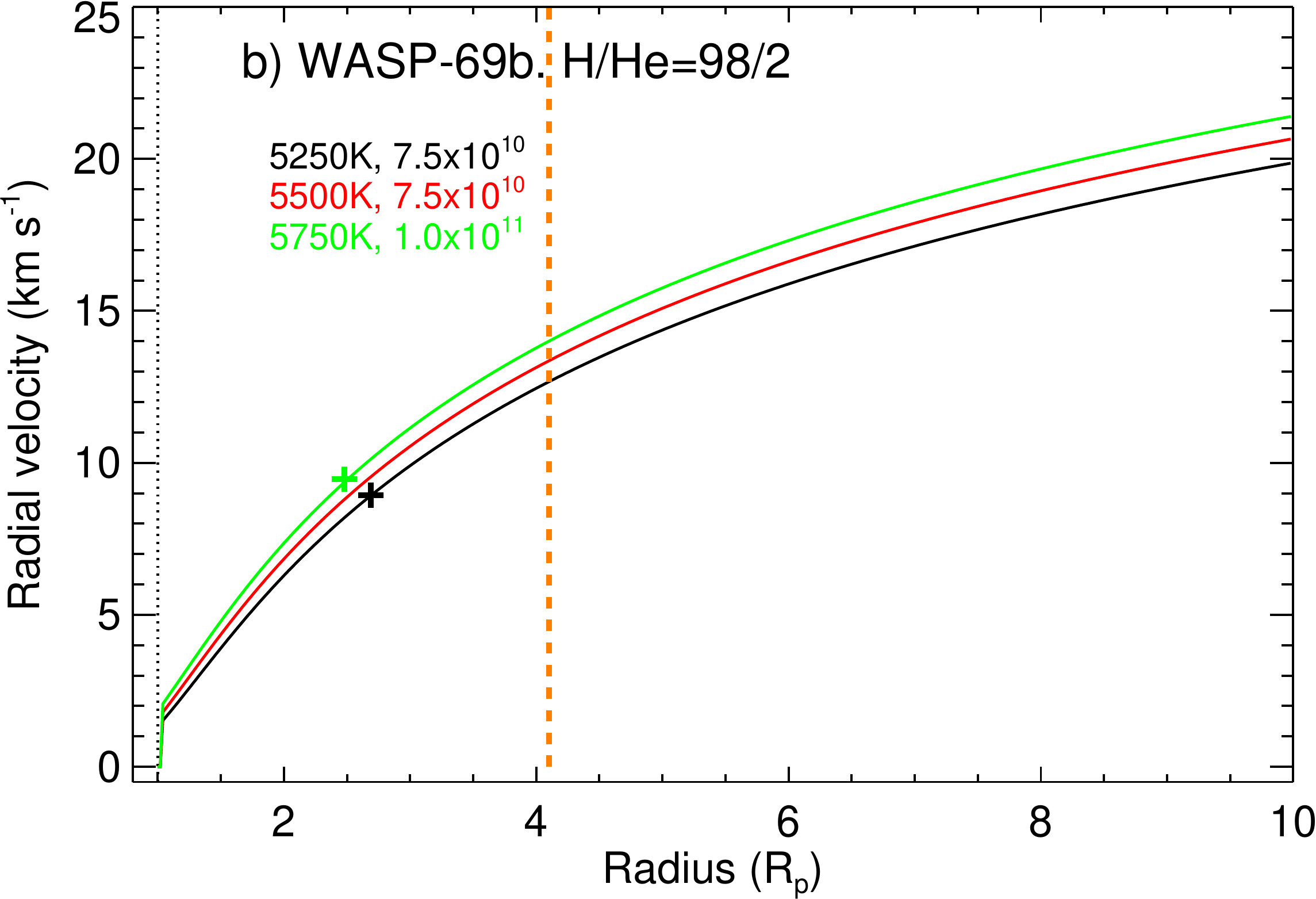}
\includegraphics[angle=90.0, width=0.88\columnwidth]{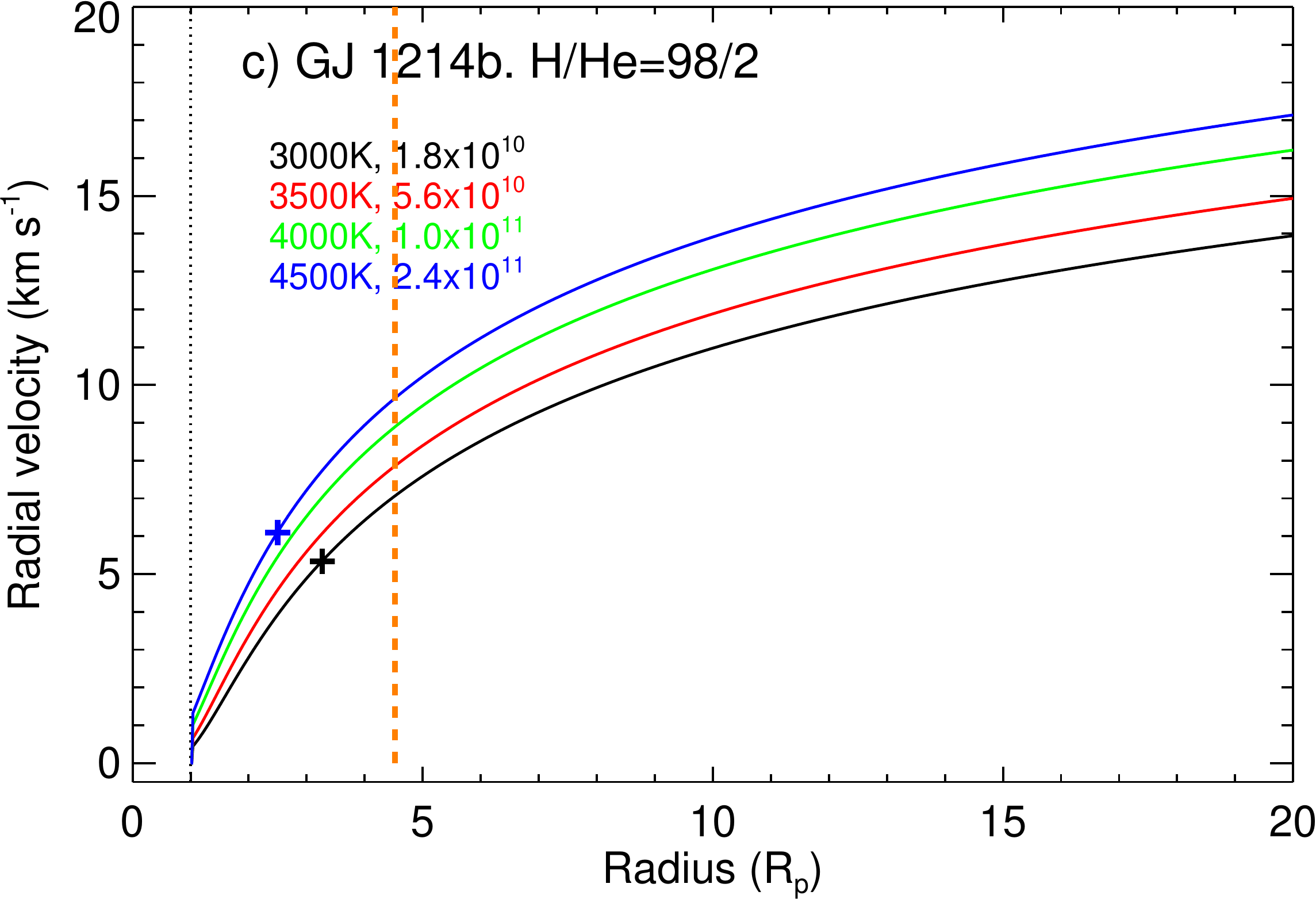}
\includegraphics[angle=90.0, width=0.88\columnwidth]{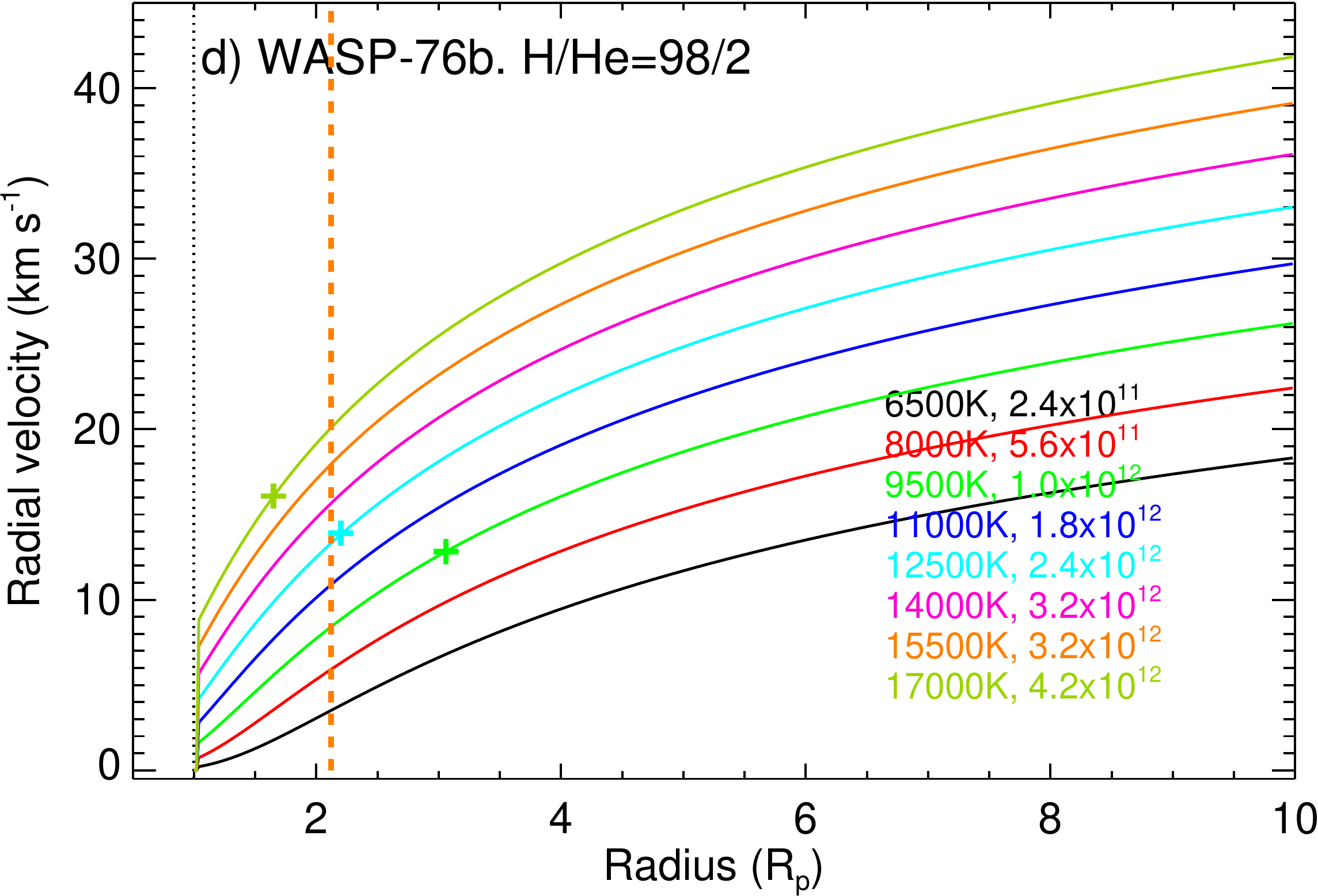}
\caption{Gas radial velocities of the model for the 
best fits of the \het\ measured absorption (for the white symbols in Fig.~\ref{chi2}). The vertical dashed orange lines indicate the mean Roche lobes and the plus signs are the sonic velocities. The scales of the x-axis are different.
} 
\label{vel} 
\end{figure}

\clearpage

\section{Maximum heating efficiency} 
\label{ap:eta_max}

The heating efficiency
can be defined as the fraction of the absorbed stellar radiative energy, {\textrm{W$_{\textrm{hv}}$}}, which is converted into kinetic energy of the gas \citep[see e.g.][]{Lampon_2021b,Shematovich_2014}:
\begin{align}
    \eta = \frac{\textrm{W}_{\textrm{hv}} - \textrm{W}_{\textrm{h}v_0} -{\textrm{W}_\textrm{c}} - \textrm{W}_{\textrm{cool}}}{\textrm{W}_{\textrm{hv}}}~.
    \label{eq:heating}
\end{align}
Here $\textrm{W}_{\textrm{hv}_{\textrm{0}}}$ is the rate of energy lost by photo-ionisation; 
W$_{\textrm c}$ is the rate of energy lost by photo-electron impact processes
\cite[as excitation and ionisation of atoms by collisions with photo-electrons, see e.g.][]{Shematovich_2014}; 
and W$_{\textrm{cool}}$ is the radiative cooling rate, mostly produced by \lya\, and free-free emissions  \citep{Salz_2015}.
By neglecting W$_{\textrm c}$ and W$_{\textrm{cool}}$,  
Eq.\,\ref{eq:heating}\, gives an upper limit of the heating efficiency: 
\begin{align}
    \eta_{\textrm {max}} = 1 - \frac{\textrm{W}_{\textrm{hv}_{\textrm{0}}}} {\textrm{W}_{\textrm{hv}}}~.
    \label{eq:heating_max}
\end{align}
Given an outflow with $K$ species and $L$ electronically excited states,  {\textrm{W$_{\textrm{hv}}$}(r)} can be expressed as \citep[see][]{Shematovich_2014}
\begin{align}
    & \textrm{W}_{\textrm{hv}}(r) = \sum_{k=1,K}\,\sum_{l=1,L} \textrm{W}_{\textrm{hv}}^{(k,l)}(r)~, {\rm with}  \nonumber \\
    & \textrm{W}_{\textrm{hv}}^{(k,l)}(r) = 
    \int_{v_{k0}}^{\infty} E_{v}\,I(v)\,\exp[-\tau(v,r)]\,\sigma_{k}^{a}\,p_{k}(v,E_{k,l})\,n_k(r)\,dv~, 
    \label{eq:heating2}
\end{align}
where $n_k$ is the neutral number density of species $k$; 
$v_{0,k}$ is the frequency corresponding to their ionisation potential;
E$_{v}$ is the energy of the photon at frequency $v$; E$_{k,l}$ is the ionisation potential of the species $k$ in the excited state $l$; $\tau$ is the optical thickness; $\sigma_{k}^{a}$ is the absorption cross-section of species $k$; and p$_{k}$ is the relative yield to form ions.

On the other hand, $\textrm{W}_{\textrm{hv}_{\textrm{0}}}(r)$ can be expressed as
\begin{align}
    & \textrm{W}_{\textrm{h}v_0}(r) = \sum_k\,\sum_l \textrm{W}_{\textrm{h}v_0}^{(k,l)}(r)~, {\rm with} \nonumber  \\
    & \textrm{W}_{\textrm{h}v_0}^{(k,l)}(r) = 
    \int_{v_{k0}}^{\infty} E_{k,l}\,I(v)\,\exp[-\tau(v,r)]\,\sigma_{k}^{i}\,p_{k}(v,E_{k,l})\,n_k(r)\,dv~, 
    \label{eq:heating3}
\end{align}
where $\sigma_{k}^{i}$ is the ionisation cross-section of species k-$th$.
Assuming every photon coming from the star ionises a ground-state hydrogen atom in the upper atmosphere, and integrating over distance \citep[see e.g. Eq.\,20 from][]{Erkaev_2007}, we obtain
\begin{align}
    \frac{\textrm{W}_{\textrm{hv}_{\textrm{0}}}} {\textrm{W}_{\textrm{hv}}} = \frac{\int_{v_{\textrm {0}}}^{\infty} E_{0}\,I(v)\,dv}{F_{XUV}}
    \label{eq:heating_max2}
\end{align}
and Eq.\,\ref{eq:heating_max} reduces to
\begin{align}
    \eta = 1 - \frac{\int_{v_{\textrm {0}}}^{\infty} E_{0}\,I(v)\,dv}{F_{XUV}}~.
\end{align}

\end{appendix}

\end{document}